%% file: paper.tex
\documentclass[10pt]{article}
\usepackage[utf8]{inputenc}
\usepackage[T1]{fontenc}
\usepackage{amssymb,amsmath,amsthm,amsfonts,amsbsy,dsfont,bm}
\usepackage{graphicx,graphics, enumerate,enumitem}
\usepackage{pdflscape}
\usepackage{afterpage}
\usepackage{booktabs, caption, multirow, float, url, array}
\usepackage{comment}
\usepackage{color}
\usepackage{indentfirst}
\usepackage[hidelinks]{hyperref} % a commenter pour Latex
\usepackage[a4paper, margin=1.25in, nohead]{geometry}
\usepackage{tcolorbox}
\newtcolorbox{mybox}{colback=red!5!white,
colframe=red!75!black}
\usepackage{natbib} % a supprimer pour Arxiv

\input{Commands} % mathematical macros.

\definecolor{darkorange}{rgb}{1.0, 0.55, 0.0}
\definecolor{black}{rgb}{0.0, 0.0, 0.0}
\colorlet{couleurModification}{black}

\newcommand{\colModification}[1]{\textcolor{couleurModification}{#1}}

\newenvironment{modification}
{	
\color{couleurModification}
}
{
\normalcolor
}

\newcolumntype{P}[1]{>{\centering\arraybackslash}p{#1}}

\title{\vspace{-0em}Bridging Methodologies: Angrist and Imbens' Contributions to Causal Identification\thanks{
We thank José~De~Sousa and an anonymous referee for their constructive feedback that helped improve greatly the presentation of the paper. 
We are responsible for any remaining error.
}}

\author{Lucas Girard\thanks{
CREST-ENSAE,
5, avenue Henry Le Chatelier,
91120 Palaiseau, France. 
lucas.girard@ensae.fr},
Yannick Guyonvarch\thanks{
INRAE, Paris-Saclay Applied Economics, Université Paris-Saclay,
place de l'Agronomie,
91120 Palaiseau, France.
yannick.guyonvarch@inrae.fr.}}

\date{22 November, 2023}

\begin{document}

\maketitle

\begin{modification}
\begin{abstract}
\noindent In the 1990s, Joshua Angrist and Guido Imbens studied the causal interpretation of Instrumental Variable estimates (a widespread methodology in economics) through the lens of potential outcomes (a classical framework to formalize causality in statistics).
Bridging a gap between those two strands of literature, they stress the importance of treatment effect heterogeneity and show that, under defendable assumptions in various applications, this method recovers an average causal effect for a specific subpopulation of individuals whose treatment is affected by the instrument.
They were awarded the Nobel Prize primarily for this Local Average Treatment Effect (LATE).
The first part of this article presents that methodological contribution in-depth: the origination in earlier applied articles, the different identification results and extensions, and related debates on the relevance of LATEs for public policy decisions.
The second part reviews the main contributions of the authors beyond the LATE.
J.~Angrist has pursued the search for informative and varied empirical research designs in several fields, particularly in education.
G.~Imbens has complemented the toolbox for treatment effect estimation in many ways, notably through propensity score reweighting, matching, and, more recently, adapting machine learning procedures.
\end{abstract}
\end{modification}

\medskip 

\noindent
\begin{small}
\textbf{Keywords:} 
instrumental variables (IV); 
Neyman-Rubin causal model;
local average treatment effect (LATE); 
natural experiments;
returns to schooling;
US educational system;
propensity score matching; 
causal machine learning.
\end{small}

\begin{comment}
en français :
variables instrumentales (IV) ;
modèle causal de Neyman-Rubin ;
effet local moyen du traitement (LATE) ;
expériences naturelles ;
rendements de l’éducation ;
système éducatif états-unien ;
appariement sur le score de propension ;
apprentissage statistique causal.

titre en français :

A la croisée de 

A la frontière de l’économétrie / économie et de la statistique 

économie plus qu'économétrie pour l'aspect interprétation, attention au sens économique

A la croisée de l'économie et de la statistique : les contributions d’Angrist et d’Imbens à l’identification causale

Entre économie et statistique : les contributions d’Angrist et d’Imbens à l’identification causale

Unifier des méthodologies : les contributions d’Angrist et d’Imbens à l’identification causale

Réunir économie et statistique : 

Rapprocher économie et statistique : 

Faire dialoguer économistes et statisticiens : les contributions d’Angrist et d’Imbens à l’identification causale

\end{comment}

\section{Introduction}
\label{sec:introduction}

In the mid-1990s, Professors Joshua Angrist and Guido Imbens contributed to a series of foundational papers \citep{imbens_angrist_1994, angrist_imbens_1995, angrist_imbens_rubin_1996} which brought together two powerful yet seemingly distant concepts, namely Neyman-Rubin's potential outcome framework \citep{rubin1974estimating, rubin1990} and instrumental variables \citep{wright_1928, Reiersl1945ConfluenceAB}.
At the time, the former was mainly popular in medical studies with random allocation of treatment to patients.
In that field, the idea that each patient is faced with two potential health statuses, one with treatment and one without, was already widely accepted.
Thanks to a random treatment allocation, comparisons across treatment groups allow for uncovering the average change in health status due to treatment, also termed the average causal effect of the treatment.
This approach does not require imposing strong {a priori} assumptions on the nature of the link between treatment and health status, which is appealing. 

Unlike medical researchers, economists were (and still are) mostly faced with observational data, namely data collected on agents who make choices in uncontrolled environments.
In this context, it is the rule rather than the exception to witness agents who make interdependent and sometimes simultaneous choices.
Researchers try to identify the impact of one choice on another in a \textit{ceteris paribus} way, that is, holding the rest of the environment fixed.
Unfortunately, co-occurring choices are often based on factors unobserved by the researcher, which makes it hard to disentangle causes and consequences leading to the well-known endogeneity issue (see \cite{Wooldridge2010} for an introduction). 
As an example, let us consider the question: 
``What is the impact of education years on potential wage?''
Intrinsic motivation, which is hardly captured in collected data, can affect agents' decisions both regarding education and labor market outcomes. 
Without more information, it is thus impossible to interpret correlations between observed educational choices and 
%(potential) 
wages as evidence of the \textit{ceteris paribus} influence of education on wages.
A long-established solution is to resort to instrumental variables (IV) plus the assumption that choices affect one another linearly, leading to the linear IV model.
In the education-wage example, IVs are, loosely speaking, variables that capture variations in education unrelated to motivation\colModification{; for instance, the distance between home and the nearest college has been used in the literature \citep{card1993using}.}
Such variables enable researchers to identify the effect of education on wages holding motivation fixed. 

\medskip

The popularity of the linear IV model among applied economists largely comes from its ease of implementation and the simplicity of the model connecting the endogenous variable (the number of years of education in the previous example) and the outcome (wages).
Having a simple-to-use estimator is an unquestionable advantage from a practical perspective.
The simplicity of the underlying model has a major downside, however.
As an illustration, a linear model of wages in terms of education implies that the \textit{ceteris paribus} effect of education on wages is assumed to be the same for every individual in the population.
%(see Section~\ref{subsubsec:comparison_LATE_SEM} for more details on IV models). 
In other words, basic linear models impose individuals be perfectly homogeneous in their response to a given (economic) stimulus.
This is the starting point of J.~Angrist and G.~Imbens's common research agenda of the 1990s.
Roughly speaking, these authors ask: ``Assume one runs a linear IV regression while the true underlying model is not linear. Does the estimated coefficient on the endogenous variable still capture an economically meaningful quantity?''
Addressing this ambitious question is not straightforward and requires, in particular, defining what is meant by economically meaningful. 
To do so, the authors propose to extend the canonical Neyman-Rubin potential outcome framework to make it compatible with the non-experimental designs in which IV models are typically applied.
The economic quantities they are primarily interested in are average treatment effects, as in the classical Neyman-Rubin setup.
Their idea is fairly simple and turns out to be extremely powerful in analyzing the properties of linear IV estimators in populations of heterogeneous agents.
As we will see in Section~\ref{sec:LATE}, the authors give a partially positive answer to the question presented above: linear IV can recover average treatment effects on some specific subgroups of individuals provided instruments are exogenous, relevant, and restrictions are placed on the direction in which instruments affect the endogenous variables; the so-called monotonicity assumption.
On the other hand, treatment effects can remain fully heterogeneous at the individual level.
Since only average effects for certain subpopulations can be identified, the authors called those \emph{local} average treatment effects (LATE), a term now ubiquitous in economics.

\medskip

Several aspects of J.~Angrist and G.~Imbens's message about their LATE-type identification results are often overlooked.
Even though they claim that LATEs are, in some sense, the best one can hope to identify using IV methods when treatment effects' heterogeneity is not restricted, they unambiguously urge empiricists to reflect upon the relevance of the instruments used in any particular application.
They emphasize the need for researchers to first focus on obtaining data with clean-enough sources of exogenous variation in the instruments, and they give detailed accounts of situations where the monotonicity of the instrument may or may not be satisfied.
%(\cite{imbens_angrist_1994}).
They also acknowledge that LATE may or may not be informative about average treatment effects in the entire population.
To get round this intrinsic limitation, they encourage applied researchers to replicate IV estimations in varied scenarios to gain insight into the potential to extrapolate from LATE to more general average treatment effects. 
%\citep{imbens2010}. 
In J.~Angrist and G.~Imbens's eyes, the LATE paradigm only makes sense when it is combined with cautious empirical practices and rich enough sources of variation in the data at hand \citep{imbens2010, angrist_pischke_2010}.
%The new ideas promoted by J.~Angrist and G.~Imbens have been welcomed with skepticism by some highly-respected econometricians and economists such as James Heckman or Angus Deaton (ADD REFS).

\medskip

As made explicit by the Nobel committee, Joshua Angrist and Guido Imbens have been awarded the prize primarily for their breakthrough findings around the LATE \citep{KVA_Nobel_2021}.
However, their contributions extend way beyond this series of papers.
J.~Angrist is a recognized labor and education economist. He is renowned in particular for his work on the link between family composition and labor market outcomes \citep{angrist_evans_1998}, the labor market prospects of war veterans \citep{angrist_1990}, the impact of class size on educational achievement \citep{angrist_lavy_1999}, or the effect of additional education years on earnings \citep{angrist_krueger_1991}.
G.~Imbens is primarily a theoretical econometrician. 
He has worked in many different fields, including calibration methods inspired by survey sampling \citep{imbens1992}, efficient estimation in models identified through moment conditions \citep{imbens_johnson_spady_1998}, and causal inference at large. 
His contributions to the latter literature are numerous.
He has worked on many methodological issues related to the estimation of average treatment effects, including propensity score reweighting \citep{hirano_et_al_2003}, matching \citep{abadie_imbens_2006}, regression discontinuity \citep{imbens_lemieux_2008}, and synthetic control \citep{arkhangelsky_et_al_2021}.
He has also worked on the identification of policy-relevant parameters in instrumental variable models \citep{imbens_newey_2009} and on quantifying treatment effects' heterogeneity \citep{athey_imbens_2015}.
The academic influence of these two authors is also to be seen in the two reference econometrics textbooks \citep{angrist_pischke_2008, imbens_rubin_2015} they have co-authored with Jörn-Steffen Pischke and Donald Rubin, respectively.

\medskip

The rest of this article is organized as follows.
Section~\ref{sec:LATE} is devoted to a detailed presentation of the ``LATE revolution'' that J.~Angrist and G.~Imbens initiated in the 1990s.
It aims at giving a comprehensive account of the rich set of results put forward by the authors in the series of articles that gave birth to the LATE paradigm.
The theory framed indeed extends way beyond the canonical LATE theorem with one binary treatment and one binary instrument traditionally presented in econometrics courses. 
In Section~\ref{sec:beyond_LATE}, we discuss the numerous other fields to which these authors have contributed throughout their careers. 
\colModification{Section~\ref{sec:conclusion} concludes highlighting the similarity between the questions (and answers) brought by J.~Angrist and G.~Imbens on linear IV models and current issues about the interpretation of the parameters identified in two-way fixed effects models.}

\newpage
\section{A LAsTing rEvolution}
\label{sec:LATE}

This section focuses on the LATE methodological contribution, by which we refer to the framework, assumptions, and identification results introduced by J.~Angrist and G.~Imbens along with D.~Rubin, mainly in the three following articles (the ``LATE trilogy''):
\begin{itemize}[noitemsep,nolistsep]

\item[--] 
``Identification and Estimation of Local Average Treatment Effects'', \textit{Econometrica},
%1994
%, Guido Imbens, Joshua Angrist 
\cite{imbens_angrist_1994} (henceforth IA94),
%\footnote{
%We preserve the ordering of the authors that happens to be distinct for these two papers.
%}

\item[--]
``Two-Stage Least Squares Estimation 
of Average Causal Effects in Models 
With Variable Treatment Intensity'',
\textit{Journal of the American Statistical Association}, 
%1995,
%Joshua Angrist, Guido Imbens
\cite{angrist_imbens_1995} (henceforth AI95),

\item[--]
``Identification of Causal Effects Using 
Instrumental Variables'',
\textit{Journal of the American Statistical Association},
%1996,
%Joshua Angrist, Guido Imbens, Donald Rubin
\cite*{angrist_imbens_rubin_1996} (henceforth AIR96),

\end{itemize}
and direct extensions.

\medskip 

The fundamental ideas of the LATE contribution are now standard in many econometrics courses and textbooks, and various references are available.
In the context of this review, a good reference, for instance, is the article of the Royal Swedish Academy of Sciences on the scientific background of the Nobel 2021 Prize \citep{KVA_Nobel_2021}.
For completeness, Appendix~\ref{appendix:sec:remindersOLSIVCausality} recalls basic notions of the Neyman-Rubin potential outcomes framework and their relations to Ordinary Least Square (OLS) and Instrumental Variables (IV) or Two-Stage Least Squares (TSLS) in econometrics.
Then, it presents the textbook LATE theorem, that is, the identification of the average effect on compliers in the setting of a binary treatment, a binary instrument, and no covariates.
That result may be familiar to many readers.
Hence our choice to report this more standard material as annexes and devote this section to 
(\textit{i})~locate the different settings and results of the three seminal articles (IA94, AI95, AIR96), which extend the textbook LATE theorem (Section~\ref{subsec:LATE_trilogy});
%\footnote{Although it is perhaps more of historical relevance, it remains interesting in this present survey-type article to be precise about the contributions of each paper.}
(\textit{ii})~investigate the origination of these ideas that can be found in earlier applied articles (Section~\ref{subsec:early_LATEs});
(\textit{iii})~discuss some specific points more in-depth, notably those related to controversies about the interest of the LATE and the opposition between ``structural'' and ``causal'' approaches (Section~\ref{subsec:discussion_comments_criticisms_answers_LATE}).
Before doing so, we briefly recall the setting, main assumptions, and insights of the LATE contribution.
Appendix~\ref{appendix:sec:remindersOLSIVCausality} presents the framework and notations with more details for curious readers or less familiar with this formalization.

\paragraph{Setting and notation}

We are interested in the causal effect of a treatment~\(D\) on an outcome variable~\(Y\).
Following Neyman-Rubin's model, we introduce the following random variables.
\begin{itemize}[noitemsep,nolistsep]

\item
\(Z\) is the instrument; it is a scalar real random variable that will be binary, namely \(\supportde{Z} = \{0,1\}\), or discrete, multi-valued with finite support; in this case, the support of~\(Z\) is \(\{z_0, z_1, \ldots, z_K\}\), also encoded as \(\{0, 1, \ldots, K\}\).

\item
For any~\(z \in \supportde{Z}\), \(D(z)\) denotes the potential treatment variable: what would be the treatment status of the unit if its instrument took value/modality~\(z\).

\item 
\(D := D(Z)\) is the \textit{observed} (as opposed to potential) \textit{treatment} variable;
like \(D(z)\), it is a scalar real random variable that will be binary (\(\supportde{D} = \{0,1\}\)), or discrete with a quantitative ordered meaning and taking values in a finite set~\(\{0, 1, \ldots, J\}\), or a continuous random variable with \(\supportde{D} = [0, +\infty)\).

\item
For any~\(z \in \supportde{Z}\) and~\(d \in \supportde{D}\), \(Y(z,d)\) is the potential outcome variable: what would be the outcome of the unit had it got \(Z = z\) for the instrument and \(D = d\) for the treatment; under the exclusion restriction (see Assumption~\eqref{eq:assumptionExclusionRestriction} below), it does not depend on the instrument so that \(Y(z,d)\) does not depend on~\(z\) and is simply denoted~\(Y(d)\).

\item 
\(Y := Y(Z, D(Z))\) or, under the exclusion restriction, \(Y := Y(D)\), is the observed outcome variable; like \(Y(d,z)\), \(Y\) is a real random variable that can be binary (qualitative outcome with two modalities) or continuous (quantitative outcome).

\item 
If present, \(X\) is a vector of observed covariates: each covariate only takes a finite number of distinct values; that is, we restrict to discrete covariates.
%a scalar or multi-dimensional random variable of observed covariates; it will be discrete covariates taking a finite number of distinct values.

\end{itemize}
We additionally define individual causal effects as differences in potential variables.
For instance, under the exclusion restriction and with a binary treatment, \(\Delta := Y(1) - Y(0)\) is the individual causal/treatment effect (used as synonyms) of~\(D\) on~\(Y\).
\(\Delta\) is a real random variable and, a priori, with an unrestricted distribution (\textit{heterogeneous} effects as opposed to \textit{homogeneous} causal effects that assume \(\Delta = \deltazero \in \Reels\) almost surely).

\medskip

We observe an independent and identically distributed (i.i.d) sample \((Y_i, D_i, Z_i, X_i)_{i=1, \ldots, n}\) %(or without~\(X\) if there is no covariate) 
drawn from the population of interest.
All the previous variables are defined at the level of an individual/agent/(statistical) unit indexed by~\(i \in \{1, \ldots, n\}\).
%individual, agent, or (statistical) unit (used as synonyms), indexed by~\(i \in \{1, \ldots, n\}\).
Thanks to i.i.d-ness and to lighten notation, we often omit the index~\(i\): a random variable not indexed by~\(i\) denotes a generic instance with the same distribution.
For instance, \((Y, D, Z, X)\) has the same joint distribution, denoted \(\Distributionde{(Y, D, Z, X)}\), as any observation \((Y_i, D_i, Z_i, X_i)\) for~\(i \in \{1, \ldots, n\}\).
Potential variables are also defined at the individual level.
For instance, \(D(1)_i\) is unit~\(i\)'s treatment had it received~\(Z_i = 1\) for the instrument; \(D(1)\) is the same for an arbitrary individual of the population of interest.
Likewise, \(\Delta_i\) is the individual causal effect for unit~\(i\).
% Note that, although counterfactual and not observed, %(except one of them in a sense)
% the potential variables are also defined at the individual level and implicitly indexed by~\(i\)
% For instance, \(D(1)_i\) is what would be unit~\(i\)'s treatment had it had~\(Z_i = 1\) for the instrument; \(D(1)\) is the same for an arbitrary individual of the population of interest.
% Likewise, \(\Delta_i\) is the causal effect of the binary treatment~\(D\) on the outcome~\(Y\) for individual~\(i\).

\medskip 

\textit{Miscellaneous notations.}
For any random variables~\(A\) and \(B\), \(\Distributionde{A}\) denotes the %(marginal, and possibly joint if~\(A\) is multi-dimensional) 
distribution of~\(A\), and \(\Distributiondecond{A}{B}\) the conditional distribution of~\(A\) knowing~\(B\); \(\supportde{A}\) denotes the support of~\(\Distributionde{A}\).
\(\indep\) denotes independence between random variables; \(A \indep B\) means that \(\Distributiondecond{A}{B} = \Distributionde{A}\).
\(\Probabilite\), \(\Esperance\), etc. denote the probability, expectation, etc. operator.
The symbol \(:=\), as opposed to \(=\), denotes equality by definition (the left-hand term is defined by the right-hand term; or the reverse with~\(=:\)).
\(f\) or \(g\) denote a function (not necessarily explicitly defined) used locally for reasoning.
%For information, Footnote~\ref{footnote:explanationsNotationsSymbols} in Appendix~\ref{appendix:sec:remindersOLSIVCausality} (page~\pageref{footnote:explanationsNotationsSymbols}) further discusses the notation used.

\paragraph{Main assumptions}

The following assumptions will prove central.
We refer to them as the ``LATE assumptions''.
\begin{itemize}[noitemsep,nolistsep]

\item 
Exclusion restriction (the instrument affects the outcome only through the treatment):
\begin{equation*}  
\label{eq:assumptionExclusionRestriction}
\tag{\textnormal{E}}
\forall z, z' \in \supportde{Z}, \,
\forall d \in \supportde{D}, \,
Y(z,d) = Y(z',d).
\end{equation*}

\item 
Independence (of the instrument; it can be considered ``as-if'' randomized):
\begin{equation*}
\label{eq:assumptionIndepedenceInstrument_bodypaper}
\tag{\textnormal{I}}
Z \indep \big\{ (D(z))_{z \in \supportde{Z}}, (Y(z, d))_{z \in \supportde{Z}, d \in \supportde{D}} \big\}.
\end{equation*}

\item 
Relevance (of the instrument; it has a causal effect on the treatment): \colModification{when \(Z\) is binary, it can be written as}
\begin{equation*}
\label{eq:assumptionRelevanceInstrumentBinaryZ_bodypaper}
\tag{\textnormal{R}}
\Esperancede{D(1) - D(0)} \neq 0.    
\end{equation*}
\colModification{With more general instruments, this assumption can be formulated in several ways. We introduce such extensions on a case-by-case basis in the rest of the article.} %depending on the setting, this assumption is formalized in different ways formulated below when they are used; 
\item Monotonicity (of the effect of the instrument on the treatment):
\begin{equation*}
\label{eq:assumptionMontonicityMultivaluedZ_bodypaper}
\tag{\textnormal{M}}
    \forall z, w \in \supportde{Z}, \,
    \text{ either } \,
    D(z) \geq D(w) \text{ almost surely (a.s)}
    \, \text{ or } \,
    D(z) \leq D(w) \text{ a.s},
\end{equation*}
\colModification{which, with a binary instrument (and a conventional choice for the ranking), amounts to}
\begin{equation*}
\label{eq:assumptionMontonicityBinaryZ}
\tag{\textnormal{M'}}
    \colModification{D(1) \geq D(0) \text{ almost surely.}}
\end{equation*}

\end{itemize}

\begin{modification}

\paragraph{Definition of relevant subpopulations}

According to their potential treatment variables, individuals or units can be partitioned into several subpopulations which play a crucial role in the definition and interpretation of LATE parameters. It is useful to focus on the case of a binary instrument and binary treatment to describe these populations in an intuitive fashion. In this canonical setting, individuals or units can be partitioned into four types depending on their potential treatment variables (Table~\ref{tab:table_NT_C_D_AT}).

\begin{table}[H]
    \centering
    \caption{Partition of the population with a binary instrument and a binary treatment.}
    \begin{tabular}{l | c |  c }
         & \({D(1)} = 0\) & \({D(1)} = 1\)  \\
        \hline 
        \({D(0)} = 0\) & never-taker (NT) & complier (C) \\
        \hline 
        \({D(0)} = 1\) & defier (D) & always-taker (AT)
    \end{tabular}
    \label{tab:table_NT_C_D_AT}
\end{table}
% Remarque: les abbréviations NT, AT, D, C ne sont a priori pas utilisées dans le corps du texte mais repris comme superscript pour nommer les delta donc à laisser.

Never- and always-takers are individuals whose treatment status is not affected by the instrument.
To resume the education-wage example of the introduction, 
%simplified into that binary set-up, 
the treatment~\(D\) could be defined as the indicator of going to college and the instrument~\(Z\) as the indicator of living less than, say, 10 miles away to a college.
An individual is an always-taker (respectively never-taker) if, whatever the distance between the nearest college and her home, she goes (does not go) to college.
In contrast, the instrument does affect the treatment of compliers and defiers.
For instance, in our example, a complier is defined by the fact that she would go to college had she lived close enough to a college but, all other things equal, would not go had she lived further. 

In the simple case of \(\supportde{Z} = \supportde{D} = \{0, 1\}\), \eqref{eq:assumptionMontonicityBinaryZ} is equivalent to the absence of defiers: if the instrument affects the treatment, it must do so in the same direction for each individual.
In our example, living close to a college cannot induce anyone not going to college.

\paragraph{Comments on the assumptions and insights}

As just explained, for compliers, the instrument has an effect on the treatment (see the relevance assumption~\eqref{eq:assumptionRelevanceInstrumentBinaryZ_bodypaper} above).
In addition, the second condition for a valid instrument relates to its exogeneity:
Can it be considered as-if randomly assigned like in a controlled experiment (see independence assumption~\eqref{eq:assumptionIndepedenceInstrument_bodypaper} above)? 
Does it affect the outcome only through the treatment, without ``direct'' effect (see exclusion assumption~\eqref{eq:assumptionExclusionRestriction})?

In a nutshell, the LATE methodological contribution presented at length below can be formulated as follows.
Without restricting the heterogeneity of treatment effects, 
(\textit{i})~the identification of a sensible causal parameter requires that the instrument (weakly) affects all individuals \textit{in the same direction}
(see the monotonicity assumption~\eqref{eq:assumptionMontonicityMultivaluedZ_bodypaper} above),
(\textit{ii}) information learned on causal effects only concerns individuals whose treatment is affected by the instrument; in that sense, the average causal effect recovered is only \emph{local}, it is the average \emph{on the subpopulation of compliers}.
That is why the complier subpopulation is central in the interpretation of LATE-type results.

\end{modification}

\subsection{Early LATEs?}
\label{subsec:early_LATEs}

The trilogy IA94, AI95, and AIR96 forms the core of the LATE methodological contribution pioneered by J.~Angrist (J.A) and G.~Imbens (G.I) along with D.~Rubin.
%These three articles are exceptionally concentrated over time, with the initial manuscript of IA94 received in December 1991, of AI95 in August 1992, and of AIR96 in June 1993.
Interestingly, some of the key ingredients of the LATE contribution can be found in earlier papers.
This subsection reviews a number of articles and ideas that were forerunners of the LATE revolution: origins of the potential outcomes paradigm and its early use in economics (Section~\ref{subsubsec:use_potential_outcomes}), concerns about possibly heterogeneous causal effects (Section~\ref{subsubsec:concern_heterogeneity}), and the attention devoted to the exclusion restriction (Section~\ref{subsubsec:caution_exclusion_restriction}).
%It is also the opportunity to present and discuss two famous articles by J.~Angrist.

\subsubsection{The (non-)use of potential outcomes in economics}
\label{subsubsec:use_potential_outcomes}

\paragraph{The tradition of potential outcomes in statistics}

As outlined above, %in the introduction (Section~\ref{sec:introduction}),
one of the main ways to present the LATE contribution is that it builds a bridge between statistics' potential outcome framework and instrumental variables used by economists in a setting of simultaneous (structural) equation models.
The authors themselves claim that connection.\footnote{
In particular, it is noteworthy that they conclude their rejoinder to AIR96's comments (following the article, the issue presents comments by James Robins and Sander Greenland, James Heckman, Robert Moffitt, and Paul Rosenbaum) on this theme: 
``{The comments on our article cover a wide range of opinions, partly reflecting the gap between competing paradigms for evaluation research in statistics and econometrics. 
We believe that the gap between the two approaches can be narrowed. 
We hope that our article will make statisticians more appreciative of the insights offered by the IV framework invented by econometricians, while making economists more aware of the benefits of causal inference conducted in the potential outcomes framework developed by statisticians}'' (\textit{Journal of the American Statistical Association} June 1996, Vol.~91, No.~434, page~472).
}
In this respect, it is interesting to remark that the first paper of the trilogy was published in a leading economics journal; in contrast, the two others were published in a leading statistics journal.

Donald Rubin's work \citep{rubin1974estimating, rubin1978} is central in the development of the potential outcome framework since, although the notion can be found in the ``potential yields'' of \cite{neyman1923}, D.~Rubin generalizes it beyond randomized experiments to include observational studies that do not involve randomization.
Causal inference in statistics (and related fields such as epidemiology) has been inherently connected to the notion of potential outcomes \textit{a la} Neyman-Rubin at least since Rubin's articles were published.
%Indeed, potential outcomes and individual causal effects of a binary treatment~\(D\) on a quantitative outcome variable~\(Y\) are jointly defined.
%For a given individual or unit, a causal effect is defined as the difference between its two potential outcomes, which represent what would have been the outcome variable in the presence of the treatment and in the absence of the treatment.
In contrast, potential outcomes were not very common in economics at the time of the redaction of the LATE trilogy, with a few notable exceptions \citep{maddala1983, bjorklund1987, heckman1990}.
We argue below this gap was meant to be filled due to the strong natural connections between potential outcomes and economic reasoning and the joint interest of G.~Imbens in statistics and economics.
%At the time of the redaction of the LATE trilogy, potential outcomes were not common in economics, although that framework had started to be adopted in economic models \citep{maddala1983, bjorklund1987, heckman1990}.
%We do not engage in retracing the history of potential outcomes in economics but simply present two remarks regarding potential outcomes and the origination of the LATE.

\paragraph{The connection with demand and supply curves}

The notion of potential outcomes naturally connects with economics.
%also has grounds in economics so that, at least ex-post (which is, of course, much easier), the pillars of the bridge can be set (on one bank at least).
Perhaps the most straightforward objects to see this are demand and supply functions, which are ubiquitous in that field.
For instance, a demand function outputs the quantity asked by customers given a prescribed vector of prices.
An interest in the elasticity of demand with respect to price can be interpreted, in the framework of potential outcomes, as an interest in the causal effect of price on demand.
A demand function, by the fact that it is a \textit{function}, shares with potential outcomes the critical feature of being \textit{potential} or counterfactual.
It formalizes the concept of conceiving what would have been the quantity asked (the outcome variable) had the price (the treatment variable) been set to some specified value.
This example shows that, although latent, the notion of potential outcomes was already present in economists' thinking from the start.
J.A, G.I and D.~Rubin quote \cite{tinbergen1930} about this example of supply and demand curves in their answer to comments to AIR96 (see \textit{Journal of the American Statistical Association}, June 1996, Vol.~91, No.~434, page~469).
They also quote \cite{haavelmo1944} for early inception of potential outcomes.
  
This example is not uniquely illustrative since the analysis of competitive markets and the estimation of demand (or supply) curves have been a case in point of the use of instrumental variables.
J.~Angrist, Kathryn Graddy, and G.~Imbens notably extend the LATE formalism to this type of analysis with their article ``The Interpretation of Instrumental Variables Estimators in Simultaneous Equations Models with an Application to the Demand for Fish'' \citep{angrist_graddy_imbens_2000}.
This article is an important extension of the LATE trilogy, and we return to it in Section~\ref{subsubsec:summary_and_other_important_extensions}.

\paragraph{An influence of G. Imbens?} 

Second, it is interesting to notice that, in two influential applied articles,
\begin{itemize}[noitemsep,nolistsep]

\item[--]
``Lifetime Earnings and the Vietnam Era Draft Lottery: Evidence from Social Security Administrative Records'',
\textit{The American Economic Review},
%1990, J.~Angrist 
\cite{angrist_1990} (henceforth A90)

\item[--]
``Does Compulsory School Attendance Affect Schooling and Earnings?''
\textit{The Quarterly Journal of Economics}, %1991, J.~Angrist, Alan Krueger
\cite{angrist_krueger_1991} (henceforth AK91)

\end{itemize}
published just before the LATE trilogy, J.~Angrist and his co-author A.~Krueger do not resort to the potential outcome framework.
Instead, causal parameters are introduced through simultaneous structural equation models.
It may suggest that G.~Imbens imported the notion of potential outcomes, even if his early works are not directly connected with causal inference issues but with another statistical theme, namely efficiency issues (see Section~\ref{subsubsec:GImbens_1990_2000_efficiency}).
%\citep{imbens1992, imbens_lancaster_1994}

\subsubsection{A concern about heterogeneity}
\label{subsubsec:concern_heterogeneity}

A second central aspect of the LATE contribution relates to the attention devoted to heterogeneity: causal effects are a priori specific to each agent.
Consequently, average causal effects over different subpopulations are expected to differ.
It is interesting to compare the two above-mentioned   applied articles in that dimension.
In particular, although contemporary, A90 and AK91 substantially differ in their study of heterogeneity.

\paragraph{Heterogeneity in A90: already LATE}

A90 uses the Vietnam-era lottery draft as an instrumental variable for the endogenous veteran status and thus proposes TSLS estimates of the effect of military service on earnings.
A90 appears remarkable as a precursor of the formalization of LATE regarding its attention to heterogeneity.
The fifth section of this article, entitled ``Caveats,'' examines potential deviations from the main model.
It encompasses treatment effect heterogeneity that ``{merely result[s] in a reinterpretation of the estimates}'' (A90, page~329), from an average treatment effect (ATE) to a LATE interpretation.
It anticipates the distinction between what will soon be called always-takers (``true volunteers'' in A90) and compliers.
Indeed, J.~Angrist writes ``{Suppose that the impact of military service on the earnings of true volunteers differs from the impact on the earnings of draftees and \textit{men who enlisted because of the draft}. Then \textit{using functions of the draft lottery as instruments will only identify the effect of military service for the latter group}.}'' (A90, pages~329-330, we italicize).
Except for the potential outcome formalization, this is already exactly the LATE interpretation.

\paragraph{AK91: heterogeneity analysis based on observed covariates versus LATE-type heterogeneity}

AK91 uses the combination of quarters of birth and mandatory school attendance laws to study the causal effect of education (measured as the accumulated time spent in studies) on earnings.
Quarters of birth - interacted with state fixed effects - are used as instruments of the number of years of education.
%Compared to A90, AK91 could be deemed less sensitive to the issue of treatment effect heterogeneity.
In line with A90, AK91 is concerned with possible sources of treatment effect heterogeneity since they run the analysis both on the overall male population and on the sub-sample of black men (see section~II.C of AK91). A major difference between the two articles is however that AK91 is concerned with heterogeneity stemming from an observed covariate, namely ethnicity, while A90 focuses on heterogeneity across \textit{unobserved} populations, namely compliers and always-takers.

\subsubsection{Beware the exclusion restriction!}
\label{subsubsec:caution_exclusion_restriction}

The exclusion restriction~\eqref{eq:assumptionExclusionRestriction} is one of the three requirements for a valid instrument.
In this series of applied works, it is interesting to see that both A90 and AK91 are very careful about this condition, anticipating in that sense one of the assumptions of the LATE theorem and, especially, the attention devoted to this assumption compared to the simultaneous equation modeling (see Section~\ref{subsubsec:comparison_LATE_SEM} for further details).

A90 devotes one subsection of the fifth ``Caveats'' section to it.
AK91 deals with this issue in the third section of the paper. The following remark from AK91 is particularly explicit: ``if season of birth influences earnings for reasons other than compulsory schooling, our approach is called into question'' (AK91, p1007).
%A90 devotes one subsection of the fifth ``Caveats'' section to it: V.C ``Earnings-Modifying Draft Avoidance Behavior''.
%AK91 devotes the final third section of the paper, ``III. Other possible effects of season of birth'', on this threat: ``if season of birth influences earnings for reasons other than compulsory schooling, our approach is called into question'' (AK91, p1007).

Although formalization still relies on traditional simultaneous equations instead of potential outcomes, the space and attention devoted to this issue suggest J.~Angrist and A.~Krueger were already well aware of the importance of this hypothesis.

\subsection{The LATE trilogy}
\label{subsec:LATE_trilogy}

Section~\ref{appendix:subsec:LATETheorem} presents identification of the LATE in the simple setting of a binary instrument, a binary treatment, and no covariates.
It corresponds to a standard econometrics textbook presentation, whose advantage is to focus on the novel (and now Nobel) ideas of the LATE.
However, the three seminal papers, IA94, AI95, and AIR96, are far richer and address more general settings, multi-valued instruments, multi-valued treatments, presence of covariates.

This subsection presents and contrasts the more general results of these three papers:\footnote{
Contrary to Appendix~\ref{appendix:subsec:LATETheorem}, for the sake of brevity, we do not enter into the complete formalization and proof of the results.
We can but encourage interested readers to read the three original papers for more details.
} \cite{imbens_angrist_1994} (Section~\ref{subsubsec:IA94}), \cite{angrist_imbens_1995} (Section~\ref{subsubsec:AI95}), \citet*{angrist_imbens_rubin_1996} (Section~\ref{subsubsec:AIR96}).
We also evoke an essential, closely connected extension (to continuous treatment) of the LATE framework, namely \citet*{angrist_graddy_imbens_2000} (Section~\ref{subsubsec:summary_and_other_important_extensions}). 
%(extension to continuous treatment). 
%We also evoke essential, closely connected extensions of the LATE framework, namely \cite{imbens_rubin_1997} (identification of the distribution of potential outcomes), \citet*{angrist_graddy_imbens_2000} (extension to continuous treatment), and \citet*{abadie_angrist_imbens_2002} (identification of a quantile effect) in Section~\ref{subsubsec:summary_and_other_important_extensions}. 
For easier comparisons, Table~\ref{table:summary_LATEs}, displayed at the end of Section~\ref{sec:LATE} on page~\pageref{table:summary_LATEs}, proposes a summary of the main results of IA94, AI95, AIR96, and extensions contrasting the different settings in terms of treatments and instruments.
%(and covariates).

\subsubsection{Identification and Estimation of Local Average Treatment Effects (IA94)}
\label{subsubsec:IA94}

This 9-page article is the canonical reference that introduces the monotonicity assumption and identification of the LATE, including the name itself.\footnote{
``We call this \textit{a local average treatment effect} (LATE).'' (IA94, p467).
}
It considers a binary treatment, \(\supportde{D} = \{0, 1\}\).
In contrast, the instrument is more general: it is a discrete random variable (possibly more than binary), either scalar or multi-dimensional (although the two are equivalent in the sense that we can return to a scalar enumerating the different possible modalities of the vector).

The potential outcomes framework is used.
The exclusion restriction~\eqref{eq:assumptionExclusionRestriction} is less formally explicit in so far as the only potential outcome variables introduced are \(Y(0)\) and \(Y(1)\), thus with only the treatment as an argument, not the instrument.
That being said, a comment of their Condition~1 (``Existence of instruments'') refers to the exclusion restriction more directly, suggesting that a ``random'' determination of the instrument is not enough to guarantee what corresponds to hypotheses~\eqref{eq:assumptionIndepedenceInstrument_bodypaper} and~\eqref{eq:assumptionExclusionRestriction}.\footnote{
``Note that random assignment of \(Z_i\) does not guarantee part~(i) is satisfied because although random assignment implies that \(Z_i\) is independent of \(D_i(w)\), it does not imply that \(Z_i\) is independent of \(Y_i(0)\), \(Y_i(1)\).'' (IA94, p468).
%The notations are almost the same as ours, or the reverse rather: we choose our notation to follow that of the LATE trilogy. 
Here \(w\) is any possible value taken by the instrument (as \(z \in \supportde{Z}\) above). 
%The index~\(i\) is not omitted; implicitly, the statement concerns any individual~\(i\).
%
}

\paragraph{The first LATE theorem}

Since the instrument is not restricted to be binary, Theorem~1 of IA94 is already more general than the textbook LATE theorem.
Under the existence of valid instruments (their Condition~1 corresponding to~\eqref{eq:assumptionExclusionRestriction}, \eqref{eq:assumptionIndepedenceInstrument_bodypaper}, and~\eqref{eq:assumptionRelevanceInstrumentBinaryZ}) and the monotonicity assumption (Condition~2 of IA94, which is exactly~\eqref{eq:assumptionMontonicityMultivaluedZ_bodypaper}),
for any pair \(z\) and~\(w\) of distinct values of~\(Z\) such that the instrument influences the treatment for these values (\(P(z) := \Esperancedecond{D}{Z = z} \neq \Esperancedecond{D}{Z = w}\)), the average causal parameter
\begin{align}
\label{eq:IA94_theorem1_deltaCzw}
    \deltaComplier_{z,w} 
    \, := \,
    \Esperancedecond{Y(1) - Y(0)}{D(z) \neq D(w)}
\end{align}
is identified from the joint distribution~\(\Distributionde{(Y, D, Z)}\).

The causal parameter \(\deltaComplier_{z,w}\) is defined as a function of the values~\(z\) and~\(w\) considered for the instrument.
Of course, similar to the usual binary-instrument, binary-treatment LATE (denoted~\(\deltaComplier\) in Appendix~\ref{appendix:subsec:LATETheorem}), it is a \textit{local} average treatment effect since the average is over the subpopulation of \colModification{compliers,} ``those who can be induced to change participation status [the treatment] by a change in the instrument [from \(z\) to \(w\)]'' (IA94, p470).

The textbook LATE theorem is a corollary of this result when the instrument is binary, \(\supportde{Z} = \{0, 1\}\).
Indeed, the only two distinct values of the instrument are then \(z = 1\) and \(w = 0\) (or the reverse, this is symmetric) and, thanks to monotonicity (with the conventional order of~\eqref{eq:assumptionMontonicityBinaryZ}), \(\{D(1) \neq D(0)\}\) is equivalent to \(\{D(1) > D(0)\}\), which defines compliers \colModification{(see Table~\ref{tab:table_NT_C_D_AT} in page~\pageref{tab:table_NT_C_D_AT})}.
Thus, 
\(\deltaComplier := \Esperancedecond{Y(1) - Y(0)}{D(1) > D(0)} = \deltaComplier_{1,0}\).

\paragraph{LATE interpretation of the IV estimand}

In a sense, Theorem~1 still focuses on a binary instrument since it only considers two distinct values of the instrument.
IA94 then moves on to study the case of a more general instrument: 
``One way to exploit a multi-valued instrument is to estimate the ratio of the covariance of \(Y\) and some scalar function \(g(Z)\), and the covariance of \(D\) and \(g(Z)\).
If \(Z\) is a scalar random variable, then the choice \(g(z) = z\) leads to the standard IV estimator.
If \(Z\) is a vector, \(g(z)\) is often an estimate of \(P(z)\) [the probability of being treated conditional on \(Z = z\)]''. (IA94, p470).

To do so, IA94 requires Condition~3 (IA94, p470).
In essence, it ensures a coherent choice of the function \(g(\cdot)\) in terms of ranking: for any \(z\), \(w \in \supportde{Z}\), \(P(z) \leq P(w) \implies g(z) \leq g(w)\), plus the relevance condition \colModification{\(\Covariancede{D}{g(Z)} \neq 0\)}.
Importantly, in frequent situations (binary~\(Z\) or \(g(\cdot) = P(\cdot)\)), the first part of this condition is automatically satisfied.

Under Conditions~1, 2 and~3 of IA94, Theorem~2 of this article shows that the IV estimand (that is, the limit in probability of the IV estimator instrumenting \(D\) by \(g(Z)\) in the regression of \(Y\) on \(D\)) is a convex combination of LATE for pairs of adjacent values of the instrument (where the values are ranked according to the conditional probability of being treated).\footnote{
Appendix~\ref{appendix:subsec:LATETheorem} (paragraph ``The monotonicity assumption'')
provides more details regarding this ranking.
Note that, under the LATE assumptions, it is only a re-labeling without loss of generality.
}
%Formally, it is formulated with a discrete finite support instrument.
If \(\supportde{Z} = \{z_0, z_1, \ldots, z_K\}\) (ordered such that \(k < m\) implies \(P(z_k) \leq P(z_m)\)),
\begin{equation}
\label{eq:IA94_theorem2}
    \underbrace{\frac{\Covariancede{Y}{g(Z)}}{\Covariancede{D}{g(Z)}}}_{\text{the identified IV estimand}}
    \!\!\!\!\! = \;\;\,    
    \sum_{k = 1}^K
    \lambda_k \,
    \underbrace{\Esperancedecond{Y(1) - Y(0)}{D(z_k) = 1, D(z_{k-1}) = 0}}_{= \, \deltaComplier_{z_k, z_{k-1}}}  
    \; =: \;
    \deltaComplier_{g, \textrm{IV}},
\end{equation}
with weights
\begin{equation*}
    \lambda_k
    \, \propto \,
    \big( P(z_k) - P(z_{k-1}) \big) \times \sum_{l = k}^K \Probabilitede{Z = z_l} \big( g(z_l) - \Esperancede{g(Z)} \big) 
\end{equation*}
that are non-negative and add up to one.

In particular, the stronger the effect of the instrument from \(z_{k-1}\) to \(z_k\) on the treatment (the higher \(P(z_k) - P(z_{k-1})\)), the more weight received by the LATE \(\deltaComplier_{z_k, z_{k-1}}\) (which is \(\deltaComplier_{z,w}\) of Theorem~1 with \(z = z_k\) and \(w = z_{k-1}\)) in the convex combination.

\paragraph{Estimation and inference}

We have only discussed identification issues so far.
In practice, we also need estimation and inference results.
That is why the third section of IA94 concludes with Theorem~3 which shows the asymptotic normality of the IV estimator and provides the expression of its asymptotic variance (when \(g(\cdot)\) is a known function; the Appendix of IA94 derives the asymptotic distribution when \(g(\cdot)\) depends on an unknown parameter that is jointly estimated).
%As of now (including Appendix~\ref{appendix:sec:remindersOLSIVCausality}), we have only discussed identification issues.
%In practice, we also need estimation and inference results.
%That is why the third section of IA94 concludes with Theorem~3, which shows the square-root asymptotic normality of the IV estimator and provides the expression of its asymptotic variance (when \(g(\cdot)\) is a known function; the Appendix of IA94 derives the asymptotic distribution when \(g(\cdot)\) depends on an unknown parameter that is jointly estimated).

\paragraph{Examples and discussion}
The fourth and final section of IA94 (``Examples'') discusses the applicability of the conditions with three examples that ``exploit the manner in which a particular program or treatment is implemented to create instruments that are exogenous. Evaluations of this type are sometimes referred to as \textit{natural experiments}.'' (IA94, p471, we italicize): 
\begin{itemize}[noitemsep,nolistsep]

\item[--]
``Draft Lottery,'' that underlies notably \citet{angrist_1990};
%A90 \citep{angrist_1990});

\item[--]
``Administrative Screening,'' with a situation where monotonicity can be doubtful;

\item[--]
``Randomization of Intention-to-Treat,'' the typical textbook presentation of a randomized experiment with a binary instrument~\(Z\) which is the randomized assignment to treatment.

\end{itemize}

\subsubsection{Two-Stage Least Squares Estimation of Average Causal Effects in Models With Variable Treatment Intensity (AI95)}
\label{subsubsec:AI95}

This 12-page article ``generalizes [their] earlier result (\citealp{imbens_angrist_1994}) to models with variable treatment intensity'' (AI95, p435).
In IA94, the treatment~\(D\) is binary, while in AI95, the treatment~\(D\) (denoted \(S\) in the article for the number of years of schooling, which corresponds to their application) is a multi-valued quantitative ordered variable with finite support \(\{0, 1, \ldots, J\}\).\footnote{
Note that considering integer values for the support is without loss of generality:
``\(S\) is assumed to take on only integer values between \(0\) and \(J\). 
It is enough, however, that \(S\) be bounded and take on a finite number of rational values.
Then one can always use a linear transformation to ensure that \(S\) takes on integer values only between \(0\) and \(J\).'' (AI95, p435).
}
This covers many applications where the treatment can be received in different doses, levels, or intensities (used as synonyms below).
%This covers many applications where the treatment, instead of being ``nothing'' or ``everything'', can be received in different doses, levels, or intensities (used as synonyms below).

In the introduction of AI95, J.A and G.I recall the widespread use of IV and TSLS estimators and the distinction between econometrics' structural equations versus statistics' potential outcomes to formalize causality.
The second section presents the underlying application based on AK91.
It provides a typical example of a treatment with various intensities, namely the number of years of schooling.

\paragraph{Average Causal Response (ACR) with a binary instrument}

The third section %, ``The average causal effect of a variable treatment,'' 
presents the main theoretical result of the article.
At this stage, the authors consider a binary instrument~\(Z\) and no control variables.
\(D(z)\) (denoted \(S_z\) in the original) still represents the potential treatment variable. %, now multi-valued in~\(\{0, 1, \ldots, J\}\).
% Dans l'original, pas tres clair entre minuscule et majuscule pour z dans S_z; on corrige avec notation utilisée ici avec un z minuscule.

As in IA94, Assumption~1 combines conditions~\eqref{eq:assumptionIndepedenceInstrument_bodypaper} and~\eqref{eq:assumptionExclusionRestriction} regarding the instrument~\(Z\).
Assumption~2 is the monotonicity assumption\colModification{~\eqref{eq:assumptionMontonicityBinaryZ}}.
A novelty of AI95, specific to the multi-valued treatment case, is to discuss some testable implications of monotonicity.
It cannot be tested as such since it involves unobserved counterfactual variables.
%\eqref{eq:assumptionMontonicityMultivaluedZ} itself cannot be formally tested since it involves unobserved counterfactual variables.
However, J.A and G.I note that \eqref{eq:assumptionMontonicityBinaryZ}, \(D(1) \geq D(0)\),
%(the formulation of~\eqref{eq:assumptionMontonicityMultivaluedZ} when \(Z\) is binary) 
implies that \(\Probabilitede{D(1) \geq j} \geq \Probabilitede{D(0) \geq j}\) for any level~\(j \in \{0, \ldots, J\}\) of the treatment.
If~\eqref{eq:assumptionIndepedenceInstrument_bodypaper} is maintained, for~\(z \in \{0, 1\}\), \(\Probabilitede{D(z) \geq j} = \Probabilitedecond{D \geq j}{Z = z}\), which is identified. 
Hence, a testable implication of~\eqref{eq:assumptionMontonicityBinaryZ} (combined with~\eqref{eq:assumptionIndepedenceInstrument_bodypaper}) is that the cumulative distribution functions of~\(D\) conditional on~\(Z = 0\) and on~\(Z = 1\) do not cross.

Theorem~1 of AI95 shows that the estimand of the IV estimator (instrumenting \(D\) by \(Z\) in the regression of~\(Y\) on~\(D\) and a constant) is equal to a convex combination of several LATEs across distinct levels of treatment. Formally, under~\eqref{eq:assumptionExclusionRestriction}, \eqref{eq:assumptionIndepedenceInstrument_bodypaper}, \eqref{eq:assumptionMontonicityBinaryZ}, and a form of the relevance condition, namely \(\Probabilitede{D(1) \geq j > D(0)} > 0\) for at least one~\(j \in \{0, \ldots, J\}\), the theorem writes
\begin{equation}
\label{eq:AI95_theorem1}
    \underbrace{\frac{\Esperancedecond{Y}{Z = 1}
    -
    \Esperancedecond{Y}{Z = 0}}{\Esperancedecond{D}{Z = 1}
    -
    \Esperancedecond{D}{Z = 0}}}_{\text{the identified IV estimand (\(\betaWald\))}}
    \;\, = \;\;\,
    \sum_{j = 1}^J
    \omega_j 
    \,
    \Esperancedecond{Y(j) - Y(j-1)}{D(1) \geq j > D(0)}
    \, =: \, 
    \deltaComplier_{\textnormal{ACR}; 1, 0}
\end{equation}
where, for any~\(j \in \{1, \ldots, J\}\), 
\begin{equation*}
    \omega_j
    :=
    \frac{\Probabilitede{D(1) \geq j > D(0)}}{\sum_{\ell = 1}^n \Probabilitede{D(1) \geq \ell > D(0)}}
    \, \propto \,
    \Probabilitede{D(1) \geq j > D(0)},
\end{equation*}
so that \(0 \leq \omega_j \leq 1\) and the weights add up to one.

\medskip

The causal parameter \(\deltaComplier_{\textnormal{ACR}; 1, 0}\) is obviously more complex than the basic LATE presented in~\eqref{eq:IA94_theorem1_deltaCzw} which illustrates the additional layer introduced by a multi-valued \colModification{treatment}. The quantities \(\Esperancedecond{Y(j) - Y(j-1)}{D(1) \geq j > D(0)}\), \(j = 1, \ldots, J\) are actually LATEs as the population of individuals who switch treatment from less than $j$ to more than $j$ when the instrument jumps from 0 to 1 cannot be observed.
%The relative complexity of the previous interpretation of \(\deltaComplier_{\textnormal{ACR}; 1, 0}\) illustrates the additional layer introduced by a multi-valued \colModification{treatment} compared to IA94.
%As averages of averages remain averages, it can be simplified as saying that the IV estimand identifies ``a weighted average per-unit treatment effect'' (AI95, p435, immediately after the statement of Theorem~1).
%It is worth stressing that two distinct averaging phenomena are involved in the definition of \textcolor{red}{I AM HEEEEEEEEERRRRRRRRREEEEEEEEE}: (\textit{a})~an average across compliers (as in IA94), and (\textit{b})~an outer average across the different levels of the treatment.
%The fact that average~(\textit{a}) is over a subpopulation of individuals whose treatment is affected by the instrument justifies the LATE interpretation.

We remark that these LATEs \(\Esperancedecond{Y(j) - Y(j-1)}{D(1) \geq j > D(0)}\), \(j = 1, \ldots, J\), that are averaged (with weights~\(\omega_j\)) can differ from one another in two dimensions:
%In average~(\textit{b}), remark that the quantities \(\Esperancedecond{Y(j) - Y(j-1)}{D(1) \geq j > D(0)}\), \(j = 1, \ldots, J\), that are averaged (with weights~\(\omega_j\)) can differ from each other in two dimensions: 
\begin{enumerate}[noitemsep,nolistsep]

\item[(\textit{i})] 
the subpopulations of compliers, because a priori \(\{D(1) \geq j > D(0)\}\) is not equivalent to \(\{D(1) \geq \ell > D(0)\}\) for \(\ell \neq j\);

\item[(\textit{ii})]
linearity of the causal effect is not assumed here; hence, the individual or average causal effect of a one-unit change from \(j-1\) to~\(j\) is a priori different than the one from \(\ell-1\) to~\(\ell\).\footnote{If linear causal effects were assumed, \eqref{eq:assumptionLinearity} (see Appendix~\ref{appendix:sec:remindersOLSIVCausality} for details), the right-hand side of \(\deltaComplier_{\textnormal{ACR}; 1, 0}\) would simplify to \(\sum_{j = 1}^J \omega_j \, \Esperancedecond{\Delta}{D(1) \geq j > D(0)}\), and only the variability of type~(\textit{i}) would arise.}
%in addition to heterogeneous effects across individuals, at the level of a given individual, the causal effect of a one-unit change from \(j-1\) to~\(j\) is a priori different than the one from \(\ell-1\) to~\(\ell\).
%
\end{enumerate}

We further note that \(\deltaComplier_{\textnormal{ACR}; 1, 0}\) can be reinterpreted as an average of individual causal effects from one level of the treatment to the next one, \(Y(j) - Y(j-1)\), which explains the characterization of \(\deltaComplier_{\textnormal{ACR}; 1, 0}\) as ``a weighted average of per-unit treatment effects along the length of a causal response function'' (IA95, p431).%\footnote{
%
%The term ``causal response function'' is another name for the notion of potential outcome variables; hence the choice of the notation~\(Y(d)\) stressing \(d\) is an argument of \(Y(\cdot)\)  (compared to \(Y_d\), which is another frequent notation; used in IA95, for instance).
%
%}
%The fact that the inner quantities, which are averaged (twice, (\textit{a}) then (\textit{b})), are individual causal effects from one level of the treatment to the next one, \(Y(j) - Y(j-1)\), explains the characterization of \(\deltaComplier_{\textnormal{ACR}; 1, 0}\) as ``a weighted average of per-unit treatment effects along the length of a causal 
%response function'' (IA95, p431).
J.A and G.I also refer to the \(\deltaComplier_{\textnormal{ACR}; 1, 0}\) as ``the average causal response (ACR)'' (AI95, p435).

\medskip 

Finally, following the point made by a referee, J.A and G.I remark that, although the ACR is a weighted average, ``it averages together components that are potentially overlapping'' (IA95, p435).
%The point connects to average~(\textit{b})
%(\textit{i})
%since, in general, nothing prevents from satisfying both \(\{D(1) \geq j > D(0)\}\) and \(\{D(1) \geq \ell > D(0)\}\) for \(\ell \neq j\).
In general, nothing indeed prevents an individual from satisfying both \(\{D(1) \geq j > D(0)\}\) and \(\{D(1) \geq \ell > D(0)\}\) for \(\ell \neq j\).
\colModification{For example, when the multi-valued treatment is the number of years of education and the binary instrument is living close enough to a college, an individual induced to go and graduate from college (thus reaching 16 years of education), but who would have completed only, say, 12 years of education (not going to college) had she lived far from a college, belongs to several complier populations: \(\{D(1) \geq j > D(0)\}\) for \(j = 13, 14, 15,\) and~\(16\).
In such a situation}, the individual contributes to several LATEs in~\(\deltaComplier_{\textnormal{ACR}; 1, 0}\).
IA95 adds ``In the schooling and other examples, however, most individuals would probably not be involved in an overlap of this sort, because the instrument would typically be expected to cause no more than a one-unit increment in treatment intensity for any particular individual.'' (IA95, p436).
There is no further comment from the authors regarding the issue.
%In our current understanding, we apologize for not being able to discuss to what extent this possible overlap damages the causal interpretation of the ACR.
In all cases, like the monotonicity assumption, the point should induce researchers to be careful about the choice of the instrument and its effects on the treatment.

\medskip

IA95 then presents a corollary of Theorem~1 for a causal interpretation of the IV estimand when a ``variable [meaning multi-valued] treatment is incorrectly parameterized as a binary treatment'' (IA95, p436).
The result is interesting because it is a common situation in various applications; for instance, looking at the effect of high school graduation or college graduation as a summary of the number of years of education.
In such cases, quantitatively, the estimation is upward biased; qualitatively, the sign remains, nonetheless, correctly estimated.

\paragraph{Extension to a multi-valued instrument and covariates}

The fourth section of IA95, ``Multiple instruments and models with covariates'', presents two other identification results that generalize Theorem~1 when there can be multiple non-binary instruments and covariates. 
These additional results have the same flavor as those presented above in the sense that TSLS still identify positively weighted combinations of LATEs.  
%Each adds another layer of averaging in the causal interpretation of the estimand. 
%The evolution is thus progressive and cumulative from Theorem~1 to Theorem~3 through Theorem~2.
%Despite these nested averages, the weights involved in all layers of averaging have the property of being non-negative.
Consequently, a concise summary of the different theorems of AI95 is the following.
Under the LATE assumptions, TSLS applied to a causal model with variable treatment intensity identifies a weighted average of per-unit causal responses in a wide variety of models.
%The average responses are average across individuals whose treatment status is affected by an instrumental variable which is required to ``affect treatment intensity in the same direction for each unit of observation.''%\footnote{
%
%IA95, p441; this summary borrows words from the introduction (p431) and the conclusion (p441) of AI95.
%
%}
We present those results below.

\medskip 

Theorem~2 of AI95 considers a non-binary instrument: \(K\) mutually exclusive binary instruments or, equivalently, a scalar instrument~\(Z\) taking values in \(\{0, 1, \ldots, K\}\).\footnote{
Following the article, the encoding in \(\{0, 1, \ldots, K\}\) is used. 
Note that, equivalently, we could denote \(\supportde{Z} = \{z_0, z_1, \ldots, z_K\}\) as in IA94, and use a generic modality/value \(z_k\) instead of~\(k\).
Behind those choices, the setting is that of a multi-valued, finitely discrete instrument~\(Z\).
}
As in IA94, the points of support of~\(Z\) are ordered such that, for any \(k, m \in \{0, \ldots, K\}\), \(k < m\) implies \(\Esperancedecond{D}{Z = k} < \Esperancedecond{D}{Z = m}\), which combines the relevance and the monotonicity assumptions.
We can then define the equivalent of \(\deltaComplier_{\textnormal{ACR}; 1, 0}\), which is identified in the binary instrument case, for any pair \((k, k-1)\) of adjacent values of~\(Z\): for any~\(k \in \{1, \ldots, K\}\),
\begin{equation*}
    \deltaComplier_{\textnormal{ACR}; k, k-1}
    \, := \,
    \sum_{j = 1}^J
    \omega_{j,k} 
    \,
    \Esperancedecond{Y(j) - Y(j-1)}{D(k) \geq j > D(k-1)},
\end{equation*}
with \( \displaystyle \omega_{j,k}
    :=
    \frac{\Probabilitede{D(k) \geq j > D(k-1)}}{\sum_{\ell = 1}^n \Probabilitede{D(k) \geq \ell  > D(k-1)}}\),
for any~\(j \in \{1, \ldots, J\}\).

\medskip  

When instrumenting the variable treatment~\(D\) by the multi-valued discrete instrument~\(Z\), Theorem~2 shows that the estimand corresponding to the IV estimator is equal to a convex combination of ACRs across the different adjacent values of the instrument:\footnote{
The first-stage of the TSLS estimation procedure is the saturated regression of~\(D\) on~\(Z\) so that the theoretical predicted value used in the second-stage is the conditional expectation \(\Esperancedecond{D}{Z}\); hence the expression of \(\betaIV\), the estimand of the IV estimator in this case (see the last paragraph of Appendix~\ref{appendix:subsec:LATETheorem} for further details).
}
\begin{equation}
\label{eq:AI95_theorem2}
    \underbrace{\frac{\Covariancede{Y}{\Esperancedecond{D}{Z}}}{\Variance({\Esperancedecond{D}{Z}})}}_{\text{the identified IV estimand (\(\betaIV\))}}
    \!\!\!\!\!\!
    =
    \;\;\;\,
    \sum_{k = 1}^K \mu_k \, \deltaComplier_{\textnormal{ACR}; k, k-1}
    \,
    =:
    \,
    \deltaComplier_{\textnormal{ACR}; Z}
    ,
\end{equation}
where the weights satisfy \(0 \leq \mu_k \leq 1\) and \(\sum_{k = 1}^K \mu_k = 1\).
%In terms of estimators, it corresponds to a linear combination of Wald estimates.

Moreover, \(\mu_k \propto (\Esperancedecond{D}{Z = k} - \Esperancedecond{D}{Z = k-1})\).
Therefore, the stronger the instrument when varying between values~\(k\) and~\(k-1\), %(in the sense of the larger its impact on the treatment~\(D\))
the more weight the corresponding ACR, \(\deltaComplier_{\textnormal{ACR}; k, k-1}\), receives in the convex combination~\(\deltaComplier_{\textnormal{ACR}; Z}\).

\medskip 

Finally, Theorem~3 of AI95 adds discrete covariates~\(X\), with a finite number of distinct values (so that, formally, the control covariates are included in the regression as the mutually exclusive indicators of each possible value~\(x \in \supportde{X}\)).
\colModification{Thanks to that restriction, for the first-stage of TSLS estimation, J.A and G.I can consider a saturated regression of \(D\) on \(Z\) and~\(X\), that is, with as many parameters as possible values for the set of explanatory variables, \((Z, X)\) here.} 
By construction, the theoretical prediction of the endogenous treatment is then \(\Esperancedecond{D}{Z, X}\).
\colModification{A \emph{saturated} model for the covariates~\(X\) happens to be crucial for the interpretation of the resulting TSLS estimand as an average causal effect on compliers. %, individuals affected by the instrument.
The next paragraph warns against possible misuses of the LATE in models with covariates that do not rely on a saturated specification.
%We elaborate on this issue in the first paragraph of Section XXX.
}
In this context, their Theorem~3 shows that the resulting TSLS estimator has a probability limit, denoted \(\betaTSLS\), equal to a weighted average of the \(\deltaComplier_{\textnormal{ACR}; Z}\) of Theorem~2 when they are defined conditionally on a value of the covariates:\footnote{
In the original paper, \(\deltaComplier_{\textnormal{ACR}; Z}\) is denoted \(\beta_Z\), and the right-hand-side term of Theorem~3 (Equation~\eqref{eq:AI95_theorem3} here) is written \(\Esperancede{\beta(X) \Theta(X)} \,/\, \Esperancede{\Theta(X)}\) where ``\(\beta(X)\) is the TSLS estimate, \(\beta_Z\), constructed using~\(Z\) as an instrument in a population where~\(X\) is fixed'' (AI95, p437).
}
\begin{equation}
\label{eq:AI95_theorem3}
    \text{the identified TSLS estimand} \;\, \betaTSLS
    \; = \;
    \frac{\Esperancede{\deltaComplier_{\textnormal{ACR}; Z}(X) \, \Theta(X)}}{\Esperancede{\Theta(X)}},
\end{equation}
with random weights
\begin{equation*}
    \Theta(X)
    : = 
    \Esperance\bigg\{
    \Esperancedecond{D}{Z, X}
    \times 
    \big(
    \Esperancedecond{D}{Z, X}
    -
    \Esperancedecond{D}{X}
    \, \big)
    \, \Big| \,
    X
    \bigg\} = \Variance\{ \Esperancedecond{D}{Z, X} \conditionnellementa X \},
\end{equation*}
and where, for any~\(x \in \supportde{X}\),  \(\deltaComplier_{\textnormal{ACR}; Z}(x)\) %(a non-stochastic number, while \(\deltaComplier_{\textnormal{ACR}; Z}(X)\) is a real stochastic variable) 
is defined as the equivalent of \(\deltaComplier_{\textnormal{ACR}; Z}\) conditional on~\(X = x\).
From Theorem~2, \(\deltaComplier_{\textnormal{ACR}; Z}(x)\) is thus identified by 
\begin{equation*}
    \frac{\Covariancedecond{Y}{\Esperancedecond{D}{Z, \colModification{X}}}{X = x}}{\Variance({\Esperancedecond{D}{Z, \colModification{X}}\,} \, \conditionnellementa \, X = x)}
    \, = \,
    \deltaComplier_{\textnormal{ACR}; Z}(x).
\end{equation*}

%\(\Theta(X)\) is a real random variable and is equal to the conditional variance of \(\Esperancedecond{D}{Z, X}\) knowing~\(X\),
%\(\Theta(X) = \Variance\{ \Esperancedecond{D}{Z, X} \conditionnellementa X \}\).
Again, the (random) weights are connected to the strength of the instrument; namely, conditional on~\(X\), the extent to which~\(Z\) affects~\(D\). 

\begin{modification}

\paragraph{Warning: identification of LATEs in models with covariates}

The articles of the LATE trilogy have opened  the study of causal interpretation of IV/TSLS estimators.
In particular, the last theorem of AI95 studies that problem in models with covariates. 
Causal interpretation of TSLS estimands with covariates as LATEs has not been entirely addressed in AI95 though and remains an active strand of the literature.
A recent contribution, \cite{blandhol2022tsls}, ``When is TSLS Actually LATE?'', notably shows that in the absence of additional parametric assumptions on the effect of covariates a causal interpretation of a TSLS estimand as a LATE requires ``saturated'' specifications, which control for covariates nonparametrically.
Without such additional assumptions, in specifications that are not saturated, the TSLS estimand mixes average causal effects for both compliers and noncompliers (always-takers or never-takers) and weighs some of the noncompliers average treatment effects negatively.
In such cases, the TSLS estimand cannot be interpreted as an average causal effect on the subpopulation of compliers.

That result is rather negative, all the more so as including covariates is frequent in practice and often important conceptually to support the validity of the instrument.
In particular, the independence assumption~\eqref{eq:assumptionIndepedenceInstrument_bodypaper} might be doubtful as such, unconditionally, but more plausible when considered conditionally on covariates.
It is therefore important to keep in mind the results of \cite{blandhol2022tsls}: either additional parametric assumptions or a nonparametric control through saturated regressions are required for the usual LATE interpretation in models with covariates.

It is worth noting that the only result of the LATE trilogy involving covariates (Theorem~3 of AI95) does use a saturated specification restricting to discrete covariates with a finite support.

\end{modification}

\paragraph{Estimation of the weights and interpretation}

In addition to the LATE-type identification results of AI95, the authors show that the weights \(\omega_j / \omega_{j,k}\) %(for a binary instrument, or \(\omega_{j,k}\) for multiple instruments more generally) 
are identified under the LATE assumptions.
Moreover, they ``can be consistently estimated from the difference between the empirical cumulative distribution functions of \(D\) given \(Z\)'' (AI95, p436).

In the fifth section, ``IV estimates of the returns to schooling: for whom?'', the authors illustrate the interest of such an estimation.
As discussed above regarding Theorem~1, the weights add another layer to the interpretation of an IV estimate.
In addition to the fact that this is a LATE (the average is over the subpopulation of compliers), the estimated weights reveal \textit{which compliers} %(that is, for which levels~\(j \in \supportde{D}\) of the treatment) 
contribute most to the weighted average.
Estimating and analyzing the weights thus permits a more comprehensive interpretation and understanding of the results.

For instance, J.~Angrist and G.~Imbens show that the weights associated with 12 to 16 %(end of college) 
years of schooling are lower for the 1920-1929 sample compared to the 1930-1939 sample.
``Therefore, men who ended up completing 
some college because they were forced to graduate high school contribute more to the estimates for men born in 1930-1939 than to the estimates for men born in 1920-1929. 
This difference may explain the higher Wald and TSLS estimates for men born in 1930-1939 [\dots] because the returns to the last year of college tend to be substantially higher than those for any single year of high school (\citealp{card_krueger_1992}).'' (AI95, p440).

\subsubsection{Identification of Causal Effects Using Instrumental Variables (AIR96)}
\label{subsubsec:AIR96}

The last article of the LATE trilogy is co-authored with Donald Rubin.
Its angle somewhat differs from IA94 and AI95.\footnote{
Also, in terms of mathematical modeling, contrary to IA94 and AI95, AIR96 uses a finite population setting: the potential variables are described as ``fixed but unknown values'' (AIR96, p446); ``\(E[g]\) denotes the average over the population of \(N\) units of any function~\(g(\cdot)\) [\dots] We emphasize that this notation simply reflects averages and frequencies in a finite population or subpopulation.'' (AIR96, p447); the independence assumption~\eqref{eq:assumptionIndepedenceInstrument_bodypaper} is formulated in a survey-type random assignment of the instrument (AI96, Assumption~2, p446). 
Interestingly, more than two decades later, G.~Imbens gets back to related questions with \cite{abadie2020sampling}, where the authors discuss sampling-based as opposed to design-based uncertainty.
}
It considers the simplest setting (binary instrument, binary treatment, no covariates), and the first identification result (Proposition~1 of AIR96 in the third section ``Causal estimands with instrumental variables'') is a particular case of Theorem~1 of IA94.
On the other hand, AIR96 appears more precise and comprehensive as regards the underlying assumptions:
\begin{itemize}[noitemsep,nolistsep]
    \item[--]
    it introduces the Stability and Unit Treatment Value Assumption (SUTVA, Assumption~1 of AIR96), which was only implicit in IA94 and AI95;
    \item[--]
    the exclusion restriction (Assumption~3) and the independence assumption (Assumption~2) are explicitly separated with the exclusion restriction expressed as a functional relationship (like in~\eqref{eq:assumptionExclusionRestriction});
    \item[--]
    the relevance condition is interpreted in terms of the causal effect of~\(Z\) on~\(D\).
\end{itemize}
Besides, AIR96 proposes a more detailed comparison with the econometrics framework of simultaneous structural equations, notably in the second section, ``Structural equation models in economics,'' and the fourth, ``Comparing the structural equation and potential outcomes frameworks'' (see Section \ref{subsubsec:comparison_LATE_SEM} for more details on this thread).
%which proposes some reviews introducing the dummy endogenous variable model, and in the fourth section of the article, entitled ``Comparing the structural equation and potential outcomes frameworks''.
\paragraph{The textbook LATE theorem}

Although the least general result among those of the three LATE papers, Proposition~1 of AIR96 is probably the most well-known.
Indeed, it is the textbook presentation of the LATE theorem: binary instrument, binary treatment, no covariates.
This result is introduced and proved in Appendix~\ref{appendix:subsec:LATETheorem}, Equation~\eqref{eq:LATEtheoremBinaryZBinaryDnoX}.
As explained above after Equation~\eqref{eq:IA94_theorem1_deltaCzw}, it is a direct corollary of Theorem~1 of IA94 when \(\supportde{Z} = \{0,1\}\): 
\begin{equation}
\label{eq:AIR96_proposition1}
    \deltaComplier := \Esperancedecond{Y(1) - Y(0)}{D(1) - D(0) = 1}
    \, \text{ coincides with } \,
    \deltaComplier_{1,0}.    
\end{equation}

\paragraph{Sensitivity to exclusion and monotonicity restrictions}

In contrast, Propositions~2 and~3 of AIR96 are two identification results that propose new results.
Indeed, the fifth section, ``Sensitivity of the IV estimand to critical assumptions,'' derives the asymptotic bias of the IV estimand (compared to the targeted LATE, \(\deltaComplier\)) when either exclusion~\eqref{eq:assumptionExclusionRestriction} or monotonicity~\eqref{eq:assumptionMontonicityBinaryZ} is violated, while the other LATE assumptions are maintained.

These theoretical identification results have a practical interest as they give qualitative and quantitative insights for sensitivity analyses.
As an illustration, the sixth section of AIR96, ``An application: the effect of military service on civilian mortality,'' ``show[s] how the sensitivity of the estimated average treatment effect to violations of the exclusion restriction and the monotonicity assumption can be explored using the results from the previous section'' (AIR96, p452).
For instance, the authors discuss the required deviations from monotonicity to reverse the sign of the LATE. This exercise anticipates sign-reversal concerns that are currently at the heart of the literature studying difference-in-differences-type estimators.

\medskip 

Proposition~2 relaxes the exclusion restriction.
Even before identification, this raises fundamental questions regarding the \textit{definition} of causal parameters.
For never-takers and always-takers, the instrument does not affect the treatment, \(D(0) = D(1)\), and AIR96 (Equation~(13) of the paper) can define the individual causal effect of the \textit{instrument}~\(Z\) on~\(Y\) for these individuals as \(H := Y(1, d) - Y(0, d)\), where \(d = 0\) (respectively \(d = 1\)) for a never-taker (resp. an always-taker).
This presentation provides another interesting interpretation of the exclusion restriction: it ensures that the causal effect of~\(Z\) on~\(Y\) is null for always-takers and never-takers.\footnote{
Remark that, with a binary instrument and a binary treatment, conditional on \(\{D(0) = 0, D(1) = 1\}\), \(D = Z\); in that sense, the causal effect of~\(D\) on~\(Y\) coincides with the causal effect of~\(Z\) on~\(Y\) for a complier under~\eqref{eq:assumptionExclusionRestriction}.
}

The situation is more complicated for compliers.
The authors explore that point through the assumption that assignment~\(Z\) and treatment~\(D\) have additive effects on the outcome~\(Y\) for all compliers.\footnote{
As far as we understood, Proposition~2 of AIR96 relaxes the exclusion restriction~\eqref{eq:assumptionExclusionRestriction} for noncompliers (always-takers and never-takers) but maintains this restriction for compliers so that the LATE parameter, \(\deltaComplier\), remains well-defined.
However, 
%as the authors stress,
``When there is a direct effect of assignment on the outcome for noncompliers, it is plausible that there is also a direct effect of assignment on outcome for complier'' (AIR96, p451).
The situation of Proposition~2 does not appear very credible; hence the relaxation of~\eqref{eq:assumptionExclusionRestriction} for anyone, including compliers, and the modeling using additively separable effects of~\(Z\) and of~\(D\) on the outcome to investigate the asymptotic bias.
}
They show that the bias relative to the average causal effect of~\(D\) on~\(Y\) for compliers positively depends on (\textit{i})~the average size of the direct effect of~\(Z\)~on~\(Y\) for noncompliers, \(\Esperancedecond{H}{D(1) = D(0)}\), and (\textit{ii})~the odds of noncompliance, \(\Probabilite\{D(1) = D(0)\} / \, \Probabilite\{D(1) > D(0)\}\); remember that~\eqref{eq:assumptionMontonicityBinaryZ} is still assumed to hold so there are no defiers.

\medskip 

On the other hand, Proposition~3 of AIR96 relaxes the monotonicity assumption.
It shows that, in this case, the IV estimand is equal to
\begin{equation*}
    (1 + \lambda) \, \times \,
    \underbrace{\Esperancedecond{Y(1) - Y(0)}{D(1) > D(0)}}_{=: \, \deltaComplier}
    \, -
    \, \lambda \, \times \, 
    \underbrace{\Esperancedecond{Y(1) - Y(0)}{D(1) < D(0)}}_{=: \, \deltaDefier}
\end{equation*}
with \( \displaystyle \lambda \, := \, 
    \frac{\Probabilite\{D(1) < D(0)\}}{
    \Probabilite\{D(1) > D(0)\} - \Probabilite\{D(1) < D(0)\}} 
\).

\medskip 

In the setting of a binary instrument and a binary treatment, under the LATE assumptions except~\eqref{eq:assumptionMontonicityBinaryZ}, the IV estimand is thus a linear combination of the average causal effect on the compliers, \(\deltaComplier\), and of the average causal effect on the defiers, \(\deltaDefier\).
However, the two weights, \(1 + \lambda\) and \(- \lambda\), are outside the unit interval~\([0,1]\) because \(\lambda > 0\) whenever there are defiers. As a result, the IV estimand can be negative ({resp.} positive) even when \(\deltaComplier\) and \(\deltaDefier\) are both positive ({resp.} negative).
\colModification{The monotonicity assumption is thus crucial to ensure the no-sign-reversal property of 
IVs.}

The asymptotic bias of IVs 
%, namely here \(\betaWald - \deltaComplier\), 
is proportional to the share of defiers in the population, \(\Probabilite\{D(1) < D(0)\}\).
Another determinant is the extent of the heterogeneity of treatment effects: 
``The less variation there is in the causal effect of \(D\) on \(Y\), the smaller the bias from violations of the monotonicity assumption'' (AIR96, p451); more specifically, in terms of heterogeneity between the average effect on compliers and on defiers.
In particular, \(\deltaComplier = \deltaDefier\) is a straightforward sufficient condition to cancel the bias.

\subsubsection{The Interpretation of Instrumental Variables Estimators in Simultaneous Equations Models with an Application to the Demand for
Fish (AGI00)}
\label{subsubsec:summary_and_other_important_extensions}

\begin{comment}
The literature on causal interpretations of TSLS is still very active.
For instance, in addition to the whole literature devoted to TWFE mentioned in the introduction, recently, \cite{blandhol2022tsls}, ``When is TSLS Actually LATE?'', examines specifications including covariates.
Absent parametric specification of the effect of covariates, it shows that the validity of the LATE interpretation crucially depends on using ``saturated'' specifications that control for covariates nonparametrically.
Otherwise, in general, the TSLS estimand mixes average causal effects for both compliers and noncompliers (always-takers or never-takers) and, furthermore, with negative weights for some of the noncompliers average treatment effects.
This is a rather negative result.
It is worth noting that the only result of the LATE trilogy involving covariates (Theorem~3 of AI95) does use a saturated specification and restricts to discrete covariates.
\end{comment}

%\paragraph{The extension to continuous treatment (AGI00)}

In \cite*{angrist_graddy_imbens_2000}, ``The Interpretation of Instrumental Variables Estimators in Simultaneous Equations Models with an Application to the Demand for Fish'', \textit{Review of Economic Studies} (henceforth AGI00), the authors extend the LATE identification results to the setting of a continuous treatment, where the causal response function (that is, the potential outcome) \(d \mapsto Y(d)\) can be assumed to have a derivative.
This section presents some of the prominent results from this paper. 
Overall, these can be seen as the continuous counterparts of Theorems~1 and~2 of AI95 (see Equations~\eqref{eq:AI95_theorem2} and~\eqref{eq:AI95_theorem3} above).

\medskip 

AGI00 develops its arguments and results in the setting of a demand-supply framework, considering, as a conventional choice, the demand function: what is the causal effect of a continuous treatment (price) on the outcome variable (quantity)?
Price is endogenous due to classical simultaneity issues.
The results are more general, and we adapt the notation for easier comparison.\footnote{
AIG00 considers demand (superscript~\(d\)) and supply (\(s\)) functions/potential variables: \(q_t^d(p, z)\) and \(q_t^s(p, z)\), where \(p\) is any price (the equivalent of our \(d\)) and \(z\) any value of the instrument.
The markets are indexed by~\(t\), which plays the role of the unit/individual index~\(i\) in our notation, and that we omit given i.i.d.-ness.
}
Also, for simplicity, we do not include covariates.\footnote{
Actually, before section~3.5 of AGI00, ``Estimation with covariates,'' covariates are formally present, but the results are only conditional on a given value~\(x\) of the covariate~\(X\) (which is equivalent to performing ``unconditional'' analyses on separate subsamples).
%Theorem~1 of AGI00 is expressed this way: ``The result is stated for a given value of the covariates'' (AGI00, p504).
%Theorems~2,~3, and~4 in section 3.4 of AGI00, ``Multiple instruments and two-stage-least-squares,'' also; ``For ease of exposition covariates are ignored in this section'' (AGI00, p510).
Covariates are more explicitly introduced in section~3.5 with a parametric specification %(to echo the previous discussion of \cite{blandhol2022tsls}) 
through AGI00's Assumption~5: ``The average equilibrium price and quantity are linear and additive in covariates'' (AGI00, p512); the corresponding result is Lemma~2 of AGI00 that we do not cover here. 
}
Remember that, for any values~\(d \in \supportde{D}\) and~\(z \in \supportde{Z}\), \(Y(z,d)\) denotes a potential outcome variable.
Under the exclusion restriction~\eqref{eq:assumptionExclusionRestriction}, it does not depend on~\(z\): \(Y(z,d) = Y(d)\), and the partial derivative with respect to~\(d\) is the derivative of \(d \mapsto Y(d)\); we denote \(d \mapsto Y'(d)\) that derivative function.
Following the application of AGI00 (non-negative prices for treatment), we consider \(\supportde{D} = [0, +\infty)\) but it could as well be the entire real line.

\medskip 

Theorem~1 of AGI00 shows that, in the case of a binary instrument, \(\supportde{Z} = \{0, 1\}\), the IV estimand is equal to
\begin{equation}
\label{eq:AGI00_theorem1}
    \deltaComplier_{\textnormal{AmarginalCR}; 1, 0}
    \,
    :=
    \,    
    \int_0^{+\infty}
    \Esperancedecond{Y'(d)}{D(1) \geq d \geq D(0)} \, \omega(d) \, \textrm{d} d,
\end{equation}
where the weighting function
\begin{equation*}    
    \omega(d)
    \, := \,
    \frac{\Probabilite\{D(1) \geq d \geq D(0)\}}{\int_0^{+\infty} \Probabilite\{D(1) \geq r \geq D(0)\} \, \textrm{d} r},
    \;\, \forall d \in \supportde{D},
\end{equation*}
is non-negative and integrates to one.
This result is the continuous counterpart of AI95's Theorem~1.\footnote{
In Equation~\eqref{eq:AI95_theorem2}, it was important that some inequalities were large (\(\geq\) some intensity~\(j\) of the treatment) and other strict (\(> j\)).
Here, the distinction does not matter since \(D\) and \(D(z)\), \(z \in \supportde{Z}\), are assumed to be continuous random variables, that is, admitting a density with respect to Lebesgue's measure.
} The derivative \(Y'(d)\) can be interpreted as the individual \textit{marginal} causal effect of~\(D\) on~\(Y\) \textit{at}~\(D = d\) (instead of the \textit{one-unit} causal effect \(Y(j) - Y(j-1)\) of IA95 with a discrete treatment, \(\supportde{D} = \{0, 1, \ldots, J\}\)).
%hence the similar notation for the causal parameter of~\eqref{eq:AGI00_theorem1} with the distinction of the subscript ``AmarginalCR'' for ``Average marginal Causal Response'' since the derivative \(Y'(d)\) can be interpreted as the individual \textit{marginal} causal effect of~\(D\) on~\(Y\) \textit{at}~\(D = d\) (instead of the \textit{one-unit} causal effect \(Y(j) - Y(j-1)\) of IA95 with a discrete treatment, \(\supportde{D} = \{0, 1, \ldots, J\}\)).

%AGI00 discusses the two types of averaging, symbolized by the integral \(\int\) and the expectation~\(\Esperance\), that are involved in~\(\deltaComplier_{\textnormal{AmarginalCR}; 1, 0}\).
%The previous comments about Theorem~1 of AI95 (Section~\ref{subsubsec:AI95}) are based on this discussion.
%The article also illustrates these two types of averaging in special cases.

\medskip 

Then, AGI00 generalizes the previous result to the case of a discrete instrument with finite support~\(\{z_0, z_1, \ldots, z_K\}\).
As in IA94 and AI95, the distinct values of the instrument are ranked such that \(\Esperancedecond{D}{Z = z_k} < \Esperancedecond{D}{Z = z_m}\) for \(k < m\). 
Theorem~2 of AGI00 shows that, in this setting and under the LATE assumptions, the IV estimand, using some scalar function \(g(Z)\) as the instrument (like in IA94's second theorem), is a weighted average of the previous causal parameters across adjacent (in terms of conditional expectations of the treatment) values of the instrument:
\begin{equation}
\label{eq:AGI00_theorem2}
    \underbrace{\frac{\Covariancede{Y}{g(Z)}}{\Covariancede{D}{g(Z)}}}_{\text{the identified IV estimand}}
    \!\!\!\!\!\!\!\! = \,    
    \sum_{k = 1}^K
    \alpha_k 
    \underbrace{
    \int_0^{+\infty}
    \Esperancedecond{Y'(d)}{D(z_k) \geq d \geq D(z_{k-1})} \, \omega_k(d) \, \textrm{d} d}_{=: \, \deltaComplier_{\textnormal{AmarginalCR}; z_k, z_{k-1}}}
    =: \deltaComplier_{\textnormal{AmarginalCR}; g(Z)}.
\end{equation}
For any~\(k \in \{1, \ldots, K\}\) and~\(d \in \supportde{D}\), the weights $\omega_k(d)$ satisfy
\begin{equation*}
    \omega_k(d)
    \, := \,
    \frac{\Probabilite\{D(z_k) \geq d \geq D(z_{k-1})\}}{\int_0^{+\infty} \Probabilite\{D(z_k) \geq r \geq D(z_{k-1})\} \, \textrm{d} r},
\end{equation*}
and the weights \((\alpha_k)_{k = 1, \ldots, K}\) add up to one and are proportional to
\begin{equation*}
    \alpha_k
    \, \propto \,
    \big( 
    \Esperancedecond{D}{Z = z_k} - \Esperancedecond{D}{Z = z_{k-1}}
    \big)
    \sum_{m = k}^K
    \Probabilitede{Z = z_m}
    \big(
    g(z_m) - \Esperancede{g(Z)}
    \big).
\end{equation*}
Moreover, the latter are non-negative for reasonable choices of~\(g(\cdot)\), in particular for the choice \(g(z) = \Esperancedecond{D}{Z = z}\) (see AGI00, p511 for details).
This is the choice made in AI95's Theorem~2 (a saturated regression for the first-stage).
Thus, this result is, again, the continuous counterpart of AI95's Theorem~2.\footnote{
On top of the discrete/continuous treatment difference, AI95 uses the simpler re-encoding \(\supportde{Z} = \{0, 1, \ldots, K\}\) (and avoids therefore the introduction of the function~\(g(\cdot)\)).
}
%``the weights are proportional to the first stage impact on prices at each successive contrast'' (AIG00, p511). -- as \(mu_k\) in AI95's Theorem~2.

\subsection{Discussions, criticisms, and answers about the LATE}
\label{subsec:discussion_comments_criticisms_answers_LATE}

Following the presentation of the main identification results in Section~\ref{subsec:LATE_trilogy}, this subsection focuses on the relevance of LATE-type parameters (Section~\ref{subsubsec:relevance_LATE_public_policy_decision}), notably in contrast with more ``structural'' simultaneous equation models (Section~\ref{subsubsec:comparison_LATE_SEM}).
It also investigates the ability of LATEs to inform public policy decisions beyond the complier subpopulation.
This question has been subject to controversies, as illustrated by the article ``Better LATE Than Nothing: Some Comments on Deaton (2009) and Heckman and Urzua (2009)'', \textit{Journal of Economic Literature}, \cite*{imbens2010} (henceforth I10), which are evoked below.

\subsubsection{What relevance of the LATE to inform public policy decisions?}
\label{subsubsec:relevance_LATE_public_policy_decision}

The most debated point likely relates to the interest of the LATE regarding the information it provides to decide whether to implement a treatment (some public policy; for instance, a training program).\footnote{
For instance, James Heckman writes in his comment to AIR96: ``LATE is a controversial parameter because it is defined for an unobservable subpopulation.
Its use as an evaluation parameter thus is of questionable value.'' (comments to AIR96, \textit{Journal of the American Statistical Association} June 1996, Vol.~91, No.~434, page~p459).
}
Related to this issue is the fact that compliers are not identified in the data.
We briefly review some arguments connected to these debates in the setting of a binary treatment.

\paragraph{Relevant causal parameters}

The starting point is: what is the relevant causal parameter to inform a public policy decision?
One answer appears indisputable: case-by-case analysis depending on the policy that is considered.

In some settings, the LATE might correspond to the exact parameter of interest, while it is the ATE (Average Treatment Effect, \(\delta := \Esperancede{Y(1) - Y(0)}\)) or the ATT (Average Treatment effect on the Treated, \(\deltatreated := \Esperancedecond{Y(1) - Y(0)}{D = 1}\)) in other cases.
For instance, in the context of A90 (effects of Vietnam military service on earnings), ``One could imagine that the policy interest is in compensating those who were involuntarily taxed by the draft, in which case the compliers are exactly the population of interest. If, on the other hand, the question concerns future drafts that may be more universal than the Vietnam era one, the overall population may be closer to the population of interest'' (I10, p414).\footnote{
In the example of this quote, however, we can wonder how such a compensation, whatever its amount, could be implemented since it would require identifying the compliers to know to whom the compensation should be sent.
Remember that, as we never observe at the same time both \(D(0)\) and \(D(1)\), the compliers are not individually identified.
Nonetheless, the point remains that the LATE could be the causal parameter of interest for the intention of specific policies.
}
\begin{comment}
% COMMENT ABOUT THE MAIN ARGUMENT OF THIS PART BUT NOT VERY GOOD EXPLANATION
\footnote{
Those settings are perhaps not the most common but could be sensible, at least in the \textit{intention}/\textit{theory} of the public policy compared to its \textit{practice}.
Indeed, the following quotation is interesting but, concretely, we can wonder how such a compensation could be implemented since it would require to identify the compliers in order to know to whom the compensation should be sent.
Yet, we cannot identify compliers individually without further assumptions. 
(It does not entirely prevent such a policy nonetheless as we could construct something like a probability of being a complier and use that as a threshold or a determinant of the compensation; but, for sure, it is not as simple as identifying the compliers.)
That first answer is therefore not really an acceptable response of the LATE about being an interesting parameter.
That being said, it is not a problem.
The aim of the discussion here is essentially to convince that the contribution regarding identification of the LATE, by itself, does not have to answer to the interest of the LATE; it is another dimension.
}
\end{comment}

\medskip

Even when the ATE or the ATT are of greater policy interest than the LATE,\footnote{
J.A and G.I ackowledge in various instances that the LATE has substantial limitations as a policy parameter; for instance,
``Note that this group [compliers] need not be representative of the population, and that the members of this group cannot be identified from the data because membership involves unobserved counterfactual treatment status.'' (AI95, p435).
} the latter may still be the best one can hope for: identifying the ATE or the ATT relies on stronger assumptions than LATE identification and these assumptions may be too strong in some situations. 
In that case, recovering the LATE can be a reasonable alternative from which one can try to extrapolate and infer something on the ATE or the ATT. 
To sum up, as G.~Imbens says, ``better LATE than nothing''.

\medskip

Another parameter often considered is the Intention-To-Treat (ITT).
In the setting of a Randomized Controlled Trial (RCT) with imperfect compliance, where the binary instrument~\(Z\) is the randomized assignment to the treatment, the ITT corresponds to the average causal effect of~\(Z\) on~\(Y\) and writes
\begin{equation*}
    \deltaITT := \Esperancede{Y(1, D(1)) - Y(0, D(0))}.
\end{equation*}
Under independence~\eqref{eq:assumptionIndepedenceInstrument_bodypaper} only, the ITT is identified by \(\Esperancedecond{Y}{Z = 1} - \Esperancedecond{Y}{Z = 0}\).
However, it is not always a relevant causal parameter for policymakers since it captures a fairly weak notion of average treatment effects: it measures how individuals react on average to being assigned into the treatment group. 
Because of imperfect compliance, this is distinct from the impact of getting treated. 
As explained above when comparing the LATE with the ATT and ATE, one could thus favor the LATE (or the ATT/ATE) over the ITT for policy-related reasons at the cost of more stringent identifying assumptions.

Overall, J.~Angrist, G.~Imbens, and D.~Rubin underscore in their works that a trade-off exists between policy relevance of a parameter and ease of identification. 
They also prompt researchers to report several sets of estimators corresponding to the different parameters of interest to better inform public policy decisions.\footnote{
``It should be stressed, however, that the assumptions needed for a causal interpretation of the instrumental variables estimand (Assumptions~1 and 3-5) are substantially stronger than those needed for the causal interpretation of 
the intention-to-treat estimand (Assumption~1). 
The plausibility of the additional assumptions (\textit{i.e}., the exclusion restriction and the monotonicity assumption) must be taken into account when facing the choice to report estimates of the intention-to-treat estimands, of the IV estimands, or both.'' (AIR96, p450).
}

\medskip

\begin{comment}
A second answer is that, arguably, the ATE or the ATT is often more relevant than the LATE.
Yet, be it ATE, ATT, or LATE, they are unknown parameters that need to be identified and estimated from the data.
The main defense of the LATE could then resemble the following.
In some situations, assumptions underlying the identification of the ATE or the ATT appear too far from reality to allow us to learn, say, \textit{directly} something on the ATE or the ATT, whereas more plausible assumptions enable us to identify/learn the LATE.
%(neglecting here statistical uncertainty).
Granted, the LATE may not be the most relevant parameter.\footnote{
%
J.A and G.I make this point clear in various instances; for instance,
``Note that this group [compliers] need not be representative of the population, and that the members of this group cannot be identified from the data because membership involves unobserved counterfactual treatment status.'' (AI95, p435).
%
}
But, \textit{then}, in a {second} step, up to you to try to learn \textit{indirectly} something on the ATE or the ATT (or any parameter relevant to the contemplated public policy) by extrapolating one way or another from the LATE.
To sum up, as G.~Imbens says, ``better LATE than nothing''.    
\end{comment}

\paragraph{Structural and causal approaches}

Under which assumptions and how to perform such an extrapolation is the next question. %, and \textit{another} one.
G.~Imbens advocates that the two sets of assumptions regarding identification of the LATE on the one hand and extrapolation to other causal parameters more related to a given public policy decision on the other, should be separated as they are of different nature:
``I would prefer to keep those assumptions separate and report both the local average treatment effect, with its high degree of internal but possibly limited external validity, and possibly add a set of estimates for the overall average effect with the corresponding additional assumptions, with lower internal, but
higher external, validity'' (I10, p415). In the previous quote, G.I distinguishes between two types of models:
\begin{itemize}[noitemsep,nolistsep]
\item LATE-type approaches (labeled in I10 as ``causal approaches'', including further extensions like regression discontinuity designs or difference-in-differences) which target one specific treatment parameter (some average of the individual causal effects) and, consequently, do not recover primitives of an underlying structural model, which is absent.
\item structural models which seek to identify the primitives behind the generic economic behavior of any individual.
Provided primitives can be recovered, structural models are therefore able to perform richer counterfactual analyses than causal ones.
\end{itemize}
In a nutshell, causal approaches put more focus on internal validity (a credible estimation of a specific causal parameter on a given population from which the data is sampled) while structural approaches target, somewhat directly, external validity (a credible generalization of a causal effect to other populations).\footnote{
We extend the comparison between ``causal'' and ``structural'' approaches in Section~\ref{subsubsec:comparison_LATE_SEM} below, which contrasts the way of expressing assumptions between the LATE framework and simultaneous equation models.
}

\paragraph{Extrapolating from the LATE}
We discuss here intuitively some general thoughts for extrapolating from the LATE.\footnote{
We come back to this question in Section~\ref{sec:beyond_LATE} when discussing~\cite{angrist_fernandezval_2010} and~\cite{angrist_rokkanen_2015}.
}
\begin{comment}
EARLIER VERSION WITH SHORT PRESENTATION OF angrist_fernandezval_2010
% For further details, we refer, for instance, to \cite{angrist_fernandezval_2010}.
% %\cite{angrist_fernandezval_2010} $\implies$ interesting article on how to give some external validity to LATE
% This article discusses how to reach some external validity for LATE estimates without imposing homogeneous effects.
% It recalls the results that results in \cite{abadie2003} -- see below -- can be used to identify virtually any moment of $(Y,D,X)$ in the compliers population.
% Then, adding some restrictions on how treatment effect heterogeneity occurs (through covariates only in a certain sense), it shows that LATE estimates can be extrapolated to ATT or ATE.
\end{comment}
In general, the LATE, \(\deltaComplier := \Esperancedecond{Y(1) - Y(0)}{D(1) > D(0)}\), is not equal to the ATE, \(\delta := \Esperancede{Y(1) - Y(0)}\), nor to the ATT, \(\deltatreated := \Esperancedecond{Y(1) - Y(0)}{D = 1}\).
Two separate dimensions contribute to this difference: (\textit{i})~individual causal effects \(\Delta := Y(1) - Y(0)\) are in general heterogeneous; (\textit{ii})~the different subpopulations of interest are, a priori, not the same.

Obviously, if effects are homogeneous (\(Y(1) - Y(0) = \deltazero \in \Reels\) almost surely), then \(\deltaComplier = \delta = \deltatreated = \deltazero\).
Beyond that polar case, a limited heterogeneity supports the extrapolation of the LATE to other causal parameters.
Provided several instruments, and thus several subpopulations of compliers, comparing LATE estimates across instruments is a way to assess the magnitude of heterogeneity.
%\footnote{
%XXX je me souviens d'une citation parler de test meême sans développer idée et test formel, à retrouver quelque part selon le temps.
%}
More generally, J.A and G.I insist on the importance of multiplying experiments or causal-approach studies to inform a public policy decision (see, for instance, the example in \cite{imbens2010}, section~6, that illustrates this logic).

\medskip 

Another type of justification to extrapolate relates to~(\textit{ii}).
Firstly, in the binary treatment binary instrument setting, extrapolation from the compliers to another population is possible in the particular case of one-sided noncompliance (in the terminology of RCT), where individuals not assigned to the treatment cannot be effectively treated.
Formally, \(D(0) = 0\) for any individual, which implies~\eqref{eq:assumptionMontonicityBinaryZ}, but also the absence of always-takers.
As a consequence, the subpopulation of treated individuals and the subpopulation of compliers coincide and \(\deltaComplier = \deltatreated\): the LATE approach identifies the ATT.\footnote{
More precisely, the subpopulation of compliers \textit{assigned to the treatment}.
Yet, given that the assignment is (as-if) random (assumption~\eqref{eq:assumptionIndepedenceInstrument_bodypaper}), compliers assigned to the treatment (thus treated) and compliers not assigned (hence not treated) are comparable, on average, in terms of potential outcomes.
} 
This case is important because of that theoretical result combined with the fact that \(D(0) = 0\) can be sensible in various applications.

More generally, it is sufficient to extrapolate that the compliers be ``representative'' of the group of interest, ideally, in terms of individual causal effects.
If \(\Distributiondecond{\Delta}{D(1) > D(0)} = \Distributionde{\Delta}\), the conditional distribution of treatment effects among compliers is the same as the marginal distribution (that is, among the entire population), then \(\deltaComplier = \delta\) (and similar reasoning applies to the ATT).
As \(\Delta\) cannot be observed for anyone, it is impossible to assess that directly.
However, the idea is to compare, in terms of some observed covariates~\(X\), compliers to the group of interest, say, the entire population, if we target the ATE.
If those covariates are thought to be informative about the causal effect of~\(D\) on~\(Y\) and compliers and the other group are similar in terms of~\(X\),\footnote{
Formally, when \(\Distributiondecond{X}{D(1) > D(0)} = \Distributionde{X}\); compliers are said to be representative of the population as regards~\(X\).
}
this supports using knowledge of the average individual causal effects among compliers to learn (partially) about the ATE.
This second justification raises another question.
\(X\) being observed, we can learn its distribution.
However, since we cannot identify compliers individually, is it possible to learn about \(\Distributiondecond{X}{D(1) > D(0)}\)?

\paragraph{Learning about the compliers}

Although we cannot identify compliers individually, learning about the subpopulation is possible.
First of all, under~\eqref{eq:assumptionIndepedenceInstrument_bodypaper} and~\eqref{eq:assumptionMontonicityBinaryZ}, the population proportion of compliers, \(\Probabilite\{D(1) > D(0)\}\), is identified and can be consistently estimated.
Evrything else equal, the more numerous the compliers, the more likely they are representative of the whole population, the more grounded extrapolating from the LATE to the ATE.
The proportion of the treated individuals who are compliers is also identified, which can be interesting, notably when trying to link the LATE to the ATT.

%Furthermore to implement the above-mentioned justification~(\textit{ii}), 
Furthermore, it is possible to identify the distribution of compliers' characteristics.
For discrete covariates, Bayes's formula shows that the relative probability that compliers have a given characteristic~\(X = x\), for any~\(x \in \supportde{X}\), (relative to the entire population)
\begin{equation*}
    \frac{\Probabilitedecond{X = x}{D(1) > D(0)}}{\Probabilitede{X = x}}
    \; \text{ is equal to } \;
    \frac{\Probabilitedecond{D(1) > D(0)}{X = x}}{\Probabilite\{D(1) > D(0)\}}.
\end{equation*}
Under the LATE assumptions with~\eqref{eq:assumptionIndepedenceInstrument_bodypaper} extended to hold conditional on~\(X\), the right-hand side is identified.
More generally, \cite{abadie2003} is an important brick to the LATE contribution.
Among other results, it shows that, for any real function \(g(\cdot)\) of \((Y, D, X)\),
%(with finite first-order moment), 
its expectation among compliers is identified since
\begin{equation}
\label{eq:Abadie2003kappa}
    \Esperancedecond{g(Y, D, X)}{D(1) > D(0)}
    \, = \,
    \frac{1}{\Probabilite\{D(1) > D(0)\}}
    \Esperancede{\kappa \, g(Y, D, X)},
\end{equation}
where 
\begin{equation*}
    \kappa 
    :=
    1 
    -
    \frac{D (1 - Z)}{\Probabilitedecond{Z = 0}{X}}
    -
    \frac{(1 - D) Z}{\Probabilitedecond{Z = 1}{X}}.
\end{equation*}
The conditions for this result are the LATE assumptions holding conditional on the covariates (see Assumption~2.1 and Theorem~3.1 of \cite{abadie2003}).
In the formal sense of Equation~\eqref{eq:Abadie2003kappa}, kappa-weighting ``detects compliers'' and thus allows to identify virtually any characteristic of the complier subpopulation.
\begin{comment}
%Yannick's thoughts:  Add other papers on characterization of compliers' profile in terms of covariates (\cite{marbach_hangartner_2020} probably not very different from Abadie's contribution, to be checked).    
\end{comment}

\paragraph{Beyond average effects}

%The previous results shows that the distribution of any function~\(f\) of the observed variables~\((Y,D,X)\) is identified (taking \(g(\cdot) = \Indicatricede{f(\cdot) \leq \tau}\) for any~\(\tau\)).
The previous result concerns any function of the observed variables~\((Y, D, X)\). We could wonder what happens regarding the potential outcomes~\(Y(0)\) and~\(Y(1)\) (still in the setting of a binary instrument).
%Another path to this interrogation is criticism from \cite{deaton2010} answered in I10.
In~\cite{deaton2010}, the author notes that in the case of an RCT with perfect compliance, identification of the ATE additionally relies on linearity of expectations.
%However, without additional assumptions, we cannot identify other features of the distribution \(\Distributionde{Y(1) - Y(0)}\).
In fact, linearity of expectations implies that if we are interested only in the expectation of \(Y(1) - Y(0)\), knowledge of the two marginal distributions \(\Distributionde{Y(0)}\) and \(\Distributionde{Y(1)}\) is sufficient for identification. 
However, this is not the case if we are interested in features beyond the average, like distributional effects of the treatment, which require in general knowledge of \(\Distributionde{Y(1) - Y(0)}\).

G.~Imbens does not contest that reasoning but argues that a policymaker's typical interest lies in the two marginal distributions instead of the distribution of the difference.\footnote{
``In many cases, average effects of (functions of) outcomes are indeed what is of interest to policymakers, not quantiles of differences in potential outcomes. The key insight is an economic one -- a social planner, maximizing a welfare function that depends on the distribution of outcomes in each state of the world, would only care about the two marginal distributions, not about the distribution of the difference.'' (I10, p409).
}
Besides, in the LATE framework, the marginal distribution of potential outcomes among compliers, \(\Distributiondecond{Y(0)}{D(1) > D(0)}\) and \(\Distributiondecond{Y(1)}{D(1) > D(0)}\), are identified.
The result is shown in ``Estimating Outcome Distributions for Compliers in Instrumental Variables Models'', \textit{The Review of Economic Studies}, \cite*{imbens_rubin_1997} (henceforth IR97).
IR97 is an important extension to the LATE trilogy as it strengthens the identification result of the LATE theorem in the setting of a binary instrument and a binary treatment.

\medskip 

Another extension beyond average effects is quantile effects, developed in ``Instrumental Variables Estimates of the Effect of Subsidized Training on the Quantiles of Trainee Earning'', \textit{Econometrica}, \cite*{abadie_angrist_imbens_2002} (henceforth AAI02).
AAI02 focuses on the binary instrument binary treatment case with covariates and is one of the leading extensions of the LATE trilogy.
%AAI02 proposes the equivalent of TSLS for OLS in the setting of quantile regressions in the context of a binary instrument and a binary treatment with covariates.
%It is one of the leading extensions of the LATE trilogy.
Their Assumption~2.1 (AAI02, p91) is the classical LATE assumptions (\eqref{eq:assumptionExclusionRestriction}, \eqref{eq:assumptionIndepedenceInstrument_bodypaper}, \eqref{eq:assumptionRelevanceInstrumentBinaryZ}, \eqref{eq:assumptionMontonicityMultivaluedZ_bodypaper})  conditional on covariates~\(X\) (plus non-trivial assignment: \(\Probabilitedecond{Z=1}{X} \in (0,1)\)).
Their Lemma~2.1 is the basis of all the LATE identification results derived in the paper: 
\begin{equation}
\label{eq:AAI02_Lemma21}
    (Y(0), Y(1)) \indep D \conditionnellementa X, D(1) > D(0).
\end{equation}
The decisive consequence of~\eqref{eq:AAI02_Lemma21} is that ``in the population of compliers, comparisons by~\(D\) conditional on~\(X\) have a causal interpretation.'' (AAI02, p94).
It was the case for averages; it is also the case for quantiles.
AAI02 focuses on a linear model for conditional quantiles, which allows considering a single treatment effect (at any given quantile~\(\tau \in (0, 1)\)), namely the difference between the \(\tau\)-quantiles of \(Y(1)\) and of \(Y(0)\) for compliers conditional on~\(X\):
\begin{equation}
\label{eqAAI02_QTE_section31}
    \gamma_\tau^{\textnormal{C}}
    \, := \,
    \Quantiledeconda{Y(1)}{D(1) > D(0), X}{\tau}
    -
    \Quantiledeconda{Y(0)}{D(1) > D(0), X}{\tau},
\end{equation}
where \(\Quantile_\tau\) denotes the \(\tau\)-quantile operator.\footnote{
To connect to the previous discussion, compared to the expectation \(\Esperancede{\cdot}\), \(\Quantile_\tau[\cdot]\) is not linear.
In particular, \(\gamma_\tau^{\textnormal{C}}\) is not the \(\tau\)-quantile, conditional on~\(X\), of~\(\Delta := Y(1) - Y(0)\).
}
Relying on \cite{abadie2003}'s kappa function, AAI02 shows that the causal parameter \(\gamma_\tau^{\textnormal{C}}\) is identified in the LATE setting and proposes the quantile treatment effects (QTE) estimator to estimate it.

\subsubsection{LATE and simultaneous equation models frameworks}
\label{subsubsec:comparison_LATE_SEM}

For now, this review has focused on the identification results of the LATE trilogy and some of its main extensions.
Another important aspect of these articles, also claimed as such by the authors, is to compare to the set-up of simultaneous (structural) equation models (SEM).\footnote{
It is in particular very present in AIR96, whose Sections~2 and~4 are entitled respectively ``Structural Equation Models in Economics'' and ``Comparing the Structural Equation and Potential Outcomes Framework''.
AGI00 also underscores this comparison focusing on a system of supply and demand equations: 
``A second contribution is the formulation of critical assumptions underlying identification of heterogeneous, time-varying demand functions in terms of potential demand and supply at different values of prices and instruments. This contrasts with most of the literature on simultaneous equations models, which casts critical assumptions in terms of unobservable functional-form-specific residuals.'' (AGI00, p500).
}
J.A, G.I, and their co-authors claim that a critical contribution of the LATE framework is to make the identifying assumptions more transparent than in SEMs.
This section briefly presents these arguments, following notably the exposition in AIR96.
In the setting of a binary treatment, AIR96 considers a basic dummy endogenous variable model:\footnote{
We note that, in a comment to this article, J.~Heckman proposes more advanced structural models. 
However, the target of this Section~\ref{subsubsec:comparison_LATE_SEM}, which is to illustrate the different types of assumptions as expressed in the LATE framework compared to the SEM formalization, remains in this simple setting.
}
\begin{align}
    \label{eq:SEM_eq_Y}
    Y & = \beta_0 + \beta_1 D + \eps, \\
    \label{eq:SEM_eq_Dlatent}
    \Dlatent & = \gamma_0 + \gamma_1 Z + \nu, \\
    \label{eq:SEM_eq_D}
    D & = \Indicatricede{\Dlatent \geq 0}.
\end{align}

\begin{comment}
OTHER WRITING (taking more space)
\begin{equation}
\label{eq:SEM_eq_D}
    D =
    \begin{cases}
      1 & \text{if } D^* \geq 0, \\
      0 & \text{if } D^* < 0.      
    \end{cases}   
\end{equation}
\end{comment}

The first major difference is the formalization of causal effects. 
The LATE framework, following Neyman-Rubin's, defines them jointly with potential outcomes.
In contrast, the parameter~\(\beta_1\) represents the causal effect of~\(D\) on~\(Y\) in the SEM defined by~\eqref{eq:SEM_eq_Y}-\eqref{eq:SEM_eq_Dlatent}-\eqref{eq:SEM_eq_D}.
It is possible to extend the model to introduce individual-specific causal effects (through random coefficients), hence heterogeneous causal effects.
Nonetheless, we could say that, by default, LATE modeling posits heterogeneous causal effects by introducing \(\Delta := Y(1) - Y(0)\) as the building block, whereas SEMs more naturally consider a homogeneous causal effect through the parameter~\(\beta_1\).

In SEMs, the relevance condition corresponds to \(\gamma_1 \neq 0\).
There is not much debate about the formulation of this assumption as it is simpler conceptually and can be easily tested in the first-stage regression.
We now consider the other LATE assumptions.

\paragraph{Exogeneity as opposed to independence and exclusion}

Much more debated is the standard SEM assumption of the ``exogeneity'' of the instrument, namely that \(Z\) is uncorrelated/orthogonal to the error terms or disturbances of the two equations~\eqref{eq:SEM_eq_Y} and~\eqref{eq:SEM_eq_Dlatent}:\footnote{
Without loss of generality due to the intercepts \(\beta_0\) and \(\gamma_0\), \(\eps\) and \(\nu\) are centered; hence \(\Esperancede{Z \eps} = \Covariancede{Z}{\eps}\) and \(\Esperancede{Z \nu} = \Covariancede{Z}{\nu}\).
}
\begin{equation}
\label{eq:SEM_exogeneityZ}
    \Esperancede{Z \eps} = 
    \Esperancede{Z \nu} =
    0.
\end{equation}

The fact that \(Z\) does not intervene in  Equation~\eqref{eq:SEM_eq_Y} and the first part of~\eqref{eq:SEM_exogeneityZ} (\(Z\) and \(\eps\) are uncorrelated) capture the idea that \(Z\) affects~\(Y\) only through~\(D\).
As a comparison, the LATE-type exclusion restriction~\eqref{eq:assumptionExclusionRestriction} is expressed through restricting the potential outcomes: \(Y(z,d) = Y(z',d)\).
Condition~\eqref{eq:SEM_exogeneityZ} also embeds the independence assumption~\eqref{eq:assumptionIndepedenceInstrument_bodypaper}.

Conceptually, \eqref{eq:assumptionExclusionRestriction} and~\eqref{eq:assumptionIndepedenceInstrument_bodypaper} are quite different assumptions.
To assess~\eqref{eq:assumptionExclusionRestriction}, 
``the researcher must consider, at the unit level, the effect of changing the value of the instrument while holding the value of the treatment fixed.'' (AIR96, p449).
In ordinary language, \eqref{eq:assumptionExclusionRestriction} relates to the individual effect of~\(Z\) on~\(Y\).
On the contrary, \eqref{eq:assumptionIndepedenceInstrument_bodypaper} relates to the assignment mechanism of~\(Z\), that is, how the instrument is determined and, most importantly, can it be considered ``as-if'' randomly determined. %(and thus ``ignorable'' in the statistical terminology also used in AIR96 or ``random'' in ordinary language)?
As a consequence, J.~Angrist, G.~Imbens, and D.~Rubin consider that ``pooling these assumptions into the single assumption of zero correlation between instruments and disturbances has led to confusion about the essence of the identifying assumptions and hinders assessment and communication of the plausibility of the underlying model'' (AIR96, p450).
It is noteworthy that \cite{deaton2010}, while globally criticizing the LATE approach, also underscores this distinction by proposing and adopting different words: ``external'' for instruments that satisfy~\eqref{eq:assumptionIndepedenceInstrument_bodypaper} only, ``exogenous'' for instruments satisfying~\eqref{eq:assumptionExclusionRestriction} (and~\eqref{eq:assumptionIndepedenceInstrument_bodypaper}).\footnote{
% 
% We are not sure to perfectly understand the distinction on pages~430-431, but the important is that that distinction is made: 
``Whether any of these instruments is exogenous (or satisfies the exclusion restrictions) depends on the specification of the equation of interest, and is not guaranteed by its externality.'' (\cite{deaton2010}, p431).
}

An additional point made in AIR96 is that assumptions such as~\eqref{eq:SEM_exogeneityZ}, cast in terms of moment conditions based on error terms, are hard to interpret from the start as those error terms are not straightforward to apprehend (without reference to potential outcomes).
%assumptions in terms of moment conditions (null covariance) made on residuals/error terms/disturbances (which, according to J.~Angrist and co, are hard to interpret without reference to potential outcomes -- see the answer to Heckman in the 1996 Angrist, Imbens, Rubins AIR96 paper in particular, pages 25 and 26 of the PDF document)

\paragraph{Monotonicity}

Likewise, AIR96 defends that the LATE framework is more explicit about the monotonicity assumption.
The argument may be less convincing here since, under homogeneous causal effects, the monotonicity is irrelevant.
It thus appears severe to blame SEMs both for homogeneous causal effects and keeping implicit the monotonicity assumption.
%severe to reproach simultaneously to SEM homogeneous causal effects and an implicit monotonicity assumption.
That being said, it remains interesting to notice that the use of a model with a constant parameter~\(\gamma_1\) for the relation between \(Z\) and \(D\) (Equations~\eqref{eq:SEM_eq_Dlatent} and~\eqref{eq:SEM_eq_D}) implies that~\eqref{eq:assumptionMontonicityMultivaluedZ_bodypaper} is automatically satisfied.
%\footnote{Other reference: Exemple avec modèle latent où montonie implicitement supposée : intéresant à mettre page 17 de Imbens JEL 2010 (better LATE than nothing, aussi dans d'autres papiers de AIR)
%``Note that this model implies the monotonicity or no-defiers condition, although, unlike in the IA set up, the assumption is implicit, rather than explicit.'' (\cite{imbens2010}, p415). 
%}

\paragraph{Equivalence and conciliation}

Let us conclude this Section~\ref{subsubsec:comparison_LATE_SEM} with one remark and one result that may (or may not) reconcile the two positions.
First, as noted by Robert Moffitt in his comments to AIR96, ``Of course, one should not expect economists and statisticians, or even different individuals within each discipline, to find their intuition in the same way, and there is no reason not to have the model translated into multiple frameworks.'' (\textit{Journal of the American Statistical Association} June 1996, Vol.~91, No.~434, page~463).
Second, \cite{vytlacil2002} shows an equivalence between the two frameworks in the sense that, given the classical LATE assumptions, it is possible to construct a latent-index model (as Equations~\eqref{eq:SEM_eq_Dlatent} and~\eqref{eq:SEM_eq_D}) that generates~\(D(0)\) and~\(D(1)\).
% Details: see paper and AAG03 footnote 4, p93 

% TABLE RECAP RESULTS END OF SECTION 2 (begin)
\newgeometry{left=5mm,
    right=5mm,
    top=5mm,
    bottom=5mm}
\afterpage{%
    \clearpage% Flush earlier floats (otherwise order might not be correct)
    \thispagestyle{empty}% empty page style (?)
    \begin{landscape}% Landscape page
        \centering % Center table
        \begin{tabular}{|p{4.2cm}||p{4.4cm}|p{5cm}|p{2.6cm}|P{9.6cm}|}
        \hline 
        % ROW FIRST
        \centering {Results and settings} 
        &
        \centering {Treatment}~\(D\)
        &
        \centering {Instrument}~\(Z\)
        &
        \centering {Covariates}~\(X\)
        &
        {Identified causal parameter}
        \\ \hline \hline 
        % ROW AIR96 Prop 1
        \textbf{AIR 1996 Proposition~1}\newline 
        {\small (see~\eqref{eq:AIR96_proposition1} and~\eqref{eq:LATEtheoremBinaryZBinaryDnoX})}
        &
        binary \newline 
        \(\supportde{D} = \{0,1\}\)
        &
        binary \newline 
        \(\supportde{Z} = \{0,1\}\)
        &
        none
        &
        \(\underbrace{\Esperancedecond{Y(1) - Y(0)}{D(1) > D(0)}}_{=: \, \deltaComplier \,=\, \deltaComplier_{1,0}, \text{ ``the'' Local Average Treatment Effect (LATE)}}\) 
        \\ \hline 
        % ROW IR97
        \textbf{IR 1997 Section~3} \newline 
        {\small (see~\ref{subsubsec:relevance_LATE_public_policy_decision})}
        &
        binary \newline 
        \(\supportde{D} = \{0,1\}\)
        &
        binary \newline 
        \(\supportde{Z} = \{0,1\}\)
        &
        none
        &
        \(
        \underbrace{\Distributiondecond{Y(0}{D(1) > D(0)} \text{ and }
        \Distributiondecond{Y(1}{D(1) > D(0)}}_{\text{the marginal distributions of potential outcomes among compliers}}
        \)
        \\ \hline 
        % ROW AAI02 
        \textbf{AAI 2002 Section~3.1} \newline 
        {\small (see~\eqref{eqAAI02_QTE_section31})}
        &
        binary \newline 
        \(\supportde{D} = \{0,1\}\)
        &
        binary \newline 
        \(\supportde{Z} = \{0,1\}\)
        &
        conditionally \newline 
        on~\(X\) 
        &      
        \(
        \underbrace{\Quantiledeconda{Y(1)}{D(1) > D(0), X}{\tau}
        - \Quantiledeconda{Y(0)}{D(1) > D(0), X}{\tau}}_{=: \, \gamma_\tau^{\textnormal{C}}, \text{ local \(\tau\)-quantile treatment effect on compliers}}
        \)
        \\ \hline         
        % ROW IA94, Thm 1
        \textbf{IA 1994 Theorem~1} \newline 
        {\small (see~\eqref{eq:IA94_theorem1_deltaCzw})}
        &
        binary \newline 
        \(\supportde{D} = \{0,1\}\)
        &
        two modalities~\(z\) and~\(w\) \newline of a discrete multi-valued~\(Z\)
        &
        none
        &
        \(
        \underbrace{\Esperancedecond{Y(1) - Y(0)}{D(z) \neq D(w)}}_{=: \, \deltaComplier_{z,w}}
        \) 
        \\ \hline 
        % ROW IA94 Them2
        \textbf{IA 1994 Theorem~2} \newline 
        {\small (see \eqref{eq:IA94_theorem2})}
        &
        binary \newline 
        \(\supportde{D} = \{0,1\}\)
        &
        discrete multi-valued, finite \newline 
        \(\supportde{Z} = \{z_0, z_1, \ldots, z_K\}\) \newline ranked by increasing conditional \newline on~\(Z\) probability of being treated 
        &
        none
        &
        %a convex combination over~\((z_k, z_{k-1})_{k=1}^{K}\) of LATE  
        a convex combination over adjacent \(Z\)-values of LATE
        \newline 
        %(\(k = 1, \ldots, K\))
        \(
        \sum_{k = 1}^K \lambda_k
        \underbrace{\Esperancedecond{Y(1) - Y(0)}{D(z_k) = 1, D(z_{k-1}) = 0}}_{= \, \deltaComplier_{z_k, z_{k-1}}}
        \)
        \\ \hline 
        % ROW IA95 Thm 1
        \textbf{AI 1995 Theorem~1} \newline 
        {\small (see \eqref{eq:AI95_theorem1})}
        &
        multi-valued, ordered finite \newline 
        \(\supportde{D} = \{0, 1, \ldots, J\}\)
        &
        binary \newline 
        \(\supportde{Z} = \{0,1\}\)
        &
        none
        &
        a convex combination over \(D\)-values of one-level-change LATE \newline 
        \(
        \underbrace{\sum_{j = 1}^J
        \omega_j 
        \,
        \Esperancedecond{Y(j) - Y(j-1)}{D(1) \geq j > D(0)}}_{=: \, \deltaComplier_{\textnormal{ACR}; 1, 0}, \text{ Average Causal Response (ACR)}}        
        \)
        \\ \hline 
        % % ROW IA95 Thm 2
        \textbf{IA 1995 Theorem 2} \newline 
        {\small (see \eqref{eq:AI95_theorem2})}
        &
        multi-valued, ordered finite \newline 
        \(\supportde{D} = \{0, 1, \ldots, J\}\)
        &
        discrete multi-valued, finite \newline 
        \(\supportde{Z} = \{0, 1, \ldots, K\}\) \newline 
        ranked by increasing conditional \newline 
        expectation of~\(D\), \(\Esperancedecond{D}{Z = \cdot}\)
        &
        none
        &
        \(\deltaComplier_{\textnormal{ACR}; Z}\):
        a convex combination over adjacent \(Z\)-values of ACR \newline 
        \(
        \sum_{k=1}^K \mu_k 
        \big(
        \underbrace{\sum_{j = 1}^J
        \omega_{j,k} 
        \Esperancedecond{Y(j) - Y(j-1)}{D(k) \geq j > D(k-1)}}_{=: \, \deltaComplier_{\textnormal{ACR}; k, k-1}}
        \big)
        \)        
        \\ \hline 
        % ROW IA 95 Thm 3
        \textbf{IA 1995 Theorem~3} \newline 
        {\small (see~\eqref{eq:AI95_theorem3})}
        &
        multi-valued, ordered finite \newline 
        \(\supportde{D} = \{0, 1, \ldots, J\}\)
        &
        discrete multi-valued, finite \newline 
        \(\supportde{Z} = \{0, 1, \ldots, K\}\) \newline 
        ranked by increasing \(\Esperancedecond{D}{Z = \cdot}\)
        &
        yes, \colModification{with a} \newline \colModification{saturated model} \newline
        {\small (discrete~\(X\) with 
        finite support)}        
        &
        \(
        \Esperancede{\deltaComplier_{\textnormal{ACR}; Z}(X) \, \Theta(X)} 
        \, / \, 
        \Esperancede{\Theta(X)}        
        \):
        a weighted average of \newline 
        \(\deltaComplier_{\textnormal{ACR}; Z}(x)\) (the equivalent of \(\deltaComplier_{\textnormal{ACR}; Z}\) conditional on~\(X = x\)) \newline 
        with random weight \(\Theta(X) = \Variance\{ \Esperancedecond{D}{Z, X} \conditionnellementa X \}\)        
        \\ \hline
        % ROW AIG00 Thm 1
        \textbf{AGI 2000 Theorem~1} \newline 
        {\small (see~\eqref{eq:AGI00_theorem1})}
        &
        continuous \newline
        \(\supportde{D} = [0, +\infty)\)
        &
        binary \newline 
        \(\supportde{Z} = \{0,1\}\)
        & 
        none \newline 
        {\small (conditionally on~\(X\) in original)}
        &
        a (continuous) convex combination of LA marginal TE \newline
        \(
        \underbrace{\int_0^{+\infty}
        \Esperancedecond{Y'(d)}{D(1) \geq d \geq D(0)} \, \omega(d) \, \textrm{d} d}_{=: \, \deltaComplier_{\textnormal{AmarginalCR}; 1, 0}, \text{ Average Marginal Causal Response (AMCR)}}        
        \)        
        \\ \hline 
        % ROW AGI00 Thm 2
        \textbf{AGI 2000 Theorem~2} \newline 
        {\small (see~\eqref{eq:AGI00_theorem2})}
        &
        continuous \newline
        \(\supportde{D} = [0, +\infty)\)
        &
        discrete multi-valued, finite \newline 
        \(\supportde{Z} = \{z_0, z_1, \ldots, z_K\}\) \newline 
        ranked by increasing conditional \newline 
        expectation of~\(D\), \(\Esperancedecond{D}{Z = \cdot}\)
        &
        none
        &
        a convex combination over adjacent \(Z\)-values of AMCR \newline 
        \(
        \sum_{k = 1}^K
        \alpha_k 
        \underbrace{
        \int_0^{+\infty}
        \Esperancedecond{Y'(d)}{D(z_k) \geq d \geq D(z_{k-1})} \, \omega_k(d) \, \textrm{d} d}_{=: \, \deltaComplier_{\textnormal{AmarginalCR}; z_k, z_{k-1}}}
        \)
        \\ \hline 
    \end{tabular}
    \captionof{table}{{Summary of the LATE trilogy results and of some extensions}.
    The results are ordered, firstly, by the richness of the treatment~\(D\): binary and qualitative; quantitative in an ordered finite set; quantitative and continuous.
    In these two latter cases, remark that the linearity of causal effects \eqref{eq:assumptionLinearity} is not assumed (for binary treatment, linearity is not applicable/automatic).
    Secondly, by the richness of the instrument~\(Z\).
    We follow the original presentation of IA95 with integer values for~\(Z\); it could be extended to a general finite set~\(\supportde{Z} = \{z_0, z_1, \ldots, z_K\}\) at the cost of considering a scalar instrument~\(g(Z)\).
    All results are stated under the LATE assumptions: \eqref{eq:assumptionExclusionRestriction}, \eqref{eq:assumptionIndepedenceInstrument_bodypaper}, \eqref{eq:assumptionRelevanceInstrumentBinaryZ}, \eqref{eq:assumptionMontonicityMultivaluedZ_bodypaper} (or \eqref{eq:assumptionMontonicityBinaryZ}), possibly conditional on~\(X\).}% Add 'table' caption
    \label{table:summary_LATEs}
    \end{landscape}
    \clearpage% Flush page
}
\restoregeometry 
% TABLE (end)

\section{The authors' influence beyond LATE}
% Tentative: The LATEr authors' influence
\label{sec:beyond_LATE}

Joshua Angrist and Guido Imbens have been awarded the Nobel Memorial Prize in Economic Sciences mostly for the articles about LATE identification they coauthored in the 1990s. 
This series of works, however, represents only a small fraction of these authors' rich input to economics research. 
Perhaps surprisingly, J.~Angrist and G.~Imbens have hardly ever coauthored beyond their LATE series of articles. 
In this section, we give a detailed account of the numerous fundamental contributions these two economists have made in their respective fields.

\subsection{Joshua Angrist}
\label{subsec:JAngrist}

Zooming out on J.~Angrist's career, the early 2000s can be seen as a turning point. In 2000 and 2002, \cite{angrist_graddy_imbens_2000} and~\cite{abadie_angrist_imbens_2002} are the last two collaborations between J.~Angrist (J.A) and G.~Imbens (G.I), both on topics related to the identification of local average treatment effects in instrumental variable frameworks.
These articles ended a decade of joint works that laid the foundations of the LATE framework. 
In this subsection, we present J.A's contributions from 2000 onwards not coauthored with G.I, a strand of works we refer to as post-LATE.

\subsubsection{2000-2010: teachers' quality, labor market issues, and much more}
\label{subsubsec:JA_2000_2010}

\paragraph{Looking for informative and varied empirical research designs}

The 2000s decade is probably the most prolific period for J.A in terms of applied topics covered by his research: between 2000 and 2010, he made impactful contributions in fields as diverse as education \citep{angrist_et_al_2002}, labor \citep{acemoglu_angrist_2001} or criminology \citep{angrist_2006}.

All these articles have in common to exploit randomized or natural experiments that display large variations in treatment exposure across populations under study.
These experiments are, therefore, adequate to measure the causal impact of economic policies by resorting to the potential outcome framework presented at length in Section~\ref{sec:LATE}.
Looking for good natural or randomized experiments is indeed at the heart of J.~Angrist's research agenda.
Quoting J.A and his coauthor Jörn-Steffen Pischke, in their celebrated review article ``The Credibility Revolution in Empirical Economics: How Better Research Design is Taking the Con out of Econometrics'', \textit{Journal of Economic Perspectives}, \cite{angrist_pischke_2010}:
``{Empirical microeconomics has experienced a credibility revolution, with a consequent increase in policy relevance and scientific impact. Sensitivity analysis played a role in this, but as we see it, the primary engine driving improvement has been a focus on the quality of empirical research designs.}'' (page~4). 

J.A not only looks for good experiments but also for varied ones which allow building knowledge on specific questions across distinct environments.
In J.~Angrist's view, this is the most efficient tool to understand how much external validity specific estimates contain.
Another illuminating quote from~\cite{angrist_pischke_2010} illustrates this well:
``{A constructive response to the specificity of a given research design is to look for more evidence, so that a more general picture begins to emerge.}'' (page~23).

To measure possible effects present in the experimental data he uses, J.A favors simple linear models, controlling for endogeneity of the policy if needed and for fixed effects in possibly multiple dimensions.
In our views, such choices are made (\textit{i}) to promote ease of interpretation of estimated effects and (\textit{ii}) to stress that it is the experiment rather than convoluted modeling restrictions that are the driving force of policy effect identification.
As we discuss below, the simple modeling approaches promoted by J.A however sometimes call for a cautious causal interpretation of the estimated effects.

\paragraph{Education and lotteries}

One applied topic, already present in J.~Angrist's works of the 1990s, is still dominant in his research effort of the 2000s: the issue of returns to education. 
J.A's interest focuses mainly on two more specific sub-questions: the impact of teachers' quality and financial incentives or aids on educational attainments. 

%\medskip 

The interplay between teachers' quality and educational outcomes is addressed in two articles, \cite{angrist_lavy_2001} and \cite{angrist_guryan_2004}.
Those two articles exploit natural experiments, in Jerusalem and in the United States, respectively.
In~\cite{angrist_lavy_2001}, the authors take advantage of a sudden increase in funds for on-the-job training of teachers that affected only a fraction of Jerusalem's schools in 1995.
This shock is then used to study the effect of teachers' training on pupils' achievement at school.
The authors conclude to a positive impact of teachers' training on pupils' results for some types of schools only (the non-religious ones).
In \cite{angrist_guryan_2004}, state-level variations in requirements to become a teacher in the US are exploited to analyze whether enforcing standards for teachers has consequences on the quality of teachers actually hired.
The conclusion of this article is rather negative on the usefulness of testing teachers prior to hiring: these tests tend to increase teachers' wages but have little effect on teachers' quality.
These results have been followed by numerous other analyses trying to assess how robust these findings were (\textit{e.g.}, \citet{boyd_2007} and \cite{goldhaber2007}).
In \cite{angrist_lavy_2001} and \cite{angrist_guryan_2004}, endogeneity concerns are limited so that no instrumental variable approach is applied, which is the exception rather than the rule in J.~Angrist's works. 

%\medskip

The question of how relevant financial incentives are to shape pupils' or students' educational trajectories is tackled in four articles published by J.A between 2000 and 2010: \cite{angrist_et_al_2002}, \cite{angrist_et_al_2006}, \cite{angrist_lavy_2009}, and \cite{angrist_lang_oreopoulos_2009}. 
Among those, \cite{angrist_et_al_2002} and \cite{angrist_et_al_2006} are of particular interest because of the nature of the experiment at play: in the 1990s, a large-scale educational program involving 125,000 pupils from low-income families took place in Colombia.
Targeted pupils were offered very generous vouchers that were renewable over the years conditional on having high enough grades.
In certain areas, the demand for vouchers was larger than the supply by local authorities, which pushed the latter to introduce a lottery system to select among eligible pupils.
As J.~Angrist and his coauthors note, some lottery winners chose not to accept the voucher so that the lottery is an instance of a randomized experiment with imperfect compliance. 
These researchers take advantage of this fairly challenging research design thanks to a very clever idea: they remark that lottery winners have a much higher probability of resorting to financial aid of any type (vouchers, scholarships, etc.) than lottery losers.
They thus interpret lottery winning as a credible instrument for financial incentives to study, which they use to analyze the effect of financial incentives on educational attainments.
In other words, the analysis is recast in the canonical LATE framework with a binary instrument (winning the lottery or not) and a binary treatment (resorting to financial aid or not).
The result primarily put forward by the authors is a positive and significant impact of financial aid on the propensity to complete secondary school. 

%\medskip

Using lotteries as an informative source of exogenous variation to identify causal relationships is not new to J.~Angrist: this idea already featured in his early works on the civil life consequences of GI conscription \citep{angrist_1990}.
\cite{angrist_et_al_2002} and \cite{angrist_et_al_2006} are however the first works by J.A that take advantage of lotteries to investigate school-related questions.
As we will see below, lottery-based experiments will be even more at the heart of J.A's research agenda on school-related topics in the 2010s.
Among J.~Angrist's other interesting contributions to school topics, we can cite \cite{angrist_lang_2004}. 
This article pioneers J.A's research interests of the 2010s: it is indeed J.~Angrist's first work that deals with an experiment aimed at improving the school prospects of children living in impoverished inner-city districts in the US (here Boston).
 
%The effects of high stakes high school achievement awards: Evidence from a randomized trial $\implies$ Israel, did, cash award for Bagrut completion.
%Incentives and services for college achievement: Evidence from a randomized trial $\implies$ randomized experiments with three treatment groups (including cash awards if a certain GPA performance met) and one control. No perfect compliance so LATE framework again. Fully linear in multiple treatments and covariates so not robust at all I believe. No mention of potential outcomes.

\paragraph{Labor market issues}

J.~Angrist also investigates several hot labor market questions in the 2000s decade, notably in these two articles: \cite{acemoglu_angrist_2001} and \cite{angrist_kugler_2003}.

In \cite{acemoglu_angrist_2001}, the authors investigate the consequences of the Americans with Disabilities Acts (ADA), a US federal ruling active since 1992 that seeks to promote employment of people with disabilities in the US.
Given that the act targets all disabled people, this group is interpreted as a treatment group, and a difference-in-difference (DID) analysis is conducted.\footnote{
DIDs are an identification and estimation strategy to recover average treatment effects in natural experiments where two groups (one treatment and one control group) are observed at two periods, and one group is affected by a treatment between the two periods. 
The idea of DIDs can be traced back to the 19th century \citep{snow1855}.
It has been very popular among applied economists since its use by David Card in the early 1990s.
David Card was awarded the Nobel Memorial Prize in Economic Sciences in 2021, along with Joshua Angrist and Guido Imbens (we refer to the dedicated article about D.~Card's contributions in the same issue of this journal).
} 
The authors report a significantly negative effect of the act on the employment of disabled men. 

\cite{angrist_kugler_2003} studies the consequences of immigration on the labor market prospects of natives in different European countries in the 1990s.
This question is difficult: immigration to and natives' employment in a country are both partly determined by the overall economic environment in that specific country, making immigration an endogenous predictor of natives' employment.
J.~Angrist and Adriana Kugler solve that puzzle by exploiting a massive and unexpected immigration shock that occurred in Europe in the 1990s. Between 1991 and 2001, the Yugoslav Wars put hundreds of thousands of civilians on the road who fled massively to nearby Western Europe. 
Instrumenting immigration with these Yugoslav Wars, the authors point to a negative impact of immigration on natives' employment, especially in countries with rigid economic institutions (high degree of protection of workers, for instance). 
This article is quite similar in spirit to \cite{card1990}.
In this path-breaking project, David Card uses an unexpected and large immigration wave from Cuba to Florida in 1980 to unveil the link between immigration and labor market outcomes in the US.
%We note that the data used in this work are not as rich as in the other applied projects J.A takes part in: they consist of a yearly panel of country-level observations. \cite{angrist2002}

\paragraph{Methodological concerns and contributions}

As highlighted above, J.A's research agenda of the 2000s places much emphasis on looking for good experiments in order to make credible causal claims.
Are there other significant methodological outputs for policy analysis that have emerged from this research agenda?
The answer is ambiguous.
In most of his articles, J.~Angrist deals with individual or group-level panel data where treatment can be suspected to be endogenous, typically because of noncompliance issues.
He centers his analysis on linear IV strategies to correct for possible endogeneity of the treatment variable of interest, and he typically adds control variables and fixed effects in multiple dimensions (time, individual/group, etc.) to partial out systematic heterogeneity, see for instance \cite{angrist_kugler_2003}.

Following the seminal LATE articles of the 1990s, we would be tempted to interpret the second-stage coefficient on the endogenous treatment as some (local) average treatment effect.
This is, however, more complex.
The data designs used by J.~Angrist in his applied works are much richer than those considered in \cite{imbens_angrist_1994}, \cite{angrist_imbens_1995} or \cite{angrist_imbens_rubin_1996}.
It is a priori unclear whether the findings of these articles extend to much more complex designs with covariates and multiple fixed effects.
This puzzle has, in fact, been solved very recently.

The ongoing two-way fixed effects (TWFE) literature addresses exactly that question and shows that running a linear regression with a treatment variable and time and individual fixed effects has some causal interpretation only if the heterogeneity of treatment effects across individuals and over time is severely restricted.\footnote{
We elaborate on the connection between LATE-type identification results and the TWFE literature in the conclusion (see Section~\ref{sec:conclusion}).
} J.~Angrist was unaware of these results in the 2000s, and the difficulty of extending the LATE formalism to richer data designs is palpable in his projects of that period. 
In \cite{angrist_lavy_2001}, the authors remark it is natural to impose a constant treatment effect assumption to rationalize the use of a standard linear DID regression with covariates.
In \cite{angrist_et_al_2002}, the experiment under study involves one binary treatment plagued with noncompliance and one binary instrument, which is similar to the seminal LATE framework.
However, the authors hardly mention the connection with LATE and the potential outcome formalism.
The discussion about the causal interpretation of the IV parameter in their application remains quite minimal as well. 
The limitations of linear IV models with covariates and fixed effects are, however, counterbalanced in J.A's works by frequent use of separate analyses on subpopulations defined by age or gender (\textit{e.g.}, \cite{angrist_et_al_2002}, \cite{angrist_kugler_2003} or \cite{angrist_lavy_2009}).
Intuitively, running separate regressions for distinct gender-age groups restores some robustness of the IV specification against heterogeneity of treatment effects along the gender-age dimension.

%\medskip

As discussed in this subsection, the topics of interest to J.~Angrist in the 2000s are more applied than purely methodological.
Still, he proposes a few methodological contributions that complement his foundational works of the 1990s.
In \cite{angrist_et_al_2006}, for instance, nonparametric bounds on the distribution of treatment effects in the presence of noncompliance are proposed and applied to the school lottery experiment rolled out in Colombia in the early 1990s.
\cite{angrist_cherno_fernandezval_2006} is another interesting econometric contribution.
This article tackles the following simple question: what is the interpretation of running a linear quantile regression of an outcome on a set of covariates if the true quantile regression is not linear?
Even though this project is not concerned with causality, there is an obvious connection with the LATE articles of the 1990s that seek to give a meaningful (causal) interpretation of the linear IV model even when the data is not generated according to such a model.
\cite{angrist_cherno_fernandezval_2006} is quite thought-provoking from an econometrician's point of view: the authors show that linear quantile regression can always be interpreted as a weighted least square estimator, with weights that depend on the true quantile regression function.

\subsubsection{From 2010 onwards: some more methodology, re-inspection of the 1990s' applied works plus an in-depth analysis of the US educational system}
\label{subsubsec:JA_2010onwards}

The research produced by J.~Angrist since 2010 has three distinctive features in our view: (\textit{i})~more emphasis on the causal interpretation of methods used in applications than in the previous decade, (\textit{ii})~a reanalysis of some applied questions addressed by the author in the 1990s, and, most importantly, (\textit{iii})~an active contribution to the public debate over education in the US.
We present each of these features in detail below.

\paragraph{Back to the LATE framework and its external validity}
% Back to the LATE framework and new extensions

In the 2000s, the connection between the IV methodologies used in practice by J.~Angrist and his coauthors and the causal framework laid down in the 1990s seemed less present.
In the 2010s, this link was unquestionably reasserted.
The articles \cite{angrist_fernandezval_2010} and \cite{angrist_rokkanen_2015} illustrate this phenomenon well.
These articles give simple conditions under which LATE parameters, identified using IV methods, can be aggregated to recover global parameters such as average treatment effects on the treated or in the general population.\footnote{
\cite{angrist_rokkanen_2015} is not solely concerned with IV setups.
This article focuses on regression discontinuity designs (RDD) which exploit thresholds (on wages, class sizes, etc.) that split otherwise similar populations into two groups, one that benefits from some policy (say above the threshold) and the other that remains unaffected. 
As explained in \cite{imbens_lemieux_2008}, RDD allows researchers to identify average treatment effects at the threshold when treatment compliance is perfect and average treatment effects for compliers at the threshold otherwise. 
\cite{angrist_rokkanen_2015} proposes a framework to recover (local) treatment effects away from the threshold in RDD.
} The conditions are necessarily restrictive: the present articles impose that all the differences between compliers and the rest of the population can be captured through observable covariates.
Still, these papers simultaneously allow for a certain form of treatment effect heterogeneity and to recover both local and ``global'' average treatment effects in situations where those quantities need not be equal.
\cite{angrist_fernandezval_2010} and \cite{angrist_rokkanen_2015} therefore contribute to improving the LATE framework on its arguably most controversial aspect, namely what external validity does a local treatment effect contain \citep{angrist_pischke_2010, imbens2010}.

%\cite{angrist_lavy_schlosser_2010}: considerable methodological extension of \cite{angrist_evans_1998}. Clear connection to potential outcome framework, but simplications made with covariates.

\paragraph{Re-inspection of 1990s' applied topics}

Angrist also reinspects topics he worked on in the 1990s: the civilian consequences of veteran status in the US \citep{angrist_chen_frandsen_2010, angrist_chen_2011}, the socio-economic impact of having extra children on parents or already-born children \citep{angrist_lavy_schlosser_2010}, and the interplay between class size and students' achievement \citep{angrist_lavy_etal_2019}.
In \cite{angrist_chen_frandsen_2010}, \cite{angrist_chen_2011}, and \cite{angrist_lavy_etal_2019}, the authors use augmented versions of the datasets used in the original papers. 
In \cite{angrist_lavy_schlosser_2010}, the analysis is conducted using Israeli data while the seminal work \cite{angrist_evans_1998} deals with American data. 

These ``replication'' exercises are necessary to assess the long-term stability of previously estimated effects.
J.A and his coauthors indeed conclude that some significant effects observed in the 1990s' papers cannot be recovered with more recent data, for instance, the positive impact of being a US army veteran on civilian earnings \citep{angrist_chen_2011} or the negative link between class size and educational attainments \citep{angrist_lavy_etal_2019}.

\paragraph{Achievement gap in the US educational system}
% Achievement gap and different types of school in the US

Educational issues were at the center of J.~Angrist's attention in the 2000s and still are in the 2010s decade.
For more than a decade, he has been especially interested in the US secondary school system. 
The latter is a much-commented topic in American public life (see \cite{angrist_etal_boston_2012} for some contextual elements thereon). 
A reason for this can be found in the huge disparities in secondary school achievement in American cities, with poor and mostly nonwhite inner cities underperforming considerably compared to more affluent and mostly white suburbs. 
This phenomenon is called the achievement gap. 
In many US cities, the standard public school system competes with various alternatives: fully-private schools, exam schools (which are elite public schools with selection based on grades), and independent public schools that target children from impoverished backgrounds (known as charter schools).
There is a heated debate over the relative merits of those alternatives in closing the achievement gap. 
In a still ongoing series of works, J.A, along with coauthors, actively takes part in this debate by providing evidence on the achievement trajectories of students attending different types of schools (see \cite{angrist_2022} for instance for an exhaustive review of these works). 
This research program is quite unique in its scope, with 13 articles coauthored by J.A on the said subject since 2010. 
It is conducted for the most part by researchers at Blueprint Labs\footnote{
Blueprint Labs is a recently created policy-evaluation lab hosted by the MIT and directed by Joshua Angrist.
} and consists in investigating the performance of school systems in many big cities in the US: Boston, New York City, Chicago, Denver, New Orleans, \dots, and the list is still growing.
To do so, researchers at Blueprint Labs have access to very rich data both on students and schools in the cities under study. 
As explained below, those researchers seek to develop a coherent set of econometric tools that ensure comparability across evaluations. 
Having that in mind, we can say that this series of works epitomize J.~Angrist and G.~Imbens' repeated calls for replicating similar experiments across environments to gain some insights into the external validity of effects measured locally \citep{angrist_pischke_2010, imbens2010}.

%\medskip

The starting point of the research conducted at Blueprint Labs is the following: local school systems in the US offer a number of (quasi-)experimental variations in the allocation of students to schools that can be exploited to reconstruct potential achievement trajectories in alternative schools for the same student. 
These variations stem from the massive over-subscription faced by the most appealing schools, in particular exam and charter schools. 
Exam schools offer seats based on test scores. 
We can expect that marginally rejected students and those who are offered the last seats are quite indistinguishable, a situation amenable to a regression discontinuity analysis. 
Charter schools do not select their students based on school achievement, but because of over-subscription, they are forced to organize lotteries to offer seats. 
Lotteries are random by construction, so students who are offered a seat should be similar to those who are not. 
In both cases, students who are offered seats do not always accept them, which creates possible endogeneity issues that can be solved by using typical IV techniques yielding LATE-type identified treatment effects. 
In some studies, the analysis is made significantly more challenging by the design of certain school systems: students and schools in Chicago, New York, or Denver are allocated to one another through a centralized Deferred Acceptance algorithm \citep{gale_shapley_1962} so that it may be difficult to trace back the origins of observed choices ultimately made by students and schools. 
J.~Angrist and his coauthors thus propose extensions to traditional potential outcome models that allow to identify and estimate clearly-defined local treatment effects in these complex designs \citep{abdulkadiroglu_etal_elite_2014, abdulkadiroglu_etal_denver_2017}. 
The methods they come up with are fairly easy to use and nicely trade-off the richness of the analysis and practical considerations.

%\medskip 

The research conducted on US secondary schools by J.~Angrist and his coauthors has produced unexpected and strikingly consistent results across a wide range of environments.
Charter schools have been shown to improve significantly student achievement (compared to standard public schools, for instance) in 
Boston \citep{adbulkadiroglu_etal_boston_2011, angrist_etal_boston_2012, angrist_etal_boston_2016},
Denver \citep{abdulkadiroglu_etal_denver_2017},
or New Orleans \citep{abdulkadiroglu_etal_nola_boston_2016}. 
The ability of charter schools to close the achievement gap between black and white students has also been demonstrated \citep{angrist_etal_boston_2013}.
On the contrary, students who enroll in elite exam schools do not experience an achievement premium, be it in Boston or New York \citep{abdulkadiroglu_etal_elite_2014}, and can even see their achievement deteriorates slightly in Chicago \citep{abdulkadiroglu_etal_chicago_2017}. 
The mechanisms behind these results are nicely summarized in \cite{angrist_2022}: exam schools are mostly selected by students to join already high-achieving peers, but these very same students would have performed better in a charter school.

\subsection{Guido Imbens}
\label{subsec:GImbens}

Unlike Joshua Angrist, Guido Imbens (G.I) is primarily a theoretical econometrician. 
His methodological contributions outside the LATE are countless and span an impressive array of different topics. 
He is especially recognized for his works in survey sampling theory, efficient estimation of moment-restricted models, development of a rich set of tools to estimate treatment effect parameters, and identification in very general econometric models. 
G.~Imbens' desire to draw deep connections across seemingly distant questions is also visible in his articles: as early as the 1990s, he underlined many parallels between survey sampling, treatment effect estimation, and efficient estimation of moment-restricted models that are still largely overlooked in the econometric community today.
G.I's taste for state-of-the-art methods developed by the statistics community is another striking feature of his career: he was, for instance, an early promoter of the use of empirical likelihood and machine learning techniques in econometrics. 
The rest of this subsection is devoted to presenting in detail G.~Imbens' rich and multifaceted career beyond the LATE.

\subsubsection{The 1990s and early 2000s: the relentless search for statistical efficiency}
\label{subsubsec:GImbens_1990_2000_efficiency}

Outside of his collaboration with Joshua Angrist and Don Rubin, Guido Imbens' early career focused on topics largely motivated by survey sampling concerns. %This interest originates in G.I's PhD dissertation (ADD REF). 
In the early 1990s, G.I is particularly interested in semiparametric models\footnote{
As a reminder, a model is called semiparametric when it is indexed by a finite-dimensional parameter that does not fully characterize the data distribution. Conditional maximum likelihood is a prototypical example thereof: the distribution of the outcome given covariates is fully characterized by a finite-dimensional parameter, but the distribution of covariates is left unrestricted.
} estimated with choice-based sampled data. 
Choice-based sampling arises when the distribution of a discrete outcome of interest (in a regression model, say) in the sample at hand differs from the true distribution of that outcome in the general population. 
This is the rule rather than the exception for researchers who work with survey data: it is indeed widespread  to oversample or downsample certain subpopulations in a survey for cost reasons or because some subpopulations are of larger interest for that specific survey. 
Choice-based sampling may also happen in an uncontrolled fashion when there is systematic nonresponse to a survey, for example. 
The question addressed by G.~Imbens in the first chapter of his dissertation and in several of his early published papers \citep{imbens1992, imbens_lancaster_1996} is more precisely the following: how to estimate efficiently the parameter of a semiparametric model when data is drawn according to a choice-based sampling design?\footnote{
In \cite{imbens_lancaster_1996}, the authors consider a more general stratified sampling mechanism. The derived results are very close to those presented for choice-based sampling. We thus stick to the latter framework in this discussion.
} 

To tackle this question, G.~Imbens and Tony Lancaster propose a novel estimator which is both efficient in the sense of \cite{chamberlain_1987} and computationally more attractive than existing alternatives such as \cite{cosslett_1981}. 
Their contribution is quite innovative: they derive their novel estimator by solving explicitly for the conditions given in \cite{chamberlain_1987} to characterize an efficient estimator. 
This may seem to be only a far-fetched theoretical refinement, but it is actually quite the contrary: this finding can be seen as an early instance of a generic approach to designing well-behaved estimators that has come back to the forefront in the recent treatment effect literature employing machine learning tools \citep{cherno2018}; a literature to which G.I actively participates as we review below.

\cite{imbens1992} and \cite{imbens_lancaster_1996} put forward another puzzling finding: if some extra moments of the data are available to the researcher, for instance, the probability that the discrete outcome of interest be equal to some value in the entire population, then it may still be useful to re-estimate these moments from the sampled data to build more efficient estimators of the parameters of the model. 
Counter-intuitive results of that form have long existed in survey sampling (see \cite{hajek_1971} for example). 
Again, this theoretical finding has surprisingly far-reaching consequences: as discussed further down, the very influential article \cite{hirano_et_al_2003} coauthored by G.~Imbens shows that it is more efficient to estimate the propensity score even when the latter is known in order to recover average treatment effects. 

\cite{imbens_lancaster_1994} is another nice article pointing in the same direction as \cite{imbens1992} and \cite{imbens_lancaster_1994}. 
In this article, the authors show that when some moments of the data distribution are exactly known, and interest lies in estimating the parameter of a conditional likelihood model, then transforming the problem into a Generalized Method of Moment one where one seeks to match both the likelihood score and the exactly known moments yields an estimator more efficient than the traditional maximum likelihood one.\footnote{
The Generalized Method of Moments (GMM) was introduced in \cite{hansen_1982} as a generic approach to estimate parameters in semiparametric models where identification stems from a set of (conditional) moment restrictions.
}

\medskip 

From the mid-1990s to the mid-2000s, G.~Imbens explores the efficiency topic beyond his initial findings in survey sampling models. 
He devotes a large share of his research effort to investigating alternatives to generalized methods of moments (GMMs) to estimate parameters in moment-restricted models. 
GMMs have indeed one major drawback: to build efficient GMM estimators, one needs to resort to a cumbersome two-step approach. 
Parallel to the GMM literature, a novel family of promising estimators blossoms in the statistics community: these are called generalized empirical likelihood (GEL) estimators.
This literature originates in \cite{owen_1990}. 
Unlike GMMs, GELs automatically yield efficient estimators in one step. 

Following in the steps of Art Owen, G.I is one of the early proponents of these methods in econometrics. 
He studies the properties of GEL techniques in a series of articles \citep{imbens1997, hellerstein_imbens_1999, imbens2002, donald_imbens_newey_2003}. 
In these works, he draws interesting links between GEL methods and older techniques not commonly viewed through this lens, such as the raking estimator in survey sampling \citep{imbens1997}. 
Despite its theoretical elegance, the GEL framework has not developed among applied researchers due to the erratic behavior of some GEL estimators in practice \citep{guggenberger_2008}. 
This may explain why G.~Imbens does not pursue this research effort beyond the mid-2000s. 
Still, the GEL framework provides a sound theoretical benchmark that G.I puts to good use to derive some fundamental properties of treatment effect estimators presented hereafter.

\subsubsection{Enlarging the econometrician's toolbox for treatment effect estimation}

G.~Imbens has contributed to the treatment effect literature way beyond the articles about LATEs he coauthored in the 1990s. 
Since the early 2000s, he has been one of the most active econometricians in the field of treatment effect estimation. 
He has considerably enriched the practitioner's toolbox to estimate treatment effects through his works on propensity score weighting, matching, doubly robust estimation, or, more recently, machine learning-assisted treatment effect estimation. 
He has also broadened our understanding of the workings of the methods previously mentioned with notable theoretical breakthroughs in matching or regression discontinuity, for instance.

\medskip 

Since the early 2000s, two topics have been particularly central to G.~Imbens' research program on the estimation of average treatment: propensity score reweighting and matching. 
To explain the rationale behind propensity score reweighting and matching, let us briefly recall the canonical Neyman-Rubin potential outcome framework described in Section~\ref{sec:LATE}.
We consider a binary treatment variable~\(D\), a set of covariates~\(X\) unaffected by the treatment, and two potential outcomes \(Y(0)\) and \(Y(1)\).
For every individual, we only have access to \(Y = DY(1) + (1-D)Y(0)\). 
Our goal is to identify and estimate the average treatment effect~\(\Esperancede{Y(1) - Y(0)}\).

Under a number of conditions, in particular independence between $D$ and $(Y(0), Y(1))$ conditional on $X$, we can show that 
\begin{equation}
\label{eq:prop_score}
    \Esperancede{Y(1) - Y(0)}
    =
    \Esperance\!\left[ \frac{DY}{\Probabilitedecond{D = 1}{X}} \right] - 
    \Esperance\!\left[ \frac{(1-D)Y}{1 - \Probabilitedecond{D = 1}{X}}\right].
\end{equation}
% and
% \begin{align}
% \label{eq:reg_adjust}
%     \mathbb{E}[Y(1)-Y(0)] = \mathbb{E}\left[ \mathbb{E}[Y \mid D = 1, X] \right] - \mathbb{E}\left[ \mathbb{E}[Y \mid D = 0, X] \right].
% \end{align}
In words, Equation~\eqref{eq:prop_score} implies that suitably reweighting observed data allows us to recover the ATE. 
The map \(x \mapsto \Probabilitedecond{D = 1}{X = x}\) is called the propensity score function, hence the concept of propensity score reweighting.

Reweighting data to construct consistent estimators is, in fact, a very old idea that has roots in survey sampling, for instance \citep{horvitz_thompson_1952}.
G.~Imbens has contributed to the literature on propensity score weighting mainly in two directions. 
In \cite{imbens2000} and \cite{hirano_imbens_2004}, he extends the notion of propensity scores to more complex experiments that involve a non-binary, possibly continuous, treatment. 
In \cite{hirano_et_al_2003}, the authors prove the counter-intuitive result that estimating the propensity score leads to a more efficient estimator of the ATE even when the propensity score is known. 
This finding has direct implications for practitioners working with randomized experiments in which the propensity score is exactly known. 
To prove their result, the authors elegantly recast their estimator as an instance of the GEL family of estimators mentioned earlier. 
This finding is also reminiscent of the results presented in \cite{imbens1992} and \cite{imbens_lancaster_1996}.

\medskip 

Matching is a very intuitive idea. 
For every treated, \(D = 1\) ({respectively} control, \(D = 0\)), individual, we only observe their potential outcome with ({resp.} without) treatment. 
To ``recreate'' the other potential outcome, a simple idea consists in identifying, for each individual~$i$, a small set of individuals in the other group with similar covariate profiles. 
This group is called the matching group of individual~$i$. 
The average observed outcome in the matching group serves as an imputed value of the unobserved potential outcome for~$i$.

The conceptual simplicity of matching is arguably a reason for its popularity among applied economists. 
However, it is not intuitive at all why matching should work from a theoretical viewpoint
As recalled in \cite{abadie_imbens_2006}, matching with a fixed number of matches can be seen as an extreme form of nonparametric regression, which has unappealing theoretical properties even when the number of individuals in the sample is large. 
In \cite{abadie_imbens_2006}, the authors succeed in showing that, despite the limitations of matching with a fixed number of matches, this method can still yield consistent and asymptotically normal estimates of the ATE when there is at most one continuous covariate.
This result is not completely favorable to matching but is still quite unexpected. 
Alberto Abadie and G.I have continued exploring the intriguing theoretical properties of matching for ATE estimation in a series of articles \citep{abadie_imbens_2008, abadie_imbens_2011, abadie_imbens_2012}.

\medskip 

A widely used extension to propensity score reweighting and matching is propensity score matching \citep{rosenbaum_rubin_1983}, which combines the two original methods as its name suggests. 
More precisely, individuals from one group are matched to individuals from the other group based on how close their propensity scores are. 
In a recent article \citep{abadie_imbens_2016}, A.~Abadie and G.~Imbens derive the asymptotic properties of this method.
The obtained results are more favorable than for standard matching.
This can be understood on the following grounds: in propensity score matching, matching is performed based on a unique variable, the propensity score, instead of possibly many covariates.

\medskip 

The list of topics to which Guido Imbens has contributed is still long. 
Among those, regression discontinuity designs (RDDs) and the use of machine learning methods for causal analysis (also termed ``causal ML'') are currently two of G.~Imbens' most active research areas.
G.I's most influential input on RDDs is his review paper coauthored with Thomas Lemieux \citep{imbens_lemieux_2008}, which provides applied researchers with simple guidelines to implement RDD methods rigorously. 
Even without covariates, implementing RDD techniques requires some care: standard RDD methods involve nonparametric regressions on each side of the treatment allocation threshold. 
As recalled in \cite{imbens_kalyanaraman_2011}, nonparametric regressions depend on hyper-parameters that are generally difficult to choose and have a strong influence on the obtained results. 
In that same article, the authors propose a practical choice of the hyper-parameter associated with local linear regression (a form of nonparametric regression) that is also theoretically justified. 
Through this article, G.I has played a crucial role in the massive adoption of local linear regression techniques for RDD estimation. 
Those initial theoretical findings have been recently complemented in \cite{gelman_imbens_2019}.

\medskip 

Since the mid-2010s, G.~Imbens has been a strong advocate of the use of machine learning algorithms to address policy questions in economics. 
In the 2010s decade, researchers working at the frontier between economics and statistics brilliantly demonstrated that modern predictive algorithms developed in the machine learning community could be adapted to improve economists' ability to solve complicated policy questions (see \cite{athey_imbens_2017} and \cite{athey_imbens_2019} for comprehensive reviews and \cite{cherno2018} for a recent theoretical contribution). 
In \cite{athey_imbens_2015}, G.I along with his coauthor Susan Athey proposes to adapt the well-known Classification And Regression Tree algorithm (CART) introduced in \cite{breiman1984} to the problem of estimating and conducting inference on conditional average treatment effects \(\Esperancedecond{Y(1) - Y(0)}{X}\). 
%(CATE). 
This work illustrates the conceptual challenge associated with adapting machine learning tools to causal estimation and inference: the original algorithms are designed to predict an observed outcome accurately given observed covariates. 
Thus, many adaptations need to be made to translate these algorithms into the potential outcome framework and ensure they deliver sensible inference claims.

% Approximate Residual Balancing: Debiased Inference of Average Treatment Effects in High Dimensions \cite{athey_imbens_wager_2018}

\subsubsection{Identifying increasingly complex economic models}

The vast majority of the projects described in the previous subsection are concerned with estimating average treatment effects. 
These average effects are just the tip of the iceberg: economists are keen on looking at an array of distributional phenomena not visible at the mean.

Recovering more than average effects is challenging, though: it requires placing more structure on the data. 
As a simple illustration, in a toy experiment with a binary treatment~\(D\), potential outcomes \(Y(0)\) and \(Y(0)\) and no covariates, the ATE is identified as soon as \(\Esperancedecond{Y(1)}{D = 1} = \Esperancede{Y(1)}\) and \(\Esperancedecond{Y(0)}{D = 0} = \Esperancede{Y(0)}\). 
On the other hand, the marginal distributions of \(Y(0)\) and \(Y(1)\) are identified only under the much more stringent condition that \(D\) be independent of~\((Y(0),Y(1))\).
Furthermore, identifying these marginal distributions is still not enough to recover the distribution of treatment effects, that is, the distribution of~\(Y(1) -Y(0)\).\footnote{
The paragraph ``Beyond average effects'' in Section~\ref{subsubsec:relevance_LATE_public_policy_decision} elaborates on this distinction.
}

%\medskip 

G.~Imbens worked on these difficult questions in the 2000s. 
One article of that period has particularly profound implications in our view, \cite{athey_imbens_2006}. 
In this article, S.~Athey and G.~Imbens search for reasonable conditions under which the marginal distribution of potential outcomes \(Y(0)\) and \(Y(1)\) can be identified in a DID setup. 
This is much more challenging than in a randomized experiment: treatment is not randomized, even conditional on covariates, so that both differences across groups and over time need to be exploited.
The authors come up with a very elegant solution that is also very easy to implement (when there are no covariates in the analysis, at least). 
Their result can be interpreted as a nonlinear extension of the classical DID identification of average treatment effects. 

G.~Imbens also considers identification in complex nonlinear models plagued with endogeneity in \cite{cherno_et_al_2007} and \cite{imbens_newey_2009}. 
In the last paper, a variety of policy-relevant parameters are shown to be identified under conditions carefully discussed by the authors on specific examples derived from economic theory.

% XXX Note to myself: in situation where (conditional) unconfoundedness is reasonable, efficiency matters since estimation of ATT/ATE is credible enough for identification and consistency not to be too much of a concern.

% Imbens: econometrician very curious about innovations in the stats community (potential outcomes, empirical likelihood, machine learning). He stands exactly in between the two communities in my views: appeal for computationally feasible estimators (stats) but also concerned with efficiency (econometrics). Many contributions in the field of identification which is 100\% econometrics.

\begin{modification}

\section{Conclusion}
\label{sec:conclusion}

In the series of works that laid the foundations of the LATE revolution \citep{imbens_angrist_1994, angrist_imbens_1995, angrist_imbens_rubin_1996}, J.~Angrist, G.~Imbens, and D.~Rubin brought together several fundamental ideas that had never been truly connected before in economics: 
(\textit{i})~the use of potential outcomes to formalize causal effects,
(\textit{ii})~the attention that should be devoted to credible sources of identification, 
and (\textit{iii})~the acknowledgment that heterogeneity of treatment effects matters when giving a causal meaning to OLS or IV estimands.
These have now become the dominant paradigm for discussing the estimation of causal effects in econometrics. 

%\medskip

The currently rapidly growing two-way fixed effects (TWFE) literature perfectly illustrates the lasting impact of the LATE revolution in applied economics. 
Linear TWFE models are a popular tool among applied economists to measure average treatment effects purged from unobserved heterogeneity in two dimensions, typically at the time and individual levels in a panel environment. 
They can be viewed as a generalization of the basic difference-in-differences model to designs with multiple groups and periods.
Nonetheless, until recently, the precise nature of the average effects recovered by TWFE models had not been studied.
Recent articles falling under the TWFE banner have typically tried to fill in this gap (see \cite{dechaisemartin2021twoway} and \cite{billinski_et_al_2022} for two exhaustive reviews of this literature).
The results that have emerged from this research effort have had a lot of impact on applied researchers: linear TWFE models are unable to capture meaningful average treatment effects in general as soon as treatment effects are not perfectly homogeneous over time and across individuals. 
Even though these results are more negative than the findings of the early LATE contributions, the ingredients underpinning the TWFE literature are fundamentally the same as those behind the LATE papers.
Indeed, the authors focus on a popular estimation technique, adapt the canonical potential outcome framework (here to accommodate multiple time periods), and conclude that, without restrictions on treatment effect heterogeneity, standard tools should be used with care.

\end{modification}

\newpage
\bibliography{biblio}

\newpage 
\appendix

\section{Better LATE than never?}
\label{appendix:sec:remindersOLSIVCausality}

This appendix presents and proves the so-called LATE theorem in the simplest setting of a binary treatment, a binary instrument, and no covariate.
This theorem is the following identification result: the IV estimand (or Wald estimand in this case) is equal to the average causal effects on the subpopulation of compliers, the individuals whose treatment is affected by the instrument.\footnote{
Here, we take the estimand, for a given estimator, as its limit in probability.
Note that, given this definition and the modeling of causal effects through unobserved counterfactual potential outcomes, the estimand is not defined as the parameter of interest and is not necessarily equal to it. 
The former is a function of the distribution of the data (for this reason, we say it is \textit{identified}), while the latter will typically be an average of causal effects.
If, under some hypotheses, the estimand is equal to the causal parameter, then this causal parameter is identified.
}
As a preamble, we review some basic notions in causal inference, in the framework of potential outcomes, and their relation with econometrics tools, notably Ordinary Least Squares (OLS), Instrumental Variables (IV), and Two-Stage Least Squares (TSLS).
This material is now widely considered textbook econometrics.\footnote{
In fact, this appendix is directly inspired by ENSAE's graduate ``Econometrics~1'' course conceived by Xavier d'Haultf\oe{}uille.
Additions are the explicit exclusion restriction following lessons of the LATE contribution (see Section~\ref{subsubsec:comparison_LATE_SEM}) and discussions of the relation with a policymaker's decision to implement or not the treatment. 
}

We focus here on identification issues and leave aside estimation issues (basic knowledge of OLS and TSLS estimators is assumed).
Upstream of estimation and inference challenges, identification reasoning forgets about statistical uncertainty: imagine we observe an infinite sample so that we know the distribution~\(\Distribution\) of the data, are we able to learn from it something about a causal parameter~\(\delta\) of interest?
If that parameter can be expressed as a function of the distribution, \(\delta = f(\Distribution)\), it is the case, and the parameter~\(\delta\) is said to be (point-)identified.\footnote{
More generally, identification is not ``all or nothing'': a parameter might not be point-identified; yet, we can learn things about it, typically we can bound it (see the partial identification approach of \cite{manski2003}).
}

\subsection{Causal effects as differences of potential outcomes}
\label{appendix:subsec:causalEffectsPotentialOutcomes}

\paragraph{Set-up and objective}

We are interested in the causal effect of a treatment variable~\(D\) on an outcome variable~\(Y\).
\(D\) can be binary (for instance, serving or not in the military during the Vietnam war); discrete with a quantitative ordered meaning (for example, number of years of education measured in quarters); continuous (for instance, prices).
\(Y\) can be binary (for instance, being employed or not) or, more frequently, also has a quantitative meaning and is a discrete or continuous variable (for example, earnings or quantities).

We assume to observe an independent and identically distributed (i.i.d) sample from a given population of interest.
\(Y\) and \(D\) are defined at the level of each individual or, more generally, a statistical unit of this population (``agents'', ``units'', ``individuals'' will be used as synonyms henceforth).
In the sample, they are indexed by \(i = 1, \ldots, n\).
But, thanks to the i.i.d modeling, to lighten notation, we can omit the \(i\) index: \((Y, D)\) is thus a couple of real random variables representing a generic instance drawn from the distribution \(\Distributionde{(Y, D)}\),
and we observe \( (Y_i, D_i)_{i = 1, \ldots, n} \simiid \Distributionde{(Y, D)}\).
For any random variable or vector~\(V\), \(\Distributionde{V}\) denotes its distribution.

We recall that when discussing \textit{identification}, as opposed to estimation and inference, we neglect statistical uncertainty in the sense that we reason as if we knew the joint distribution \(\Distributionde{(Y, D)}\) that generates the data.

\paragraph{Potential outcomes}

Following the influence of the LATE contribution in econometrics, and previously in statistics works by D.~Rubin \citep{rubin1974estimating, rubin1990} that generalize \cite{neyman1923}, we define causal effects in terms of potential outcomes.\footnote{
Section~\ref{subsubsec:comparison_LATE_SEM} discusses the distinction between this framework of potential outcome variables and the framework of structural equation models, which was more traditional in econometrics before the LATE contribution.
}
For any possible value~\(d\) in the support of~\(D\), we posit the existence of the potential outcome variable \(Y(d)\).
\(Y(d)\) is what \textit{would have been} the value of the outcome variable~\(Y\) if the individual had the value~\(d\) for the treatment~\(D\).\footnote{
\label{footnote:explanationsNotationsSymbols}
We try to stick to the following conventions for consistency and easier understanding. 
Stand-alone Roman uppercase letters such as \(D, Y, Z\), or \(X\) are used to denote random variables \textit{observed} by the econometrician as opposed to a function-type notation such as \(Y(d)\) or \(D(z)\) for \textit{potential} variables, which are random variables, one of which is observed by the econometrician while the others are counterfactual.}
%
%Lowercase Roman letters are typically used as ``free variables''; beware of this terminology: they are not stochastic but denote arbitrary values, for instance, any possible value~\(d \in \supportde{D}\) taken by the treatment.
%For examples, if the treatment is binary, \(d \in \{0,1\}\); if the treatment is a (non-negative) price, \(d \in [0, +\infty)\).
%
%Lowercase Greek letters generally denote non-stochastic unknown parameters or estimands.
%
%Following standard notation, the exception is the letters \(\eps\), \(\nu\), or sisters that denote disturbances or error terms: random variables not observed by the econometrician.
%
%Also, \(\lambda\), \(\mu\), \(\omega\), and \(\alpha\) are used to denote non-stochastic weights.
%
%The uppercase Greek letter \(\Delta\) denotes the individual causal effect; it is a random variable (hence the uppercase).
%Lowercase \(\delta\), possibly with sub- and superscripts, is used to refer to causal average effects, which are the primarily parameters of interest here (and \(\gamma\) for quantile effects).
%In contrast, the letter \(\beta\) is typically used to denote estimands, automatically identified from the data, as opposed to the \(\delta\) for which this requires identifying assumptions.
%A paragraph at the beginning of Section~\ref{sec:LATE} lists other general and miscellaneous notations used.
%
Such a construction is closely connected to the assignment mechanism design: how is the observed treatment status~\(D\) generated?
The attention devoted to this assignment is typical of the ``design-based'' approach pioneered by the 2021 Nobel laureates.

Causal effects are then defined as differences between potential outcomes. 

\paragraph{Individual causal effects (binary treatment)}

When \(D\) is binary, the individual \textit{causal} 
% the: deja une hypothese en fait que de 0 à 1, l'oppose de 1 à 0 mais je ne connais pas de présentation où cette distinction est faite.
\textit{effect} (or \textit{treatment effect}, the two terms are used as synonyms) is defined as 
\begin{equation*}
\Delta := Y(1) - Y(0).    
\end{equation*}
It is an individual effect as it is specific to each individual. %(with the index~\(i\), \(\Delta_i = Y(1)_i - Y(0)_i\)).
A priori, there is no reason that the effect of the treatment is the same across individuals: the effects are heterogeneous, and \(\Delta\) is modeled as a non-degenerate real random variable (that is, \(\Variancede{\Delta} > 0\)).
In contrast, the assumption of homogeneous causal effects asserts that there exists a non-stochastic real number \(\deltazero\) such that \(\Delta = \deltazero\) almost surely (a.s): the effect is the same for all individuals.
%\footnote{We neglect in this article measure theory technicalities omitting ``almost surely'' in the previous equality and later.}

\paragraph{Individual causal effects (multi-valued treatment)}

When \(D\) is multi-valued, without further assumption, there is not anymore a unique causal effect of \(D\) on \(Y\).
Indeed, for two distinct values \(d\) and \(d'\) in the support of~\(D\), we could define the causal effect on~\(Y\) of a change of~\(D\) from~\(d\) to~\(d'\): \(\Delta_{d', d} := Y(d') - Y(d)\).
%\footnote{
%
%This modeling implicitly assumes that it is 
%equal to the opposite of the effect of a change from~\(d'\) to~\(d\).
%
%}

A typical simplification assumes linear causal effects\footnote{
If \eqref{eq:assumptionLinearity} holds for some \(d_0\), it holds for any \(d_0\); \(d_0\) merely plays the role of a reference point. 
Note that \eqref{eq:assumptionLinearity} is less restrictive than one can think at first sight, as we could have linearity for known transformations of \(Y\) or \(D\) instead of the initial variables themselves.
For instance, a log-level model (with \(Y\) replaced by \(\log(Y)\)) postulates a linear relative effect instead of a linear absolute effect in a level-level model. 
%(\(Y\) and \(D\) as such for the explained and explanatory variables).
%
}: 
% In text
%there exists \(d_0 \in \supportde{D}\) and a real random variable~\(\Delta\) such that for any \(d \in \supportde{D}\), \(Y(d) - Y(d_0) = \Delta (d - d_0)\)}    
\begin{align*}
\label{eq:assumptionLinearity}
    \text{there exists }
    & d_0 \in \supportde{D}
    \text{ and a real random variable } \Delta 
    \text{ such that } \nonumber \\
    & \forall \, d \in \supportde{D}, \,
    Y(d) - Y(d_0) = \Delta (d - d_0).
    \tag{\textnormal{L}}
\end{align*}
Assumption~\eqref{eq:assumptionLinearity} allows us to get back to a single, yet still random and individual-specific, \textit{linear} causal effect \(\Delta\) of \(D\) on~\(Y\).
Besides, if \(D\) is continuous, \(\Delta\) can be interpreted as a marginal causal effect being the derivative of \(d \mapsto Y(d)\) with respect to~\(d\).

\paragraph{The fundamental problem} 

We only observe \(Y := Y(D)\), the outcome variable corresponding to the realized treatment~\(D\).
The remaining potential outcomes are counterfactual.
Consequently, we can never observe for a given unit its individual causal effect~\(\Delta\).

Consequently, we move from an individual level to the level of (sub)populations: parameters of interest become averages of individual causal effects and are called \textit{causal parameters}.
Adapting the notation to the current setting, J.A and G.I write, in a setting of multi-valued treatment (see Section~\ref{subsubsec:AI95} for a presentation of AI95)
``We view estimates of \(\rho\) in Equation~(1) [the coefficient associated with the treatment in the equation of the outcome; expression in terms of simultaneous equation models] as having a causal interpretation when they have probability limit [estimand] equal to a weighted average of \(\Esperancede{Y(d) - Y(d-1)}\) for 
all~\(d\) in some subpopulation or subpopulations of interest'' (AI95, p433).
Remark that AI95 does not assume~\eqref{eq:assumptionLinearity}; hence keeps the one-unit increase causal effects \(Y(d) - Y(d-1)\), for~\(d \in \supportde{D}\).

Two typical average causal parameters of interest are the so-called ATE and ATT.
Without loss of generality for a binary treatment or under~\eqref{eq:assumptionLinearity} more generally, the 
average treatment effect (ATE) is defined as\footnote{
When~\(D\) is binary,
%linearity~\eqref{eq:assumptionLinearity} is somewhat not applicable or automatic, and the ATE can be defined directly as \(\delta := \Esperancede{Y(1) - Y(0)} = \Esperancede{Y(1)} - \Esperancede{Y(0)}\).
the ATE can be defined directly as \(\delta := \Esperancede{Y(1) - Y(0)}\) and $\Delta$ is equal to the individual treatment effect \(Y(1) - Y(0)\).
%In some sense, the second expression (difference in averages) is conceptually closer to the interest of a policymaker than the first (average of the difference), although the two are, of course, equal by linearity of the expectation (see Section~\ref{subsubsec:relevance_LATE_public_policy_decision}, paragraph ``Beyond average effects'').
%
}
\begin{equation*}
    \delta 
    := \Esperancede{\Delta}.
    %= f(\Distributionde{\Delta})
\end{equation*}
It is thus the average individual (linear) causal effect over the entire population (without weights, same weight for all).
In comparison, the average treatment effect on the treated (ATT) writes in the binary treatment case (\(\supportde{D} = \{0, 1\}\))
\begin{align*}
    \deltatreated := &
    \Esperancedecond{Y(1) - Y(0)}{D = 1} \\
    = &
    \Esperancedecond{\Delta}{D = 1} \\
    = &
    \Esperance\!\left[
    \frac{\Indicatricede{D = 1}}{\Probabilitede{D = 1}} \, \Delta
    \right].
\end{align*}
Contrary to the ATE, it is thus an average over a subpopulation only, the subpopulation of treated units.\footnote{
Formally, the ATT can be seen as an average over the entire population but a weighted average with peculiar weights: \(0\) for untreated individuals (\(D = 0\)), a constant weight of \(1 / \Probabilitede{D=1}\) for treated individuals (\(D = 1\)).
}

\subsection{Identification of causal parameters in linear regressions (OLS)}
\label{appendix:subsec:CausalParametersOLS}

In the linear regression of \(Y\) on \(D\) (and a constant), the theoretical slope coefficient %(that is, under standard moment conditions, the probability limit, or estimand, of the OLS estimator)
is \(\betaD := \Covariancede{Y}{D} / \Variancede{D}\).
Although this quantity is always identified and can be consistently estimated, without further assumptions \(\betaD\) cannot be interpreted as an average causal parameter, namely an average %(weighted or not, over the entire population or some subpopulation) 
of the individual causal effect~\(\Delta\).
The reason behind this is the fundamental problem of ``endogeneity'' of the treatment, which can also be interpreted as a selection bias (see also Simpson's paradox in statistics).

\paragraph{Binary treatment and selection bias} 

We consider first the binary treatment case. 
In that case, it can be shown that \(\betaD=\Esperancedecond{Y}{D = 1}-\Esperancedecond{Y}{D = 0}\).
A priori, without specific manipulation of the treatment assignment, the individuals who are treated (\(D = 1\)) are on average different from the individuals who are not treated (\(D = 0\)) in terms of potential outcomes.
Consequently, the simple difference between treated and untreated mixes a causal effect and a selection bias: absent the treatment, it is likely that the treated individuals would anyway have different outcomes than the untreated!

Formally, simple manipulations show that 
\begin{equation*}
%label{eq:OLSbinary}
    \betaD
    =
    \Esperancedecond{Y}{D = 1}
    - \Esperancedecond{Y}{D = 0}
    =
    \underbrace{\Esperancedecond{Y(1) - Y(0)}{D = 1}}_{=: \, \deltatreated}
    +
    \underbrace{
    \Esperancedecond{Y(0)}{D = 1}
    -
    \Esperancedecond{Y(0)}{D = 0}
    }_{=: \, B},    
\end{equation*}
where \(B\) is called the selection bias.
The previous equation shows that a necessary and sufficient condition for a simple linear regression of the outcome~\(Y\) on the treatment~\(D\) to identify the ATT is\footnote{
Because~\(D\) is binary, mean-independence with respect to~\(D\) and non-correlation with~\(D\) are equivalent.
}
\begin{equation}
\label{eq:noselection_Y0_Dbinary}
    \Esperancedecond{Y(0)}{D = 1}
    =
    \Esperancedecond{Y(0)}{D = 0}
    \iff 
    \Covariancede{Y(0)}{D} = 0.
\end{equation}
Equation~\eqref{eq:noselection_Y0_Dbinary} requires the treatment status not to be correlated with what would have been the outcome absent the treatment.
In general, typically because individuals (partially) choose to enter treatment~\(D\), there is no reason to believe~\eqref{eq:noselection_Y0_Dbinary} be credible.

Furthermore, \eqref{eq:noselection_Y0_Dbinary} is not enough to identify the ATE.
A sufficient condition to do so is:
\begin{equation*}
\label{eq:assumptionAbsenceSelectionYdD}
\forall d \in \supportde{D}, \Covariancede{Y(d)}{D} = 0,
\tag{\textnormal{AS}}
\end{equation*}
which imposes the treatment status not to be correlated with what would have been the outcome with or without treatment.
In the binary treatment case, \eqref{eq:assumptionAbsenceSelectionYdD} writes
\begin{equation}
\label{eq:noselection_Y0_and_Y1_Dbinary}
    \forall d \in \{0, 1\}, \,
    \Covariancede{Y(d)}{D} = 0
    \iff 
    \forall d \in \{0,1\}, \,
    \Esperancedecond{Y(d)}{D = 1}
    =
    \Esperancedecond{Y(d)}{D = 0}, 
\end{equation}
and under this assumption, \(\betaD = \deltatreated = \delta\).

\paragraph{The special case of randomized experiments}

%A particular critical case, in practice and even more conceptually, is randomized controlled experiments (RCT).
Randomized controlled experiments (RCTs) are a particular critical case.
Let \(Z\) be an individual-level indicator variable equal to~1 if the individual is \textit{assigned} to the treatment, and to~0 otherwise.
\(Z\) is determined exogenously by some randomization process (a workable version of tossing a coin).
Thus, it is reasonable to assume that \(Z \indep ((Y(0), Y(1))\), where \(\indep\) denotes independence between random variables.

In contrast, remember that \(D\) is an indicator equal to~1 if the individual is \textit{effectively treated}.
In social sciences and economics, individuals often preserve some leeway so that \(D \neq Z\), a situation called \textit{imperfect compliance}.
Moreover, they typically make their decision based on some unobserved characteristics also affecting the outcome so that we cannot sensibly assume that \(D\) is independent of \((Y(0), Y(1))\) (again the selection bias or endogeneity of the treatment).

However, if perfect compliance happens, that is, \(D = Z\), then the effective treatment is randomized and \(D \indep ((Y(0), Y(1))\).
This implies~\eqref{eq:assumptionAbsenceSelectionYdD}; hence the ATE and ATT are identified by a simple comparison of the average observed outcome~\(Y\) between treated and untreated individuals.

\paragraph{Multi-valued treatment and weighted average causal effects} 

We now consider the general case where the treatment is not restricted to be binary.
Of course, the endogeneity or selection bias issue remains crucial.
Interestingly, even neglecting this issue, 
leaving the heterogeneity of treatment effects unrestricted already entails some complications.
Indeed, under~\eqref{eq:assumptionLinearity} and~\eqref{eq:assumptionAbsenceSelectionYdD},
it is possible to show that the regression of the outcome on the treatment (and a constant) identifies 
\begin{equation*}
    \betaD
    = 
    \underbrace{\Esperancede{W \Delta}}_{=: \, \deltaW},
    \text{ where }
    W := \frac{(D - \Esperancede{D})^2}{\Variancede{D}}.
\end{equation*}
Thus, it is possible to identify a causal parameter: an average causal effect over the entire population, similar in that respect to the ATE, but it is a \textit{weighted} average.
The random weights~\(W\) are non-negative and normalized in the sense that \(\Esperancede{W} = 1\).

Non-negativity is appealing as it implies the so-called no-sign-reversal property: if \(\Delta \geq 0\) (respectively \(\leq 0\)) almost surely, then \(\betaD \geq 0\) (resp. \(\leq 0\))~a.s.
That property has been extensively studied in the recent literature devoted to two-way fixed effects (TWFE).
It is an important yet rather weak requirement of a causal parameter in the perspective of a social planner contemplating the decision to implement or not the treatment.
%Nonetheless, they are less directly interesting when considering the decision ultimately connected to the investigation of causal effects: should we implement or not the treatment?
Indeed, for a given individual, her contribution to the average is proportional to the distance between her treatment and the average treatment, \(W \propto (D - \Esperancede{D})^2\).
Such a weighting scheme is likely to be unrelated to the considerations of the social planner.
In particular, one could imagine situations where \(\deltaW > 0\) but \(\delta < 0\); in such cases, \(\betaD\) is misleading when the welfare criterion used by the social planner involves uniform weights.
%The weighted average effect suggests a positive impact of a policy (assuming that the social planner and agents enjoy higher values of the outcome), but implementing the treatment (say, increasing \(D\) by one unit for each individual) is negative in the perspective of a utilitarian social planner.

\subsection{Identification of the LATE}
\label{appendix:subsec:LATETheorem}

\paragraph{Potential treatment variables}
We introduce the LATE parameter in the simple setting of a binary treatment~\(D\) and a binary instrumental variable (IV)~\(Z\).
Again, the setting of a randomized experiment is both of practical and conceptual interest in this case, but, this time, with \textit{imperfect compliance} (\(D \neq Z\)).

Heterogeneity is not limited to treatment effects here.
%Heterogeneity arises in treatment effects but is more general. 
In particular, people may react differently to the assignment to the treatment.
To take that into account, a key idea of G.~Imbens and J.~Angrist is to introduce, in addition to potential outcome variables, \textit{potential treatment} variables: \(D(z)\) for \(z \in \{0,1\}%\supportde{Z}
\), \textit{i.e.}, what would the treatment have been with the assignment set to~\(z\).\footnote{
Footnote~2 of IA94 indicates that the notation was suggested to G.I and J.A by Gary Chamberlain.
}
In the binary case, \(D(0) = 1\), for example, represents the fact that even if not assigned to the treatment, this individual would be effectively treated; \(D(1) = 1\) that assigned to the treatment, this individual would follow the assignment and get effectively treated.
Similarly to the outcome, we have only access to the \textit{observed} treatment \(D := D(Z)\).

\paragraph{Exclusion restriction (the instrument should affect the outcome only through the treatment)}

The notation \(Y(d)\) %(before introducing the instrument~\(Z\))
for the potential outcome variable hides a fundamental assumption of a valid IV, namely that the instrument should affect the outcome \textit{only through} its effect on the treatment.
It is the exclusion restriction.
In the traditional presentation with simultaneous equations, the instrument is indeed excluded from the equation of the outcome.\footnote{
Section~\ref{subsubsec:comparison_LATE_SEM} compares the potential framework and simultaneous equations models.
Actually, the notation \(Y(d)\), or likewise \(Y(z,d)\) below, hides another assumption: the so-called SUTVA (Stable Unit Treatment Value Assumption) that rules out interference or externalities across individuals. 
We remark that AIR96 explicitly introduces SUTVA as an assumption while IA94 and AI95 implicitly assume it. 
}

A key general point of the LATE contribution is that identifying assumptions deserve to be made explicit.
To do so, let \(Y(z,d)\) for any \(z \in \{0,1\}%\supportde{Z}
\) and \(d \in \{0,1\}
%\supportde{D}
\) denote the potential outcome had the assignment been equal to~\(z\) and the treatment to~\(d\).
A priori, there may exist ``direct'' effects of the assignment on the outcome.
For instance, as discussed in Section~\ref{subsec:early_LATEs}, having a low lottery number in the Vietnam draft may induce going to college to avoid the draft. If so, there is an effect of the instrument besides veteran status.

Furthermore, %upstream to the issue of \textit{identifying} a causal effect, 
the exclusion restriction is useful to \textit{define} precisely \textit{the} causal effect of \(D\) on~\(Y\).\footnote{
Remember the core of Neyman-Rubin's model: causal effects are defined as differences in potential outcomes.
Before introducing the instrument~\(Z\), we thus naturally looked at \(Y(1) - Y(0)\).
Which difference is to be considered with the more general notation and modeling \(Y(z,d)\) in order to define the causal effect of~\(D\) on~\(Y\)?
It is unclear without the exclusion restriction.
}
It writes 
\begin{equation*}
    \label{eq:assumptionExclusionRestrictionBinary}
    \tag{\textnormal{E'}}
    \forall z, z' \in \{0,1\},%\supportde{Z}, 
    \,
    \forall d \in \{0,1\},%\supportde{D}, 
    \,
    Y(z,d) = Y(z',d).
\end{equation*}
As a consequence, since \(Y(z,d)\) does not vary with~\(z\), we can denote it again with possible values~\(d\) of the treatment as the unique argument: \(Y(d)\).
``It captures the notion underlying instrumental variables procedures that any effect of \(Z\) on \(Y\) must be via an effect of 
\(Z\) on \(D\)'' AIR96, page~447. Put differently, the individual causal effect of \(D\) on \(Y\) can again be defined as \(\Delta := Y(1) - Y(0)\), whatever the value of the assignment to treatment.

%%In the setting of a binary instrument, \eqref{eq:assumptionExclusionRestriction} means \(Y(0,d) = Y(1,d)\) and, thanks to it, the individual causal effect of \(D\) on \(Y\) can again be defined as \(\Delta := Y(1) - Y(0)\), whatever the value of the assignment to treatment.

\paragraph{(As-if) ``Random'' instruments} 

The next essential assumption is that, in ordinary language, the instrument is ``random''. %(mathematically, any instrument is ``random'' in the sense of being a random variable).
In the setting of an experiment, this results from the randomization of the assignment to treatment.
In probabilistic language, it translates into the fact that \(Z\) can be considered independent of all potential variables:
\begin{equation*}
\label{eq:assumptionIndepedenceInstrumentBinaryDZsimplification}
\tag{\textnormal{I'}}
    Z \indep \{D(0), D(1), Y(0), Y(1)\}.
\end{equation*}

As a consequence, comparisons between individuals assigned to the treatment and those not assigned identify the average causal effect of \(Z\) on~\(D\),
\begin{equation*}
    \Esperancedecond{D}{Z = 1}
    -
    \Esperancedecond{D}{Z = 0}
    =
    \Esperancede{D(1) - D(0)},    
\end{equation*}
and the average causal effect of \(Z\) on \(Y\), which is the so-called average \textit{Intention-To-Treat} (ITT) effect,
\begin{equation*}
    \Esperancedecond{Y}{Z = 1}
    -
    \Esperancedecond{Y}{Z = 0}
    =
    \Esperancede{Y(1, D(1)) - Y(0, D(0))}.
\end{equation*}
Note that we do not use~\eqref{eq:assumptionExclusionRestrictionBinary} to express the ITT.
\eqref{eq:assumptionExclusionRestrictionBinary} restricts the effect of~\(Z\) on~\(Y\): it says there is no (direct) causal effect of~\(Z\) on~\(Y\).
Under~\eqref{eq:assumptionExclusionRestrictionBinary}, 
\begin{equation*}
    Y(1, D(1)) - Y(0, D(0))
    =
    Y(D(1)) - Y(D(0)),
\end{equation*}
which is \textit{not} \(Y(1) - Y(0)\) since it depends on the values of~\(D(1)\) and~\(D(0)\).
To anticipate %(or it would be another path to motivate it) 
the introduction of compliers, remark that it is equal to the individual causal effect \(\Delta\) of~\(D\) on~\(Y\) if and only if \(D(1) = 1\) and~\(D(0) = 0\); that is, exactly for the complier population who will be characterized by this expression.

\paragraph{Wald estimand and relevance of the instrument} 

At this stage, let us recall the expression of the limit in probability (under standard moment conditions) of the Wald estimator, which coincides with the more general TSLS estimator in this setting without covariate and a single binary instrument,
\begin{equation*}
    \betaWald := 
    \frac{\Esperancedecond{Y}{Z = 1}
    -
    \Esperancedecond{Y}{Z = 0}}{\Esperancedecond{D}{Z = 1}
    -
    \Esperancedecond{D}{Z = 0}}
    =
    {f(\Distributionde{Y,D,Z})}, 
    \text{ hence \(\betaWald\) is identified}.
\end{equation*}

Properly defining \(\betaWald\) requires a non-zero denominator.
Under (I), this corresponds to the relevance condition of the instrument (with a binary instrument):
\begin{equation*}
\label{eq:assumptionRelevanceInstrumentBinaryZ}
\tag{\textnormal{R}}
\Esperancede{D(1) - D(0)} \neq 0.    
\end{equation*}
\eqref{eq:assumptionRelevanceInstrumentBinaryZ} ensures the average causal effect of \(Z\) on \(D\) is not null. We remark that \eqref{eq:assumptionRelevanceInstrumentBinaryZ} implies $\Probabilitede{D(1) \neq D(0)} > 0$ meaning that at least a subset of the population reacts to the instrument in terms of treatment choice. The compliers which are at the heart of the identification of the LATE based on \(\betaWald\) belong to that subset of the population.

\begin{comment}
In more general settings, the formalization of the relevance condition may slightly differ.
Nonetheless, its intuition remains that the instrument~\(Z\) should have a non-zero causal effect on the treatment~\(D\).
It is also known as the ``first-stage'' condition.
In AI95 where the treatment variable is multi-valued, \(\supportde{D} = \{0, 1, \ldots, J\}\), the relevance condition is for instance written as
\begin{equation*}
\label{eq:assumptionRelevanceInstrumentAI95}
\tag{\textnormal{R-AI95}}
\exists j  \in \{0, 1, \ldots, J\}: \,
\Probabilitede{D(1) \geq j > D(0)} > 0.
\end{equation*}
Anticipating the monotonicity assumption, under~\eqref{eq:assumptionMontonicityBinaryZ}, 
we can show that 
\eqref{eq:assumptionRelevanceInstrumentBinaryZ} is equivalent to \(\Probabilitede{D(1) > D(0)} > 0\), because, under monotonicity, \(D(1) - D(0)\) takes only values in \(\{0, 1\}\)
and \(D(1) > D(0)\) is the only way to differ from~0 for that difference.
Thus, when the treatment is binary (\(J = 1\) formally), \eqref{eq:assumptionRelevanceInstrumentBinaryZ} is indeed a particular case of \eqref{eq:assumptionRelevanceInstrumentAI95} (for \(j = 1\)) since the event \(\{D(1) \geq 1 > D(0)\}\) is the same as \(\{D(1) = 1, D(0) = 0\}\).
In other words, it says there are some compliers: individuals whose treatment reacts to the instrument.
\end{comment}

\paragraph{Towards monotonicity}

Satisfying \eqref{eq:assumptionExclusionRestrictionBinary}, \eqref{eq:assumptionIndepedenceInstrumentBinaryDZsimplification}, and \eqref{eq:assumptionRelevanceInstrumentBinaryZ} means that the instrument is \textit{valid}.
However, the starting point of the LATE contribution is that a valid instrument ``is not enough to identify \textit{any} average treatment effect'' (IA94, page~468). 

Let us present the original equations of the LATE trilogy to have a sense of this issue.%explaining that and then the proof of the LATE theorem.
\footnote{
See 
AI95, Equations~(3), (4), and~(5), page~434;
AIR96, Equations~(9) and~(10), p448; with multi-valued instruments, IA94, Equation~(1), p469.
}
To begin with, using~\eqref{eq:assumptionExclusionRestrictionBinary} for the first equality, the fact that \(D(z)\) is binary for the second, and straightforward computations for the third, we have
\begin{align*}
    Y(1, D(1)) - Y(0, D(0))
    & 
    =
    %\overset{\text{(E)}}{=}
    Y(D(1)) - Y(D(0)) \\
    &
    =
    %\overset{D(z) \text{ is binary}}{=}
    \big[ Y(1) D(1) + Y(0) (1 - D(1)) \big]
    - \big[ Y(1) D(0) + Y(0) (1 - D(0)) \big] \\
    &
    =
    %\overset{\text{computations}}{=}
    [Y(1) - Y(0)] \times [D(1) - D(0)].
\end{align*}

Since the treatment is binary, the difference \(D(1) - D(0)\) takes value in \(\{-1, 0, 1\}\).
Thus, applying the expectation in the previous equality, we obtain
\begin{align}
\label{eq:computation_ITT_without_M}
    & \overbrace{\Esperancede{Y(1, D(1)) - Y(0, D(0))}}^{\text{identified by } \Esperancedecond{Y}{Z = 1} - \Esperancedecond{Y}{Z = 0}}
    = \nonumber \\ 
    & \qquad \quad \Esperancedecond{Y(1) - Y(0)}{D(1) - D(0) = 1}
    \times 
    \Probabilitede{D(1) - D(0) = 1}
    \nonumber \\ 
    & \qquad \qquad + \Esperancedecond{Y(1) - Y(0)}{D(1) - D(0) = -1}
    \times 
    \big( - \Probabilitede{D(1) - D(0) = -1} \big)   
\end{align}
which is a linear combination of average causal effects on some subpopulations but not a convex one. 
The weights, \(\Probabilitede{D(1) - D(0) = 1}\) and \(- \Probabilitede{D(1) - D(0) = -1}\), are not non-negative (for the latter) and they do not sum to one.
As a consequence, the quantity in~\eqref{eq:computation_ITT_without_M} ``can be zero or even negative despite a strictly positive causal effect of \(D\) on \(Y\) for all individuals'' (IA94, p469).
``Intuitively, the problem here is that the treatment effect for those who shift from nonparticipation to participation [to being effectively treated] when \(Z\) is switched from [0] to [1] can be cancelled out by the treatment effect of those who shift from participation to nonparticipation.'' (IA94, p469).

What to do, then? One possibility is to rule out heterogeneity and assume a homogeneous causal effect: \(\Delta = \deltazero \in \Reels\).
\(\deltazero\) is then identified. This is however not the direction taken by G.I and J.A.
%(\textit{i})~One possibility is to rule out heterogeneity and assume a homogeneous causal effect: \(\Delta = \deltazero \in \Reels\).
%\(\deltazero\) is then identified.
%(\textit{ii})~Another one would be to have the combination of a situation (for instance, one-sided noncompliance in a classical binary treatment binary instrument RCT setting) and an instrument (for instance, a sufficiently rich/continuous instrument) such that for some value~\(z \in \supportde{Z}\), the probability of being treated conditional on that value, \(\Esperancedecond{D}{Z = z}\), is equal to zero.
%Then, the ATT is identified as recalled in IA94, p469.

\paragraph{The monotonicity assumption} 

%Maybe, the core of the LATE contribution comes from the following assumption and thinking.
These authors try to answer the following question: are there reasonable assumptions which allow to identify an average causal effect from \(\betaWald\) without imposing strong homogeneity restrictions on treatment effects and treatment assignment? %so that (\textit{i})~is not an acceptable solution.
%Furthermore, (\textit{ii})~is often too restrictive in practice.
%Is it possible to find another solution with alternative assumptions to identify an average causal effect?

It would be the case if one of the two weights in \eqref{eq:computation_ITT_without_M} is null, for example, if the event \(\{D(1) - D(0) = -1\}\) occurs with null probability. In the binary instrument and treatment case, the latter is in fact equivalent to the so-called monotonicity assumption
\begin{equation*}
\label{eq:assumptionMontonicityBinaryZ_appendix}
\tag{\textnormal{M'}}
    D(1) \geq D(0) \text{ almost surely}.
\end{equation*}
In the setting of an RCT with imperfect compliance, \eqref{eq:assumptionMontonicityBinaryZ_appendix} means there is no defier, \textit{i.e.}, individuals who would be treated when not assigned to the treatment but who would not be treated when assigned to the treatment. It is worth noting that~\eqref{eq:assumptionMontonicityBinaryZ_appendix}~does limit heterogeneity across individuals.
However, instead of limiting the heterogeneity related to the causal effect of the treatment on the outcome, it restricts the direction of the effect of the instrument on the treatment.\footnote{
For instance, J.A and G.I write the following to present the monotonicity condition:
``Instead of restricting treatment effect heterogeneity, in this article we impose a nonparametric restriction on the process determining [\(D\)] as a function of \(Z\)'', AI95, p434 (the treatment is denoted~\(S\) instead of~\(D\) in this article; see Section~\ref{subsubsec:AI95} for more details on~AI95).
}
The argument and strength of the LATE identification result is that, in various applied settings, the latter restriction is more credible than the former. Indeed, assumption~\eqref{eq:assumptionMontonicityBinaryZ_appendix}, which could be seen as an ad hoc assumption at first sight, captures sensible (economic) assumptions in many settings, as illustrated for instance in the various applied works of J.~Angrist.

\paragraph{The textbook LATE Theorem}

We can finally state the so-called LATE theorem, in the binary treatment binary instrument framework (with no covariates).

Under the assumptions~\eqref{eq:assumptionExclusionRestrictionBinary}, \eqref{eq:assumptionIndepedenceInstrumentBinaryDZsimplification}, \eqref{eq:assumptionRelevanceInstrumentBinaryZ}, and~\eqref{eq:assumptionMontonicityBinaryZ_appendix},
\begin{equation}
\label{eq:LATEtheoremBinaryZBinaryDnoX}
    \underbrace{\frac{\Esperancedecond{Y}{Z = 1}
    -
    \Esperancedecond{Y}{Z = 0}}{\Esperancedecond{D}{Z = 1}
    -
    \Esperancedecond{D}{Z = 0}}}_{=: \, \betaWald}
    \, = \,
    \underbrace{\Esperancedecond{Y(1) - Y(0)}{D(1) - D(0) = 1}}_{=: \, \deltaComplier},
\end{equation}
where \(\deltaComplier\) is the Local Average Treatment Effect (LATE), \textit{local} in the sense that it is the average causal effect among a \textit{subpopulation}.
This subpopulation is made of agents whose treatment reacts to the instrument.
They are characterized by \(D(0) = 0\) and \(D(1) = 1\), which is the only possible reaction given monotonicity, and named \textit{compliers} since they comply to the assignment (in the typical setting of an RCT with imperfect compliance).
For this reason, \(\deltaComplier\) is also known as the CATE (or CACE), Complier Average Treatment/Causal Effect.

\medskip 

The end of the proof works as follows.
Thanks to~\eqref{eq:assumptionMontonicityBinaryZ_appendix}, the quantity in \eqref{eq:computation_ITT_without_M}, which is the numerator of~\(\betaWald\), reduces to
\begin{equation*}
    \Esperancedecond{Y}{Z = 1}
    -
    \Esperancedecond{Y}{Z = 0}
    =
    \underbrace{\Esperancedecond{Y(1) - Y(0)}{D(1) - D(0) = 1}}_{\deltaComplier}
    \times  \, 
    \Probabilitede{D(1) - D(0) = 1}.
\end{equation*}
It remains to show that the denominator of~\(\betaWald\) is equal to this probability \(\Probabilitede{D(1) - D(0) = 1}\).

To do so, remark that, with binary treatment and binary instrument, there are four exclusive possibilities for the values of the couple \((D(0), D(1))\), which define the \textit{type} of each individual:\footnote{
Some remarks about the terminology: 
IA94 uses the verb only: ``The treatment effect identified here is the average treatment effect for those who always comply with their assignment.'' (IA94, p472);
AI95 (p434) talks about ``switcher-in'' and ``switcher-out'';
AIR96 (p448) uses the now standard AT, NT, C, and D and notes that the term ``defiers'' was suggested originally by A.~Balke and J.~Pearl.
}
\begin{center}
\begin{tabular}{l | c |  c }
         & \({D(1)} = 0\) & \({D(1)} = 1\)  \\
        \hline 
        \({D(0)} = 0\) & never-taker (NT) & complier (C) \\
        \({D(0)} = 1\) & defier (D) & always-taker (AT)
\end{tabular}
\end{center}
Assumption~\eqref{eq:assumptionMontonicityBinaryZ_appendix} rules out the presence of defiers.
Consequently, by the law of total probability, 
\( \Probabilitede{\eventC} := \Probabilitede{D(1) - D(0) = 1} = 1 - \Probabilitede{\eventAT} - \Probabilitede{\eventNT}\).
Furthermore, a unit is an always-taker if and only if they would be treated when not assigned to the treatment, that is,
\(\{\eventAT\} = \{D(0) = 1\}\), and, similarly, \(\{\eventNT\} = \{D(1) = 0\}\).
In addition, by definition of \(D := D(Z)\) and by assumption~\eqref{eq:assumptionIndepedenceInstrumentBinaryDZsimplification}, for any~\(z \in \{0,1\}\), \(\Esperancedecond{D}{Z = z} = \Esperancede{D(z)}\).
Finally, since \(D(z)\) is a Bernoulli variable,
\(\Esperancede{D(z)} = \Probabilitede{D(z) = 1}\).
Therefore,
\begin{align*}
    \Probabilitede{\eventC}
    :=
    \Probabilitede{D(1) - D(0) = 1}
    =
    1 - \Esperancede{D(0)} - \big\{ 1 - \Esperancede{D(1)} \big\}
    =
    \Esperancedecond{D}{Z = 1}
    -
    \Esperancedecond{D}{Z = 0},
\end{align*}
which is indeed the denominator of~\(\betaWald\).

\paragraph{IV and TSLS estimators}
\(\betaWald\) is a special case (when \(\supportde{Z} = \{0, 1\}\)) of the probability limit of the standard Instrumental Variable (IV) or Two-Stage Least Squares (TSLS) estimator of the coefficient associated with~\(D\) in the simple linear regression of~\(Y\) on \(D\) and a constant when \(D\) is instrumented by~\(Z\):
\begin{equation*}
    \underbrace{\frac{\Covariancede{Y}{Z}}{\Covariancede{Y}{D}}}_{=: \, \betaIV}
    \, = \,
    \frac{\Esperancedecond{Y}{Z = 1}
    -
    \Esperancedecond{Y}{Z = 0}}{\Esperancedecond{D}{Z = 1}
    -
    \Esperancedecond{D}{Z = 0}}
    \, \text{ when \(Z\) is binary}.
\end{equation*}

\(\betaIV\) is the estimand of the IV estimator used in the setting of a \textit{single} endogenous regressor (possibly multi-valued, not binary) and a single instrument (possibly multi-valued) without covariates.
One can show that, denoting \(\Dstar\) the theoretical prediction of~\(D\) from the first-stage regression, \(\betaIV = \Covariancede{Y}{\Dstar} \,/\, \Variancede{\Dstar}\).

It is possible to extend the IV/TSLS estimation procedure with the inclusion of covariates~\(X\) (and possibly also multi-dimensional instruments and regressors more generally).

\medskip 

We recall these elements and names to clarify the terminology that we use in this review, following the LATE articles.
Overall (it fluctuates sometimes), in the LATE trilogy, the authors reserve the name ``IV estimator'' when there is a single instrument, a single endogenous regressor, and no control variables; its limit in probability is~\(\betaIV\) under mild moment conditions.
In contrast, ``TSLS'' tends to refer to more general settings that include control variables, which are typically denoted by~\(X\), and possibly multi-dimensional instruments.
For instance, J.~Angrist and G.~Imbens write
``As a practical matter, this assumption [independence] may be true only after conditioning on covariates. (In fact, \textit{the need to control for covariates sometimes motivates the use of TSLS instead of IV}.)'' (AI95, p434; we italicize).

\end{document}

%% file: Commands.tex
\newcommand*{\simiid}{\overset{\text{i.i.d.}}{\sim}}

\newcommand*{\Indicatricede}[1]{\mathds{1}\!\left\{#1\right\}}

\newcommand*{\Reels}{\mathbb{R}}

\newcommand*{\eps}{\varepsilon}

\DeclareMathOperator{\Esperance}{\mathbb{E}}
\newcommand*{\Esperancede}[1]{\Esperance[#1]}

\DeclareMathOperator{\Quantile}{\mathbb{Q}}

\newcommand*{\Quantiledeconda}[3]{\Quantile_{#3}[#1 \conditionnellementa #2]}

\newcommand*{\Esperancedecond}[2]{\Esperance[#1 \conditionnellementa #2]}

\newcommand*{\Variance}{\mathbb{V}}
\newcommand*{\Variancede}[1]{\Variance[#1]}

\newcommand*{\Probabilite}{\mathbb{P}}
\newcommand*{\Probabilitede}[1]{\Probabilite\!\left(#1\right)}
\newcommand*{\Probabilitedecond}[2]{\Probabilite\!\left(#1 \conditionnellementa #2 \right)}

\newcommand*{\conditionnellementa}{\,|\,}

\newcommand*{\Distribution}{\mathrm{P}}
\newcommand*{\Distributionde}[1]{\Distribution_{#1}}
\newcommand*{\Distributiondecond}[2]{\Distribution_{#1 \conditionnellementa #2}}

\newcommand*{\indep}{\perp\mkern-9.5mu\perp}

\newcommand*{\deltazero}{\delta_0}

\newcommand*{\deltaW}{\delta^W}

\newcommand*{\Covariance}{\mathbb{C}\mathrm{ov}}
\newcommand*{\Covariancede}[2]{\Covariance(#1, #2)}

\newcommand*{\Covariancedecond}[3]{\Covariance\!\left(#1, #2 \conditionnellementa #3 \right)}

\newcommand*{\support}{\textrm{Support}}
\newcommand*{\supportde}[1]{\support(#1)}

\newcommand*{\betaWald}{\beta_{\textnormal{Wald}}}
\newcommand*{\betaIV}{\beta_{\textnormal{IV}}}
\newcommand*{\betaTSLS}{\beta_{\textnormal{TSLS}}}

\newcommand*{\betaD}{\beta_D}

\newcommand*{\Dstar}{D^{*}}

\newcommand*{\Dlatent}{D_{\textnormal{latent}}}

\newcommand*{\deltaComplier}{\delta^\textnormal{C}}

\newcommand*{\deltaDefier}{\delta^\textnormal{D}}

\newcommand*{\deltatreated}{\delta^{\textnormal{T}}}

\newcommand*{\deltaITT}{\delta^{\textnormal{ITT}}}

\newcommand*{\eventC}{\textnormal{C}}

\newcommand*{\eventAT}{\textnormal{AT}}
\newcommand*{\eventNT}{\textnormal{NT}}

%% file: paper.bbl
\begin{thebibliography}{104}
\newcommand{\enquote}[1]{``#1''}
\expandafter\ifx\csname natexlab\endcsname\relax\def\natexlab#1{#1}\fi

\bibitem[\protect\citeauthoryear{Abadie}{Abadie}{2003}]{abadie2003}
\textsc{Abadie, A.} (2003): \enquote{Semiparametric instrumental variable
  estimation of treatment response models,} \emph{Journal of Econometrics},
  113, 231--263.

\bibitem[\protect\citeauthoryear{Abadie, Angrist, and Imbens}{Abadie
  et~al.}{2002}]{abadie_angrist_imbens_2002}
\textsc{Abadie, A., J.~Angrist, and G.~Imbens} (2002): \enquote{Instrumental
  variables estimates of the effect of subsidized training on the quantiles of
  trainee earnings,} \emph{Econometrica}, 70, 91--117.

\bibitem[\protect\citeauthoryear{Abadie, Athey, Imbens, and Wooldridge}{Abadie
  et~al.}{2020}]{abadie2020sampling}
\textsc{Abadie, A., S.~Athey, G.~Imbens, and J.~Wooldridge} (2020):
  \enquote{Sampling-Based versus Design-Based Uncertainty in Regression
  Analysis,} \emph{Econometrica}, 88, 265--296.

\bibitem[\protect\citeauthoryear{Abadie and Imbens}{Abadie and
  Imbens}{2006}]{abadie_imbens_2006}
\textsc{Abadie, A. and G.~Imbens} (2006): \enquote{Large Sample Properties of
  Matching Estimators for Average Treatment Effects,} \emph{Econometrica}, 74,
  235--267.

\bibitem[\protect\citeauthoryear{Abadie and Imbens}{Abadie and
  Imbens}{2008}]{abadie_imbens_2008}
---\hspace{-.1pt}---\hspace{-.1pt}--- (2008): \enquote{On the Failure of the
  Bootstrap for Matching Estimators,} \emph{Econometrica}, 76, 1537--1557.

\bibitem[\protect\citeauthoryear{Abadie and Imbens}{Abadie and
  Imbens}{2011}]{abadie_imbens_2011}
---\hspace{-.1pt}---\hspace{-.1pt}--- (2011): \enquote{Bias-Corrected Matching
  Estimators for Average Treatment Effects,} \emph{Journal of Business \&
  Economic Statistics}, 29, 1--11.

\bibitem[\protect\citeauthoryear{Abadie and Imbens}{Abadie and
  Imbens}{2012}]{abadie_imbens_2012}
---\hspace{-.1pt}---\hspace{-.1pt}--- (2012): \enquote{A Martingale
  Representation for Matching Estimators,} \emph{Journal of the American
  Statistical Association}, 107, 833--843.

\bibitem[\protect\citeauthoryear{Abadie and Imbens}{Abadie and
  Imbens}{2016}]{abadie_imbens_2016}
---\hspace{-.1pt}---\hspace{-.1pt}--- (2016): \enquote{Matching on the
  Estimated Propensity Score,} \emph{Econometrica}, 84, 781--807.

\bibitem[\protect\citeauthoryear{Abdulkadiroglu, Angrist, Dynarski, Kane, and
  Pathak}{Abdulkadiroglu et~al.}{2011}]{adbulkadiroglu_etal_boston_2011}
\textsc{Abdulkadiroglu, A., J.~Angrist, S.~Dynarski, T.~Kane, and P.~Pathak}
  (2011): \enquote{Accountability and Flexibility in Public Schools: Evidence
  from Boston's Charters And Pilots,} \emph{The Quarterly Journal of
  Economics}, 126, 699--748.

\bibitem[\protect\citeauthoryear{Abdulkadiroglu, Angrist, Hull, and
  Pathak}{Abdulkadiroglu et~al.}{2016}]{abdulkadiroglu_etal_nola_boston_2016}
\textsc{Abdulkadiroglu, A., J.~Angrist, P.~Hull, and P.~Pathak} (2016):
  \enquote{Charters without Lotteries: Testing Takeovers in New Orleans and
  Boston,} \emph{American Economic Review}, 106, 1878--1920.

\bibitem[\protect\citeauthoryear{Abdulkadiroglu, Angrist, Narita, and
  Pathak}{Abdulkadiroglu
  et~al.}{2017{\natexlab{a}}}]{abdulkadiroglu_etal_denver_2017}
\textsc{Abdulkadiroglu, A., J.~Angrist, Y.~Narita, and P.~Pathak}
  (2017{\natexlab{a}}): \enquote{Research Design Meets Market Design: Using
  Centralized Assignment for Impact Evaluation,} \emph{Econometrica}, 85,
  1373--1432.

\bibitem[\protect\citeauthoryear{Abdulkadiroglu, Angrist, Narita, Pathak, and
  Zarate}{Abdulkadiroglu
  et~al.}{2017{\natexlab{b}}}]{abdulkadiroglu_etal_chicago_2017}
\textsc{Abdulkadiroglu, A., J.~Angrist, Y.~Narita, P.~Pathak, and R.~Zarate}
  (2017{\natexlab{b}}): \enquote{Regression Discontinuity in Serial
  Dictatorship: Achievement Effects at Chicago's Exam Schools,} \emph{The
  American Economic Review}, 107, 240--245.

\bibitem[\protect\citeauthoryear{Abdulkadiroglu, Angrist, and
  Pathak}{Abdulkadiroglu et~al.}{2014}]{abdulkadiroglu_etal_elite_2014}
\textsc{Abdulkadiroglu, A., J.~Angrist, and P.~Pathak} (2014): \enquote{The
  Elite Illusion: Achievement Effects at Boston and New York Exam Schools,}
  \emph{Econometrica}, 82, 137--196.

\bibitem[\protect\citeauthoryear{Acemoglu and Angrist}{Acemoglu and
  Angrist}{2001}]{acemoglu_angrist_2001}
\textsc{Acemoglu, D. and J.~Angrist} (2001): \enquote{Consequences of
  Employment Protection? The Case of the Americans with Disabilities Act,}
  \emph{Journal of Political Economy}, 109, 915--957.

\bibitem[\protect\citeauthoryear{Angrist}{Angrist}{1990}]{angrist_1990}
\textsc{Angrist, J.} (1990): \enquote{Lifetime Earnings and the Vietnam Era
  Draft Lottery: Evidence from Social Security Administrative Records,}
  \emph{The American Economic Review}, 80, 313--336.

\bibitem[\protect\citeauthoryear{Angrist}{Angrist}{2006}]{angrist_2006}
---\hspace{-.1pt}---\hspace{-.1pt}--- (2006): \enquote{Instrumental variables
  methods in experimental criminological research: what, why and how,}
  \emph{Journal of Experimental Criminology}, 2, 23--44.

\bibitem[\protect\citeauthoryear{Angrist}{Angrist}{2022}]{angrist_2022}
---\hspace{-.1pt}---\hspace{-.1pt}--- (2022): \enquote{Empirical Strategies in
  Economics: Illuminating the Path from Cause to Effect,} \emph{Econometrica},
  90, 2509--2539.

\bibitem[\protect\citeauthoryear{Angrist, Bettinger, Bloom, King, and
  Kremer}{Angrist et~al.}{2002}]{angrist_et_al_2002}
\textsc{Angrist, J., E.~Bettinger, E.~Bloom, E.~King, and M.~Kremer} (2002):
  \enquote{Vouchers for Private Schooling in Colombia: Evidence from a
  Randomized Natural Experiment,} \emph{American Economic Review}, 92,
  1535--1558.

\bibitem[\protect\citeauthoryear{Angrist, Bettinger, and Kremer}{Angrist
  et~al.}{2006{\natexlab{a}}}]{angrist_et_al_2006}
\textsc{Angrist, J., E.~Bettinger, and M.~Kremer} (2006{\natexlab{a}}):
  \enquote{Long-Term Educational Consequences of Secondary School Vouchers:
  Evidence from Administrative Records in Colombia,} \emph{American Economic
  Review}, 96, 847--862.

\bibitem[\protect\citeauthoryear{Angrist and Chen}{Angrist and
  Chen}{2011}]{angrist_chen_2011}
\textsc{Angrist, J. and S.~Chen} (2011): \enquote{Schooling and the Vietnam-Era
  GI Bill: Evidence from the Draft Lottery,} \emph{American Economic Journal:
  Applied Economics}, 3, 96--118.

\bibitem[\protect\citeauthoryear{Angrist, Chen, and Frandsen}{Angrist
  et~al.}{2010{\natexlab{a}}}]{angrist_chen_frandsen_2010}
\textsc{Angrist, J., S.~Chen, and B.~Frandsen} (2010{\natexlab{a}}):
  \enquote{Did Vietnam veterans get sicker in the 1990s? The complicated
  effects of military service on self-reported health,} \emph{Journal of Public
  Economics}, 94, 824--837.

\bibitem[\protect\citeauthoryear{Angrist, Chernozhukov, and
  Fernández-Val}{Angrist
  et~al.}{2006{\natexlab{b}}}]{angrist_cherno_fernandezval_2006}
\textsc{Angrist, J., V.~Chernozhukov, and I.~Fernández-Val}
  (2006{\natexlab{b}}): \enquote{Quantile Regression under Misspecification,
  with an Application to the U.S. Wage Structure,} \emph{Econometrica}, 74,
  539--563.

\bibitem[\protect\citeauthoryear{Angrist, Cohodes, Dynarski, Pathak, and
  Walters}{Angrist et~al.}{2016}]{angrist_etal_boston_2016}
\textsc{Angrist, J., S.~Cohodes, S.~Dynarski, P.~Pathak, and C.~Walters}
  (2016): \enquote{Stand and Deliver: Effects of Boston’s Charter High
  Schools on College Preparation, Entry, and Choice,} \emph{Journal of Labor
  Economics}, 34, 275--318.

\bibitem[\protect\citeauthoryear{Angrist, Dynarski, Kane, Pathak, and
  Walters}{Angrist et~al.}{2012}]{angrist_etal_boston_2012}
\textsc{Angrist, J., S.~Dynarski, T.~Kane, P.~Pathak, and C.~Walters} (2012):
  \enquote{Who Benefits from KIPP?} \emph{Journal of Policy Analysis and
  Management}, 31, 837--860.

\bibitem[\protect\citeauthoryear{Angrist and Evans}{Angrist and
  Evans}{1998}]{angrist_evans_1998}
\textsc{Angrist, J. and W.~Evans} (1998): \enquote{Children and Their Parents'
  Labor Supply: Evidence from Exogenous Variation in Family Size,} \emph{The
  American Economic Review}, 88, 450--477.

\bibitem[\protect\citeauthoryear{Angrist and Fernandez-Val}{Angrist and
  Fernandez-Val}{2010}]{angrist_fernandezval_2010}
\textsc{Angrist, J. and I.~Fernandez-Val} (2010): \enquote{ExtrapoLATE-ing:
  External Validity and Overidentification in the LATE Framework,} Working
  Paper 16566, National Bureau of Economic Research.

\bibitem[\protect\citeauthoryear{Angrist, Graddy, and Imbens}{Angrist
  et~al.}{2000}]{angrist_graddy_imbens_2000}
\textsc{Angrist, J., K.~Graddy, and G.~Imbens} (2000): \enquote{The
  Interpretation of Instrumental Variables Estimators in Simultaneous Equations
  Models with an Application to the Demand for Fish,} \emph{The Review of
  Economic Studies}, 67, 499--527.

\bibitem[\protect\citeauthoryear{Angrist and Guryan}{Angrist and
  Guryan}{2004}]{angrist_guryan_2004}
\textsc{Angrist, J. and J.~Guryan} (2004): \enquote{Teacher Testing, Teacher
  Education, and Teacher Characteristics,} \emph{American Economic Review}, 94,
  241--246.

\bibitem[\protect\citeauthoryear{Angrist and Imbens}{Angrist and
  Imbens}{1995}]{angrist_imbens_1995}
\textsc{Angrist, J. and G.~Imbens} (1995): \enquote{Two-Stage Least Squares
  Estimation of Average Causal Effects in Models with Variable Treatment
  Intensity,} \emph{Journal of the American Statistical Association}, 90,
  431--442.

\bibitem[\protect\citeauthoryear{Angrist, Imbens, and Rubin}{Angrist
  et~al.}{1996}]{angrist_imbens_rubin_1996}
\textsc{Angrist, J., G.~Imbens, and D.~Rubin} (1996): \enquote{Identification
  of Causal Effects Using Instrumental Variables,} \emph{Journal of the
  American Statistical Association}, 91, 444--455.

\bibitem[\protect\citeauthoryear{Angrist and Krueger}{Angrist and
  Krueger}{1991}]{angrist_krueger_1991}
\textsc{Angrist, J. and A.~Krueger} (1991): \enquote{Does Compulsory School
  Attendance Affect Schooling and Earnings?} \emph{The Quarterly Journal of
  Economics}, 106, 979--1014.

\bibitem[\protect\citeauthoryear{Angrist and Kugler}{Angrist and
  Kugler}{2003}]{angrist_kugler_2003}
\textsc{Angrist, J. and A.~Kugler} (2003): \enquote{Protective or
  Counter-Productive? Labour Market Institutions and the Effect of Immigration
  on EU Natives,} \emph{The Economic Journal}, 113, F302--F331.

\bibitem[\protect\citeauthoryear{Angrist, Lang, and Oreopoulos}{Angrist
  et~al.}{2009}]{angrist_lang_oreopoulos_2009}
\textsc{Angrist, J., D.~Lang, and P.~Oreopoulos} (2009): \enquote{Incentives
  and Services for College Achievement: Evidence from a Randomized Trial,}
  \emph{American Economic Journal: Applied Economics}, 1, 136--63.

\bibitem[\protect\citeauthoryear{Angrist and Lang}{Angrist and
  Lang}{2004}]{angrist_lang_2004}
\textsc{Angrist, J. and K.~Lang} (2004): \enquote{Does School Integration
  Generate Peer Effects? Evidence from Boston's Metco Program,} \emph{American
  Economic Review}, 94, 1613--1634.

\bibitem[\protect\citeauthoryear{Angrist and Lavy}{Angrist and
  Lavy}{1999}]{angrist_lavy_1999}
\textsc{Angrist, J. and V.~Lavy} (1999): \enquote{Using Maimonides' Rule to
  Estimate the Effect of Class Size on Scholastic Achievement,} \emph{The
  Quarterly Journal of Economics}, 114, 533--575.

\bibitem[\protect\citeauthoryear{Angrist and Lavy}{Angrist and
  Lavy}{2001}]{angrist_lavy_2001}
---\hspace{-.1pt}---\hspace{-.1pt}--- (2001): \enquote{Does Teacher Training
  Affect Pupil Learning? Evidence from Matched Comparisons in Jerusalem Public
  Schools,} \emph{Journal of Labor Economics}, 19, 343--369.

\bibitem[\protect\citeauthoryear{Angrist and Lavy}{Angrist and
  Lavy}{2009}]{angrist_lavy_2009}
---\hspace{-.1pt}---\hspace{-.1pt}--- (2009): \enquote{The Effects of High
  Stakes High School Achievement Awards: Evidence from a Randomized Trial,}
  \emph{The American Economic Review}, 99, 1384--1414.

\bibitem[\protect\citeauthoryear{Angrist, Lavy, Leder-Luis, and Shany}{Angrist
  et~al.}{2019}]{angrist_lavy_etal_2019}
\textsc{Angrist, J., V.~Lavy, J.~Leder-Luis, and A.~Shany} (2019):
  \enquote{Maimonides' Rule Redux,} \emph{American Economic Review: Insights},
  1, 309--24.

\bibitem[\protect\citeauthoryear{Angrist, Lavy, and Schlosser}{Angrist
  et~al.}{2010{\natexlab{b}}}]{angrist_lavy_schlosser_2010}
\textsc{Angrist, J., V.~Lavy, and A.~Schlosser} (2010{\natexlab{b}}):
  \enquote{Multiple Experiments for the Causal Link between the Quantity and
  Quality of Children,} \emph{Journal of Labor Economics}, 28, 773--824.

\bibitem[\protect\citeauthoryear{Angrist, Pathak, and Walters}{Angrist
  et~al.}{2013}]{angrist_etal_boston_2013}
\textsc{Angrist, J., P.~Pathak, and C.~Walters} (2013): \enquote{Explaining
  Charter School Effectiveness,} \emph{American Economic Journal: Applied
  Economics}, 5, 1--27.

\bibitem[\protect\citeauthoryear{Angrist and Pischke}{Angrist and
  Pischke}{2008}]{angrist_pischke_2008}
\textsc{Angrist, J. and J.-S. Pischke} (2008): \emph{Mostly Harmless
  Econometrics: An Empiricist's Companion}, Princeton University Press.

\bibitem[\protect\citeauthoryear{Angrist and Pischke}{Angrist and
  Pischke}{2010}]{angrist_pischke_2010}
---\hspace{-.1pt}---\hspace{-.1pt}--- (2010): \enquote{The Credibility
  Revolution in Empirical Economics: How Better Research Design Is Taking the
  Con out of Econometrics,} \emph{Journal of Economic Perspectives}, 24, 3--30.

\bibitem[\protect\citeauthoryear{Angrist and Rokkanen}{Angrist and
  Rokkanen}{2015}]{angrist_rokkanen_2015}
\textsc{Angrist, J. and M.~Rokkanen} (2015): \enquote{Wanna Get Away?
  Regression Discontinuity Estimation of Exam School Effects Away From the
  Cutoff,} \emph{Journal of the American Statistical Association}, 110,
  1331--1344.

\bibitem[\protect\citeauthoryear{Arkhangelsky, Athey, Hirshberg, Imbens, and
  Wager}{Arkhangelsky et~al.}{2021}]{arkhangelsky_et_al_2021}
\textsc{Arkhangelsky, D., S.~Athey, D.~Hirshberg, G.~Imbens, and S.~Wager}
  (2021): \enquote{Synthetic Difference-in-Differences,} \emph{American
  Economic Review}, 111, 4088--4118.

\bibitem[\protect\citeauthoryear{Athey and Imbens}{Athey and
  Imbens}{2006}]{athey_imbens_2006}
\textsc{Athey, S. and G.~Imbens} (2006): \enquote{Identification and Inference
  in Nonlinear Difference-in-Differences Models,} \emph{Econometrica}, 74,
  431--497.

\bibitem[\protect\citeauthoryear{Athey and Imbens}{Athey and
  Imbens}{2015}]{athey_imbens_2015}
---\hspace{-.1pt}---\hspace{-.1pt}--- (2015): \enquote{Recursive partitioning
  for heterogeneous causal effects,} \emph{Proceedings of the National Academy
  of Sciences}, 113, 7353 -- 7360.

\bibitem[\protect\citeauthoryear{Athey and Imbens}{Athey and
  Imbens}{2017}]{athey_imbens_2017}
---\hspace{-.1pt}---\hspace{-.1pt}--- (2017): \enquote{The State of Applied
  Econometrics: Causality and Policy Evaluation,} \emph{Journal of Economic
  Perspectives}, 31, 3--32.

\bibitem[\protect\citeauthoryear{Athey and Imbens}{Athey and
  Imbens}{2019}]{athey_imbens_2019}
---\hspace{-.1pt}---\hspace{-.1pt}--- (2019): \enquote{Machine Learning Methods
  That Economists Should Know About,} \emph{Annual Review of Economics}, 11,
  685--725.

\bibitem[\protect\citeauthoryear{Bj{\"o}rklund and Moffitt}{Bj{\"o}rklund and
  Moffitt}{1987}]{bjorklund1987}
\textsc{Bj{\"o}rklund, A. and R.~Moffitt} (1987): \enquote{The estimation of
  wage gains and welfare gains in self-selection models,} \emph{The Review of
  Economics and Statistics}, 42--49.

\bibitem[\protect\citeauthoryear{Blandhol, Bonney, Mogstad, and
  Torgovitsky}{Blandhol et~al.}{2022}]{blandhol2022tsls}
\textsc{Blandhol, C., J.~Bonney, M.~Mogstad, and A.~Torgovitsky} (2022):
  \enquote{When is TSLS actually late?} Tech. rep., National Bureau of Economic
  Research.

\bibitem[\protect\citeauthoryear{Boyd, Goldhaber, Lankford, and Wyckoff}{Boyd
  et~al.}{2007}]{boyd_2007}
\textsc{Boyd, D., D.~Goldhaber, H.~Lankford, and J.~Wyckoff} (2007):
  \enquote{The Effect of Certification and Preparation on Teacher Quality,}
  \emph{The Future of Children}, 17, 45--68.

\bibitem[\protect\citeauthoryear{Breiman, Friedman, Stone, and Olshen}{Breiman
  et~al.}{1984}]{breiman1984}
\textsc{Breiman, L., J.~Friedman, C.~Stone, and R.~Olshen} (1984):
  \emph{Classification and Regression Trees}, Chapman and Hall/CRC.

\bibitem[\protect\citeauthoryear{Card}{Card}{1990}]{card1990}
\textsc{Card, D.} (1990): \enquote{The Impact of the Mariel Boatlift on the
  Miami Labor Market,} \emph{ILR Review}, 43, 245--257.

\bibitem[\protect\citeauthoryear{Card}{Card}{1993}]{card1993using}
---\hspace{-.1pt}---\hspace{-.1pt}--- (1993): \enquote{Using Geographic
  Variation in College Proximity to Estimate the Return to Schooling,} Working
  Paper 4483, National Bureau of Economic Research.

\bibitem[\protect\citeauthoryear{Card and Krueger}{Card and
  Krueger}{1992}]{card_krueger_1992}
\textsc{Card, D. and A.~Krueger} (1992): \enquote{Does school quality matter?
  Returns to education and the characteristics of public schools in the United
  States,} \emph{Journal of political Economy}, 100, 1--40.

\bibitem[\protect\citeauthoryear{Chamberlain}{Chamberlain}{1987}]{chamberlain_1987}
\textsc{Chamberlain, G.} (1987): \enquote{Asymptotic efficiency in estimation
  with conditional moment restrictions,} \emph{Journal of Econometrics}, 34,
  305--334.

\bibitem[\protect\citeauthoryear{Chernozhukov, Chetverikov, Demirer, Duflo,
  Hansen, Newey, and Robins}{Chernozhukov et~al.}{2018}]{cherno2018}
\textsc{Chernozhukov, V., D.~Chetverikov, M.~Demirer, E.~Duflo, C.~Hansen,
  W.~Newey, and J.~Robins} (2018): \enquote{{Double/debiased machine learning
  for treatment and structural parameters},} \emph{The Econometrics Journal},
  21, C1--C68.

\bibitem[\protect\citeauthoryear{Chernozhukov, Imbens, and Newey}{Chernozhukov
  et~al.}{2007}]{cherno_et_al_2007}
\textsc{Chernozhukov, V., G.~Imbens, and W.~Newey} (2007):
  \enquote{Instrumental variable estimation of nonseparable models,}
  \emph{Journal of Econometrics}, 139, 4--14.

\bibitem[\protect\citeauthoryear{Cosslett}{Cosslett}{1981}]{cosslett_1981}
\textsc{Cosslett, S.} (1981): \enquote{Maximum Likelihood Estimator for
  Choice-Based Samples,} \emph{Econometrica}, 49, 1289--1316.

\bibitem[\protect\citeauthoryear{de~Chaisemartin and
  D'Haultfoeuille}{de~Chaisemartin and
  D'Haultfoeuille}{2022}]{dechaisemartin2021twoway}
\textsc{de~Chaisemartin, C. and X.~D'Haultfoeuille} (2022): \enquote{Two-Way
  Fixed Effects and Differences-in-Differences with Heterogeneous Treatment
  Effects: A Survey,} \emph{to appear in Econometrics Journal}.

\bibitem[\protect\citeauthoryear{Deaton}{Deaton}{2010}]{deaton2010}
\textsc{Deaton, A.} (2010): \enquote{Instruments, randomization, and learning
  about development,} \emph{Journal of economic literature}, 48, 424--455.

\bibitem[\protect\citeauthoryear{Donald, Imbens, and Newey}{Donald
  et~al.}{2003}]{donald_imbens_newey_2003}
\textsc{Donald, S., G.~Imbens, and W.~Newey} (2003): \enquote{Empirical
  likelihood estimation and consistent tests with conditional moment
  restrictions,} \emph{Journal of Econometrics}, 117, 55--93.

\bibitem[\protect\citeauthoryear{Gale and Shapley}{Gale and
  Shapley}{1962}]{gale_shapley_1962}
\textsc{Gale, D. and L.~Shapley} (1962): \enquote{College Admissions and the
  Stability of Marriage,} \emph{The American Mathematical Monthly}, 69, 9--15.

\bibitem[\protect\citeauthoryear{Gelman and Imbens}{Gelman and
  Imbens}{2019}]{gelman_imbens_2019}
\textsc{Gelman, A. and G.~Imbens} (2019): \enquote{Why High-Order Polynomials
  Should Not Be Used in Regression Discontinuity Designs,} \emph{Journal of
  Business \& Economic Statistics}, 37, 447--456.

\bibitem[\protect\citeauthoryear{Goldhaber}{Goldhaber}{2007}]{goldhaber2007}
\textsc{Goldhaber, D.} (2007): \enquote{Everyone’s Doing It, But What Does
  Teacher Testing Tell Us About Teacher Effectiveness?} \emph{The Journal of
  Human Resources}, XLII, 765 -- 794.

\bibitem[\protect\citeauthoryear{Guggenberger}{Guggenberger}{2008}]{guggenberger_2008}
\textsc{Guggenberger, P.} (2008): \enquote{Finite Sample Evidence Suggesting a
  Heavy Tail Problem of the Generalized Empirical Likelihood Estimator,}
  \emph{Econometric Reviews}, 27, 526--541.

\bibitem[\protect\citeauthoryear{Haavelmo}{Haavelmo}{1944}]{haavelmo1944}
\textsc{Haavelmo, T.} (1944): \enquote{The probability approach in
  econometrics,} \emph{Econometrica: Journal of the Econometric Society},
  iii--115.

\bibitem[\protect\citeauthoryear{Hajek}{Hajek}{1971}]{hajek_1971}
\textsc{Hajek, J.} (1971): \enquote{Comment on a paper by D. Basu,}
  \emph{Foundations of Statistical Inference}, 236.

\bibitem[\protect\citeauthoryear{Hansen}{Hansen}{1982}]{hansen_1982}
\textsc{Hansen, L.~P.} (1982): \enquote{Large Sample Properties of Generalized
  Method of Moments Estimators,} \emph{Econometrica}, 50, 1029--1054.

\bibitem[\protect\citeauthoryear{Heckman}{Heckman}{1990}]{heckman1990}
\textsc{Heckman, J.} (1990): \enquote{Varieties of selection bias,} \emph{The
  American Economic Review}, 80, 313--318.

\bibitem[\protect\citeauthoryear{Hellerstein and Imbens}{Hellerstein and
  Imbens}{1999}]{hellerstein_imbens_1999}
\textsc{Hellerstein, J. and G.~Imbens} (1999): \enquote{Imposing Moment
  Restrictions from Auxiliary Data by Weighting,} \emph{The Review of Economics
  and Statistics}, 81, 1--14.

\bibitem[\protect\citeauthoryear{Hirano and Imbens}{Hirano and
  Imbens}{2004}]{hirano_imbens_2004}
\textsc{Hirano, K. and G.~Imbens} (2004): \emph{The Propensity Score with
  Continuous Treatments}, John Wiley \& Sons, Ltd, chap.~7, 73--84.

\bibitem[\protect\citeauthoryear{Hirano, Imbens, and Ridder}{Hirano
  et~al.}{2003}]{hirano_et_al_2003}
\textsc{Hirano, K., G.~Imbens, and G.~Ridder} (2003): \enquote{Efficient
  Estimation of Average Treatment Effects Using the Estimated Propensity
  Score,} \emph{Econometrica}, 71, 1161--1189.

\bibitem[\protect\citeauthoryear{Horvitz and Thompson}{Horvitz and
  Thompson}{1952}]{horvitz_thompson_1952}
\textsc{Horvitz, D. and D.~Thompson} (1952): \enquote{A Generalization of
  Sampling Without Replacement From a Finite Universe,} \emph{Journal of the
  American Statistical Association}, 47, 663--685.

\bibitem[\protect\citeauthoryear{Imbens}{Imbens}{1992}]{imbens1992}
\textsc{Imbens, G.} (1992): \enquote{An Efficient Method of Moments Estimator
  for Discrete Choice Models With Choice-Based Sampling,} \emph{Econometrica},
  60, 1187--1214.

\bibitem[\protect\citeauthoryear{Imbens}{Imbens}{1997}]{imbens1997}
---\hspace{-.1pt}---\hspace{-.1pt}--- (1997): \enquote{{One-Step Estimators for
  Over-Identified Generalized Method of Moments Models},} \emph{The Review of
  Economic Studies}, 64, 359--383.

\bibitem[\protect\citeauthoryear{Imbens}{Imbens}{2000}]{imbens2000}
---\hspace{-.1pt}---\hspace{-.1pt}--- (2000): \enquote{The Role of the
  Propensity Score in Estimating Dose-Response Functions,} \emph{Biometrika},
  87, 706--710.

\bibitem[\protect\citeauthoryear{Imbens}{Imbens}{2002}]{imbens2002}
---\hspace{-.1pt}---\hspace{-.1pt}--- (2002): \enquote{Generalized Method of
  Moments and Empirical Likelihood,} \emph{Journal of Business \& Economic
  Statistics}, 20, 493--506.

\bibitem[\protect\citeauthoryear{Imbens}{Imbens}{2010}]{imbens2010}
---\hspace{-.1pt}---\hspace{-.1pt}--- (2010): \enquote{Better LATE Than
  Nothing: Some Comments on Deaton (2009) and Heckman and Urzua (2009),}
  \emph{Journal of Economic Literature}, 48, 399--423.

\bibitem[\protect\citeauthoryear{Imbens and Angrist}{Imbens and
  Angrist}{1994}]{imbens_angrist_1994}
\textsc{Imbens, G. and J.~Angrist} (1994): \enquote{Identification and
  Estimation of Local Average Treatment Effects,} \emph{Econometrica}, 62,
  467--475.

\bibitem[\protect\citeauthoryear{Imbens and Kalyanaraman}{Imbens and
  Kalyanaraman}{2011}]{imbens_kalyanaraman_2011}
\textsc{Imbens, G. and K.~Kalyanaraman} (2011): \enquote{{Optimal Bandwidth
  Choice for the Regression Discontinuity Estimator},} \emph{The Review of
  Economic Studies}, 79, 933--959.

\bibitem[\protect\citeauthoryear{Imbens and Lancaster}{Imbens and
  Lancaster}{1994}]{imbens_lancaster_1994}
\textsc{Imbens, G. and T.~Lancaster} (1994): \enquote{Combining Micro and Macro
  Data in Microeconometric Models,} \emph{The Review of Economic Studies}, 61,
  655--680.

\bibitem[\protect\citeauthoryear{Imbens and Lancaster}{Imbens and
  Lancaster}{1996}]{imbens_lancaster_1996}
---\hspace{-.1pt}---\hspace{-.1pt}--- (1996): \enquote{Efficient estimation and
  stratified sampling,} \emph{Journal of Econometrics}, 74, 289--318.

\bibitem[\protect\citeauthoryear{Imbens and Lemieux}{Imbens and
  Lemieux}{2008}]{imbens_lemieux_2008}
\textsc{Imbens, G. and T.~Lemieux} (2008): \enquote{Regression discontinuity
  designs: A guide to practice,} \emph{Journal of Econometrics}, 142, 615--635.

\bibitem[\protect\citeauthoryear{Imbens and Newey}{Imbens and
  Newey}{2009}]{imbens_newey_2009}
\textsc{Imbens, G. and W.~Newey} (2009): \enquote{Identification and Estimation
  of Triangular Simultaneous Equations Models Without Additivity,}
  \emph{Econometrica}, 77, 1481--1512.

\bibitem[\protect\citeauthoryear{Imbens and Rubin}{Imbens and
  Rubin}{1997}]{imbens_rubin_1997}
\textsc{Imbens, G. and D.~Rubin} (1997): \enquote{Estimating outcome
  distributions for compliers in instrumental variables models,} \emph{The
  Review of Economic Studies}, 64, 555--574.

\bibitem[\protect\citeauthoryear{Imbens and Rubin}{Imbens and
  Rubin}{2015}]{imbens_rubin_2015}
---\hspace{-.1pt}---\hspace{-.1pt}--- (2015): \emph{Causal Inference for
  Statistics, Social, and Biomedical Sciences: An Introduction}, Cambridge
  University Press.

\bibitem[\protect\citeauthoryear{Imbens, Spady, and Johnson}{Imbens
  et~al.}{1998}]{imbens_johnson_spady_1998}
\textsc{Imbens, G., R.~Spady, and P.~Johnson} (1998): \enquote{Information
  Theoretic Approaches to Inference in Moment Condition Models,}
  \emph{Econometrica}, 66, 333--357.

\bibitem[\protect\citeauthoryear{KVA}{KVA}{2021}]{KVA_Nobel_2021}
\textsc{KVA} (2021): \enquote{Answering Causal Questions Using Observational
  Data,} \emph{Scientific background on the Sveriges Riksbank Prize in Economic
  Sciences in Memory of Alfred Nobel by the Kungl Vetenskapsakademien (KVA)
  [The Royal Swedish Academy Of Sciences]}.

\bibitem[\protect\citeauthoryear{Maddala}{Maddala}{1983}]{maddala1983}
\textsc{Maddala, G.} (1983): \emph{Limited-dependent and qualitative variables
  in econometrics}, 3, Cambridge university press.

\bibitem[\protect\citeauthoryear{Manski}{Manski}{2003}]{manski2003}
\textsc{Manski, C.} (2003): \emph{Partial identification of probability
  distributions}, vol.~5, Springer.

\bibitem[\protect\citeauthoryear{Neyman}{Neyman}{1923}]{neyman1923}
\textsc{Neyman, J.} (1923): \enquote{On the application of probability theory
  to agricultural experiments. Essay on principles,} \emph{Ann. Agricultural
  Sciences}, 1--51.

\bibitem[\protect\citeauthoryear{Owen}{Owen}{1990}]{owen_1990}
\textsc{Owen, A.} (1990): \enquote{Empirical Likelihood Ratio Confidence
  Regions,} \emph{The Annals of Statistics}, 18, 90--120.

\bibitem[\protect\citeauthoryear{Reiers{\"o}l}{Reiers{\"o}l}{1945}]{Reiersl1945ConfluenceAB}
\textsc{Reiers{\"o}l, O.} (1945): \enquote{Confluence analysis by means of
  instrumental sets of variables,} Ph.D. thesis, Almqvist \& Wiksell.

\bibitem[\protect\citeauthoryear{Rosenbaum and Rubin}{Rosenbaum and
  Rubin}{1983}]{rosenbaum_rubin_1983}
\textsc{Rosenbaum, P. and D.~Rubin} (1983): \enquote{The Central Role of the
  Propensity Score in Observational Studies for Causal Effects,}
  \emph{Biometrika}, 70, 41--55.

\bibitem[\protect\citeauthoryear{Roth, Sant'Anna, Bilinski, and Poe}{Roth
  et~al.}{2023}]{billinski_et_al_2022}
\textsc{Roth, J., P.~Sant'Anna, A.~Bilinski, and J.~Poe} (2023):
  \enquote{What's Trending in Difference-in-Differences? A Synthesis of the
  Recent Econometrics Literature,} \emph{Journal of Econometrics}, 235,
  2218--2244.

\bibitem[\protect\citeauthoryear{Rubin}{Rubin}{1974}]{rubin1974estimating}
\textsc{Rubin, D.} (1974): \enquote{Estimating causal effects of treatments in
  randomized and nonrandomized studies,} \emph{Journal of Educational
  Psychology}, 66, 688--701.

\bibitem[\protect\citeauthoryear{Rubin}{Rubin}{1978}]{rubin1978}
---\hspace{-.1pt}---\hspace{-.1pt}--- (1978): \enquote{Bayesian inference for
  causal effects: The role of randomization,} \emph{The Annals of statistics},
  34--58.

\bibitem[\protect\citeauthoryear{Rubin}{Rubin}{1990}]{rubin1990}
---\hspace{-.1pt}---\hspace{-.1pt}--- (1990): \enquote{{[On the Application of
  Probability Theory to Agricultural Experiments. Essay on Principles. Section
  9.] Comment: Neyman (1923) and Causal Inference in Experiments and
  Observational Studies},} \emph{Statistical Science}, 5, 472 -- 480.

\bibitem[\protect\citeauthoryear{Snow}{Snow}{1855}]{snow1855}
\textsc{Snow, J.} (1855): \emph{On the Mode of Communication of Cholera, Second
  Edition}, John Churchill, New Burlington Street.

\bibitem[\protect\citeauthoryear{Tinbergen}{Tinbergen}{1930}]{tinbergen1930}
\textsc{Tinbergen, J.} (1930): \enquote{Determination and interpretation of
  supply curves: an example,} \emph{Zeitschrift fur Nationalokonomie}, 1,
  669--679.

\bibitem[\protect\citeauthoryear{Vytlacil}{Vytlacil}{2002}]{vytlacil2002}
\textsc{Vytlacil, E.} (2002): \enquote{Independence, monotonicity, and latent
  index models: An equivalence result,} \emph{Econometrica}, 70, 331--341.

\bibitem[\protect\citeauthoryear{Wooldridge}{Wooldridge}{2010}]{Wooldridge2010}
\textsc{Wooldridge, J.} (2010): \emph{Econometric Analysis of Cross Section and
  Panel Data, Second Edition}, MIT Press.

\bibitem[\protect\citeauthoryear{Wright}{Wright}{1928}]{wright_1928}
\textsc{Wright, P.~G.} (1928): \emph{Appendix to \textit{The Tariff on Animal
  and Vegetable oils}}, The Macmillan Company, New York.

\end{thebibliography}
